\documentclass[aps,prx,reprint,amsmath,amssymb,superscriptaddress]{revtex4-2}

\usepackage{subfiles}

\usepackage{graphicx}
\usepackage[export]{adjustbox}
\usepackage{dcolumn}
\usepackage{bm}
\usepackage{amsmath}
\usepackage{amssymb}
\usepackage{amsthm}
\usepackage{mathcomp}
\usepackage{mathdots}
\usepackage{comment}
\usepackage{titlesec}
\usepackage{url}
\usepackage{natbib}
\usepackage{here}

\begin{document}
{
\titleformat{\section}{\fontsize{12pt}{12pt}\selectfont\bfseries}{}{0pt}{}
\titlespacing{\section}{0pt}{6pt}{3pt}
\titleformat{\subsection}{\fontsize{11pt}{11pt}\selectfont\bfseries}{}{0pt}{}
\titlespacing{\subsection}{0pt}{6pt}{3pt}

\title{Higher-order exceptional points unveiled by nilpotence and mathematical induction}



\author{Kenta Takata}
\email[]{kenta.takata@ntt-research.com}
\altaffiliation[Currently at ]{NTT Research, Inc., Sunnyvale, California 94085, USA}
\affiliation{Nanophotonics Center, NTT, Inc., Atsugi, Kanagawa 243-0198, Japan}
\affiliation{NTT Basic Research Laboratories, NTT, Inc., Atsugi, Kanagawa 243-0198, Japan}

\author{Adam Mock}
\altaffiliation[Currently at ]{Central Michigan University, Mount Pleasant, Michigan 48859, USA}
\affiliation{NTT Basic Research Laboratories, NTT, Inc., Atsugi, Kanagawa 243-0198, Japan}

\author{Masaya Notomi}
\affiliation{Nanophotonics Center, NTT, Inc., Atsugi, Kanagawa 243-0198, Japan}
\affiliation{NTT Basic Research Laboratories, NTT, Inc., Atsugi, Kanagawa 243-0198, Japan}
\affiliation{Department of Physics, Institute of Science Tokyo, Ookayama, Meguro 152-8550, Japan}

\author{Akihiko Shinya}
\affiliation{Nanophotonics Center, NTT, Inc., Atsugi, Kanagawa 243-0198, Japan}
\affiliation{NTT Basic Research Laboratories, NTT, Inc., Atsugi, Kanagawa 243-0198, Japan}


\date{\today}

\begin{abstract}
	Non-Hermitian systems can have peculiar degeneracies of eigenstates called exceptional points (EPs). An EP of $n$ degenerate states is said to have order $n$, and higher-order EPs (HEPs) with $n \ge 3$ exhibit intrinsic order-scaling responses potentially applied to superior sensing and state control. However, traditional eigenvalue-based searches for HEPs are facing fundamental limitations in terms of complexity and implementation. Here, we propose a design paradigm for HEPs based on a simple property for matrices termed nilpotence and concise inductive procedure. The nilpotence guarantees a HEP with desired order and helps divide the problem. Our inductive scheme repeatedly extends a system and doubles its EP order, starting with a known design. Based on the nilpotence, we systematically design photonic cavity arrays operating at chiral, passive, and active HEPs with $n = 3, 6, 7$ and show their peculiar directional radiation, induced transparency, and enhanced transmittance and spontaneous emission, respectively. We inductively find lattice systems with diverging EP order originating from a well-known $2 \times 2$ parity-time-symmetric Hamiltonian. We also extend the active HEP system with $n = 7$ to another with $n = 14$ and have further magnified responses. Our work pushes the investigation and application of HEPs to previously unexplored regimes in various physical systems.
\end{abstract}


\maketitle



\section{Introduction}
The principle of superposition holds in a large part of our physical world. When considering such a linear system, one can model its response with a matrix representation. The eigenvalue equation analysis of this system matrix tells a lot about the steady behavior and stability of, such as optical, electrical, and vibrational waves \cite{Georgi1993}. 

Every practical system is also more or less open. Thus, one usually focuses on finite degrees of freedom in it. Their interaction with the exterior environment may be simplified as locally energy-nonconserving processes such as gain and loss. Correspondingly, the system matrix, or effective Hamiltonian $\hat{\rm H}$, becomes non-Hermitian, i.e. $\hat{\rm H} \ne \hat{\rm H}^{\dagger}$, where $\hat{\rm H}^{\dagger}$ is the Hermitian conjugate of $\hat{\rm H}$ \cite{Moiseyev2011}.

It is within the past few decades that systems with engineered non-Hermiticity has gained prominence \cite{Bender1998,Makris2008,Miri2019,Ozdemir2019par,Parto2021}. One of their peculiar features is the exceptional point (EP) \cite{Kato1995,Berry2004,Heiss2012}, where not only the eigenvalues but also the eigenstates coalesce. This degeneracy accompanies a phase transition and singularity because of the eigenvalue landscape described by complex multi-valued radical functions. As a result, many EP-based phenomena potential for applications have been revealed and elaborated. These include chiral, directional, and topological responses \cite{Dembowski2001,Lin2011,Feng2013,Lawrence2014,Peng2016,Leykam2017,Gao2019,Wang2020,Tang2020,delPino2022,Patil2022}, control of lasing modes \cite{Feng2014sin,Hodaei2014,Brandstetter2014,Miao2016,Liao2023}, spectral and spatial anomaly of radiation decay \cite{Yoo2011,Makris2015,Lin2016,Pick2017gen,Takata2021,Yulaev2022}, enhanced sensitivity to perturbation \cite{Wiersig2014,Hodaei2017enh,Chen2017exc,Park2020,Kononchuk2022} and nonadiabatic mode conversion \cite{Doppler2016,Schumer2022}.

The coalescence of $n$ eigenstates is termed an $n$th-order EP (EP$n$). Higher-order EPs (HEPs) with $n \ge 3$ have been explored actively \cite{Tang2020,Patil2022,Hodaei2017enh,Lin2016,Graefe2008,Heiss2008,Nada2017,XiaoYX2019,XiaoZ2019,Zhong2020,Yang2021,Mandal2021,Delplace2021,Sayyad2022,Kaltsas2022,Kullig2023}, because they have intrinsic properties scaling with the order $n$ potential for applications. For example, they can exhibit an $n$th-root and thus larger eigenvalue variation with perturbation \cite{Heiss2008,Hodaei2017enh}, a narrower and higher spectral lineshape along the $n$th power of the Lorentzian function (LF$n$) \cite{Lin2016,Pick2017gen}, and $n$ intersecting Riemann surfaces allowing more diverse state control by winding around EPs in parameter space \cite{Tang2020,Patil2022}.

However, persistent challenges remain in the design and demonstration of HEPs. Traditionally, these studies have been based on the search for multiple-root eigenvalues \cite{Berry2004,Heiss2012}. As the desired $n$ increases, we shortly face the difficulty that neither the eigenvalues nor their degeneracy conditions are given in simple analytic forms. It could be eased by considering the coefficients and discriminant of the characteristic equation, and such approaches may be useful to discuss how non-Hermitian symmetry reduces the constraints to be satisfied \cite{Delplace2021,Mandal2021,Sayyad2022}. However, the number of degenerate eigenvalues (algebraic multiplicity) is generally distinct from the order of an EP (geometrical multiplicity) \cite{Nada2017, Miyake2022}. In other words, conventional eigenvalue-based schemes only deal with necessary\textemdash but not sufficient\textemdash conditions for EPs. They are hence ineffective at the scenarios where HEPs are found to be absent or have reduced order, such as diabolic points and multiple separate EPs. In these cases, different eigenstates share a single eigenvalue, and the impact of HEPs is lost or significantly compromised.

Another issue is that most known HEP designs suffer the complexity or difficulty in implementation. One powerful and trivially extensible strategy is the use of unidirectional couplings \cite{XiaoYX2019,Zhong2020,Yang2021,Kullig2023}. However, it needs cavities or waveguides coupled by effectively nonreciprocal wave injections, where any impact of reverse processes, such as scattering and reflection, has to be strictly suppressed. The other strategy is finding particular solutions with the help of symmetry as mentioned above \cite{Delplace2021,Mandal2021,Sayyad2022,Kaltsas2022}, including one valid for any $n$ \cite{Graefe2008}. Unfortunately, it tends to require different irrational parameters as $n$ rises. Each of them needs an extreme accuracy, drastically complicating any of such system implementation at scale. To our knowledge, there has been no HEP reported based on a balance among parameters with equal or even a few integer levels of magnitude for $n \ge 5$.

Here we change the paradigm of designing HEPs by introducing two powerful algebraic approaches. One is the direct application of nilpotence of matrices, $\hat{\rm H}^{m} = \hat{\rm O}$ with $\hat{\rm O}$ being zero matrix and $m \ge 2$ a positive integer, to practical designs of HEPs. The nilpotence always guarantees a HEP with desired order as a sufficient condition and especially covers all possible cases for maximally degenerate EPs. In addition, it enables us to discuss systematically how the EP order can be reduced. Its straightforward corollaries are identical to the foregoing necessary conditions and help simplify the problem. The other tool developed is an inductive scheme to repeatedly double the EP order by combining two inverted copies of systems with maximally degenerate EPs. We prove this as a theorem valid for a certain class of symmetric matrices. Starting from a system with an EP$N$ ($N \ge 2$) and applying the process $h$ times ($h \in \mathbb{N}$), we can directly obtain a series of effective Hamiltonians with HEPs having order up to $N2^{h}$, which diverges as $h \rightarrow \infty$.

We demonstrate photonic HEPs with unique features by applying our schemes. First, we determine all the $3 \times 3$ symmetric matrices having an EP3 with 3-vector chirality. We numerically find one of them in coupled photonic crystal cavities, which exhibits significantly directional radiation. Next, we design a passive EP6 and a \textit{minimally} active EP7 in systems of reciprocally coupled ring cavities and waveguides with uniform parameter magnitude. The EP6 here behaves as a dark mode for inputs from the waveguides. Instead, the transmission along one of the waveguides indicates a peculiar variant of induced transparency effects \cite{Wang2020,Smith2004,Waks2006}. The EP7 is obtained just by compensating for a part of outcoupling loss of a single cavity with an active medium, and its eigenmode does not have any net gain. However, its spectral responses exhibit particularly narrowed lineshapes and peak intensity amplifications. These passive and active HEPs also have contrasting response dynamics. Finally, we inductively derive HEP systems with both exponentially scaling $n$ and implementability from a well-known $2 \times 2$ parity-time-symmetric (PT-symmetric) Hamiltonian. We also extend the above-mentioned photonic system with the active EP7 to create an EP14 that exhibits further enhanced responses.
\begin{figure}[t]
	\includegraphics[width=\linewidth]{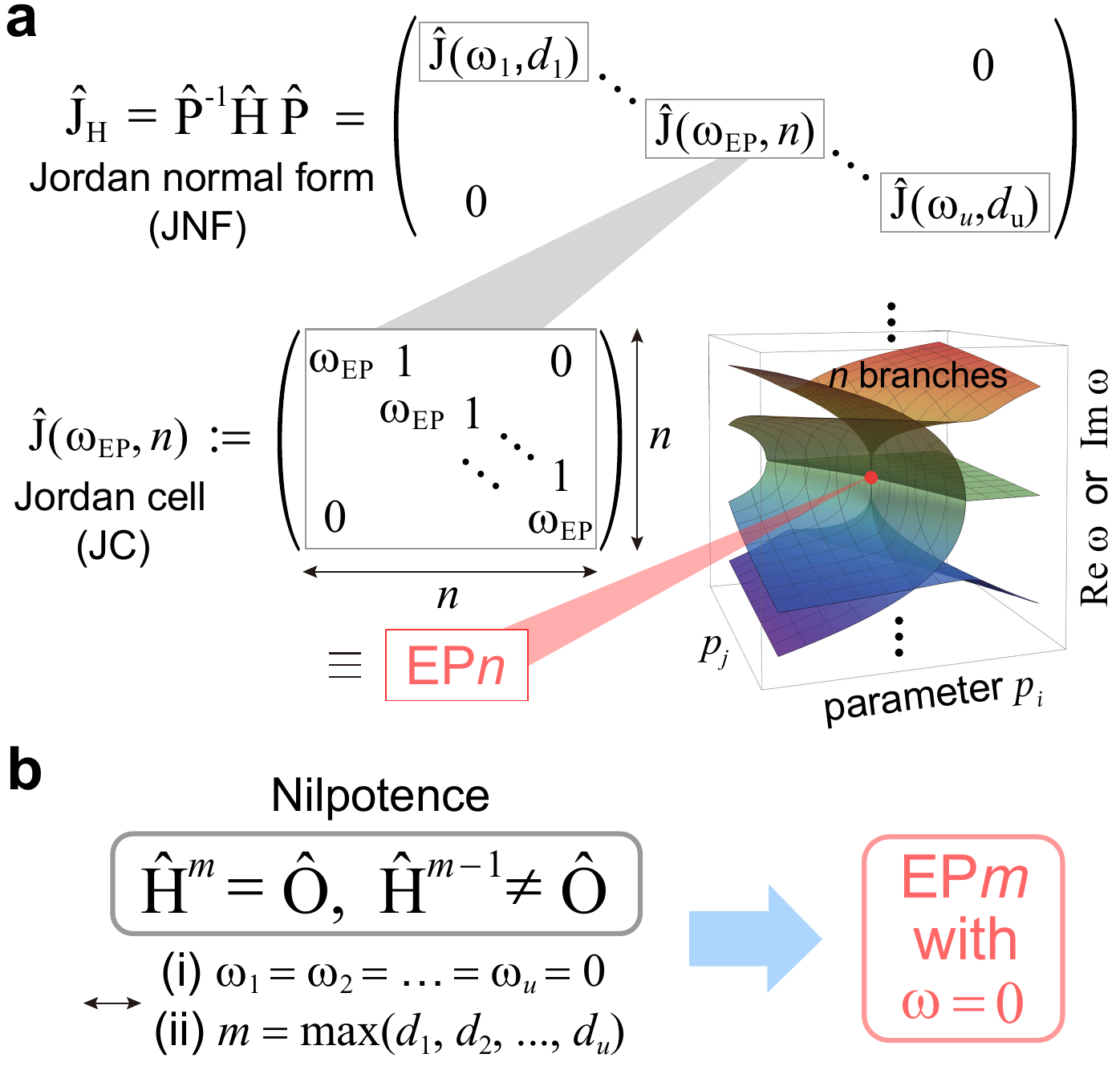}
	\caption{\label{fig:JNF_nilpotence} EPs, JNF, and nilpotence of a system matrix $\hat{\rm H}$. (a) The JNF $\hat{\rm J}_{\rm H}$ is a block-diagonal similarity transformation of $\hat{\rm H}$ comprising specific semi-diagonal square matrices called JCs. Here, $\omega_j$ and $d_j$ ($j = 1, 2, \dots, u \in \mathbb{N}$) are the eigenvalue and dimension (size) of the $j$th JC, which is denoted as $\hat{\rm J}(\omega_j, d_j)$. A JC $\hat{\rm J}(\omega_{\rm EP}, n)$ in $\hat{\rm J}_{\rm H}$ means that $\hat{\rm H}$ has an EP$n$ with an eigenvalue $\omega = \omega_{\rm EP}$. (b) When $\hat{\rm H}$ has nilpotence for an index of $m$, it follows that all eigenvalues of $\hat{\rm H}$ are null, and that $m$ is the largest dimension of $\hat{\rm H}$'s JCs. Therefore, it is a sufficient condition for the existence of an EP$m$ with $\omega = 0$ of $\hat{\rm H}$, and this can be used for the design of HEP systems.}
\end{figure}

\section{Results}
\subsection{Principles of design by nilpotence}
EPs appear in the linear eigenvalue problem
\begin{equation}
	\hat{\rm H}\bm{\alpha} = \omega \bm{\alpha}, \label{eigenvalue_eq}
\end{equation}
where $\hat{\rm H} \in \mathbb{C}^{N \times N} \ (N \in \mathbb{N})$, $\bm{\alpha} \in \mathbb{C}^{N}$ is the eigenvector, and $\omega \in \mathbb{C}$ the eigenvalue. The system matrix $\hat{\rm H}$ can be the effective Hamiltonian of a Schr\"odinger-type equation or can describe a discrete process like the scattering matrix. Thus, Eq. (\ref{eigenvalue_eq}) is widely used in photonics, plasmonics, phononics, electronics, condensed-matter physics, and so on.  In linear algebra, the existence of an EP$n$ means that the Jordan normal (canonical) form of $\hat{\rm H}$ has an ${n \times n}$ specific non-diagonalizable block component termed Jordan cell (JC) \cite{Graefe2008,Heiss2008}, as illustrated in Fig. \ref{fig:JNF_nilpotence}(a). Thus, examining such a condition mathematically may lead to the simplest and most essential way to design EPs. The basics related to the Jordan normal form (JNF) are summarized in Supplementary Information (SI).

The nilpotence of $\hat{\rm H}$ with the index of $m$ is defined as 
\begin{equation}
	\hat{\rm H}^m = \hat{\rm O}, \ \ \hat{\rm H}^{m - 1} \ne \hat{\rm O}, \label{nilpotence_eq}
\end{equation}
for $m \in \mathbb{N}$, $2 \le m \le N$. Equation (\ref{nilpotence_eq}) lets us trivially derive that all eigenvalues of nilpotent matrices are null \cite{Miyake2022}. By considering the JNF of such $\hat{\rm H}$, we can show that $m$ is identical to the largest order of $\hat{\rm H}$'s EPs. This is implied in a few theoretical studies on general perturbation responses of HEPs \cite{Wiersig2023,Schomerus2024}; see Methods for its proof. The nilpotence for an index of $m$ is hence sufficient to the existence of an EP$m$ with $\omega = 0$ and can hence be used directly for the design, as summarized in Fig. \ref{fig:JNF_nilpotence}(b). Particularly, the sufficiency turns into equivalence for $m = N$, namely systems with \textit{maximally} degenerate EPs, because a sole EP$N$ in the eigenspace is the only possibility that an $N \times N$ matrix has an EP$N$. Moreover, such a system with an arbitrary EP eigenvalue $\Omega_{\rm EP} \in \mathbb{C}$ can also be covered by $\hat{\rm H} + \Omega_{\rm EP} \hat{\rm I}_N$, where $\hat{\rm H}$ is nilpotent for $m = N$ and $\hat{\rm I}_N$ the $N \times N$ identity matrix, thereby restoring generality. 

Remarkably, $\hat{\rm H}^2, \hat{\rm H}^3, \dots, \hat{\rm H}^{m - 1}$ are always nilpotent when $\hat{\rm H}$ is so. As a result, we have
\begin{equation}
	{\rm tr} \, \hat{\rm H}^l = 0 \ (l = 1, 2, \dots, m - 1) \label{eq:trace_condition},
\end{equation}	
where ${\rm tr} \, \cdot$ denotes the trace, as necessary conditions for the nilpotence of $\{\hat{\rm H}^l\}$. Unless trivially satisfied, Eq. (\ref{eq:trace_condition}) provides $2 (m - 1)$ constraints on real parameters of $\hat{\rm H}$ and hence helps simplify the problem, as has been discussed especially for $m = N$ \cite{Sayyad2022}. The sufficient condition, Eq. (\ref{nilpotence_eq}), involves additional equality constraints for any $m$, including conditions for nulling the product of $\hat{\rm H}$'s eigenvalues (sufficient for ${\rm det} \, \hat{\rm H} = 0$, where ${\rm det} \, \cdot$ is the determinant). Moreover, its inequality part excludes any possibilities that the EP order becomes less than $m$. On the other hand, finding an $m$th-root eigenvalue generally does not mean an EP$m$. A good class of system matrices can miss maximally degenerate EPs, despite that their eigenvalues are all identical; see SI for an example.

\subsection{EP3s with 3-vector chirality}
The nilpotence provides practical designs of HEPs that have been beyond the reach of conventional searches. A good application is finding HEPs with multi-dimensional vector chirality of eigenstates, namely consistent and directional phase current among their elements \cite{Heiss2008}. For an EP2, the chiral eigenvector available $\propto (1, \pm i)^{\rm T}$ only gives an orientation in one dimension. In contrast, when $N \ge 3$ we can arrange basis components such as resonators in a two- or three-dimensional space. They enable a rich variety of the configuration and degree of chiral behavior. Designing such a HEP will draw out its pure potential, suppressing the impact of other eigenstates present otherwise with no and reversed chirality. It will pave the way for, for example, coupled lasers, active gratings, and metamaterials with controlled orbital angular momenta of light and directional emission.

However, this problem is open even for $n = 3$. Reference \cite{Heiss2008} discussed systems described by $\hat{\rm H}_0 + \lambda \hat{\rm H}_1$, where $\hat{\rm H}_0, \ \hat{\rm H}_1 \in \mathbb{R}^{3 \times 3}$ and $\lambda \in \mathbb{C}$ is a single complex parameter, in terms of their eigenvalues and eigenstates including an EP3. Despite that the work covered a broad class of system matrices, it concluded that chiral states with a uniform absolute amplitude and equally spaced phases, $(1, e^{\pm i 2 \pi/3}, e^{\pm i 4 \pi/3})^{\mathrm{T}}$, appeared rather off from the EP3. In addition, famous HEPs with non-Hermitian symmetry have eigenvector elements with non-uniform absolute values and fixed phase differences ($\pm \pi$) \cite{Graefe2008,Hodaei2017enh}.

We look for systems with chiral EP3s at $\omega = 0$ by considering generic $3 \times 3$ symmetric Hamiltonian $\hat{\rm H}_{\rm 3S} (= \hat{\rm H}_{\rm 3S}^{\rm T})$
\begin{equation}
	\hat{\rm H}_{\rm 3S} = \left(\,
	\begin{matrix}
		d & a & b \\
		a & e & c \\
		b & c & f
	\end{matrix}\right), \ \ \hat{\rm H}_{\rm 3S} \ne \hat{\rm O}, \label{eq:H3S}
\end{equation}
where $a, b, c, d, e, f \in \mathbb{C}$, so that it can be implemented by reciprocally coupled optical cavities with gain and loss. Steady responses of such systems are estimated by solving Eq. (\ref{eigenvalue_eq}) derived using the temporal coupled-mode theory (TCMT) \cite{Suh2004,Takata2022}. Here, $\omega$ denotes the complex eigen-detuning with reference to the average resonance frequency $\Omega_0 \in \mathbb{C}$ of the considered cavity modes, and $\bm{\alpha} = (\{\alpha_j \}) \ (j \in \mathbb{N})$ their complex amplitudes including phases. When the $(j, l)$ element of $\hat{\rm H}$ is symbolized as $[\hat{\rm H}]_{j l} \ (l \in \mathbb{N})$, $\delta_j := {\rm Re} \, [\hat{\rm H}]_{j j}$, $\gamma_j := {\rm Im} \, [\hat{\rm H}]_{j j}$, $\mu_{j l} := {\rm Re} \, [\hat{\rm H}]_{j l}$, and $\eta_{j l} := {\rm Im} \, [\hat{\rm H}]_{j l} \ (j \ne l)$ indicate the cavity frequency detuning, on-site gain or loss, evanescent coupling, and dissipative coupling, respectively (see Methods). Mathematically the same model will also be applicable to coupled waveguides.

We assume $\bm{\alpha}_{3 +} := (1, e^{i 2 \pi/3}, e^{i 4 \pi/3})^{\mathrm{T}}/\sqrt{3}$ as our chiral eigenstate to emerge as an EP3. Clearly, the dual state with opposite chirality is $\bm{\alpha}_{3 -} = \bm{\alpha}_{3 +}^{*}$. Although Eq. (\ref{eq:H3S}) has six free parameters, the eigenvalue equation $\hat{\rm H}_{\rm 3S} \bm{\alpha}_{3 +} = \bm{0}$ actually provides additional constraints
\begin{align}\label{eq:constraints_H3S}
	\begin{split}
		d =& -a \, e^{i \frac{2 \pi}{3}} - b \, e^{i \frac{4 \pi}{3}}, \\
		e =& -a \, e^{i \frac{4 \pi}{3}} - c \, e^{i \frac{2 \pi}{3}}, \\
		f =& -b \, e^{i \frac{2 \pi}{3}} - c \, e^{i \frac{4 \pi}{3}},
	\end{split}
\end{align}
which let us cut them down. The sufficient condition for the EP3 is Eq. (\ref{nilpotence_eq}) for the resultant $\hat{\rm H}_{\rm 3S}$ and $m = 3$. It can be specified just by matrix multiplication and simplified by Eq. (\ref{eq:trace_condition}), since $\{{\rm tr} \, \hat{\rm H}^l\}$ should appear as common factors in elements of $\hat{\rm H}_{\rm 3S}^3$. With Eq. (\ref{eq:constraints_H3S}) at hand, it is indeed straightforward to have $\hat{\rm H}_{\rm 3S}^3 = (a+b+c) \hat{\rm H}_{\rm 3S}^2 = ({\rm tr} \, \hat{\rm H}_{\rm 3S}) \hat{\rm H}_{\rm 3S}^2$. By putting $a + b + c = 0$ together, we explicitly obtain the two-parameter Hamiltonian 
\begin{equation}
	\hat{\rm H}_{\rm EP3S} = \left(\,
	\begin{matrix}
		e^{-i \frac{\pi}{3}} a + e^{i \frac{\pi}{3}} b & a & b \\
		a & \sqrt{3} i a + e^{i \frac{2\pi}{3}} b & -a-b \\
		b & -a-b & e^{i \frac{4\pi}{3}} a - \sqrt{3} i b
	\end{matrix}\,\right), \label{eq:H3S_EP3}
\end{equation}
which always satisfies $\hat{\rm H}_{\rm EP3S}^3 = \hat{\rm O}$ and has $\bm{\alpha} = \bm{\alpha}_{3 +}$ for $\omega = 0$.
\begin{figure*}[t]
	\includegraphics[width=\linewidth]{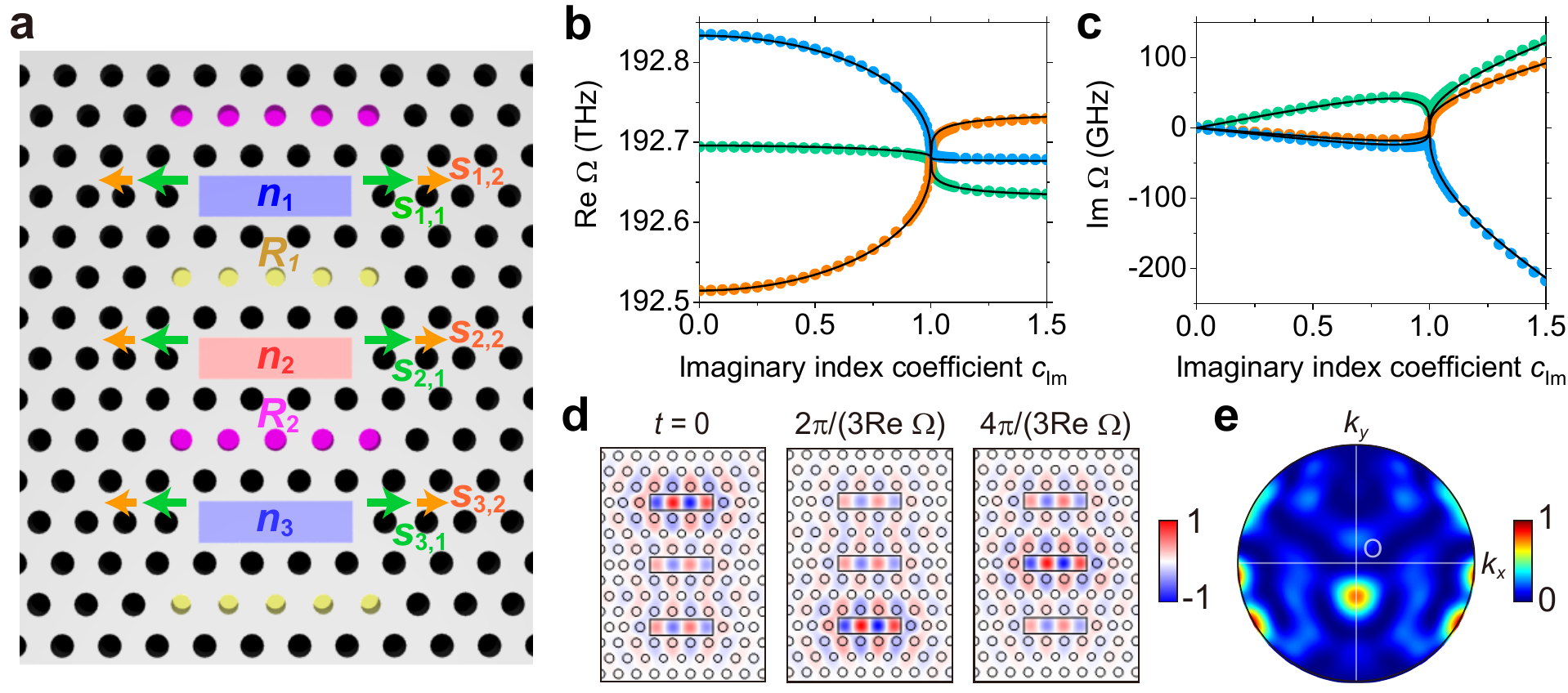}
	\caption{\label{fig:chiral_EP3} Demonstration of an EP3 with 3-vector chirality by using coupled PhC cavities in an air-suspended slab. (a) System schematic. Each cavity has an active region whose complex refractive index is $n_j$ ($j = 1, 2, 3$). Two couples of air holes on its ends are displaced outwards by $s_{j, 1}$ and $s_{j, 2}$ for the variation in its resonance frequency with its radiation loss restricted. The rows of holes colored yellow (magenta) have an radius of $R_1$ ($R_2$), so that the evanescent coupling between cavities 1 and 2 (2 and 3) is controlled. (b) Real and (c) imaginary parts of the eigenfrequencies $\Omega$ for the optimized system. Here, the imaginary parts of the active medium indices are varied as ${\rm Im} \, (n_1, n_2, n_3) = (0.00502, -0.00871, 0.00366) \times c_{\rm Im}$. The TE-like eigenmodes come close to the EP3 when $c_{\rm Im} = 1$. Markers: simulation result. Black lines: analytic fitting based on Eq. (\ref{eq:H3S_EP3}) for $(a, b) = (-111.1, 9.744) \ {\rm GHz}$ with linearly introduced on-site imaginary potential. (d) Snapshots of magnetic fields $H_z(x, y) \ (z = 0)$ for the eigenmode nearest to the EP3 indicating the chiral state $\bm{A}(t) \propto (1, e^{i 2 \pi/3}, e^{i 4 \pi/3})^{\mathrm{T}} e^{i \Omega_{\rm EP} t}$. (e) Reciprocal Fourier intensity of transverse electric fields $|E_y (k_x, k_y)|^2 \ (z = 0)$ within the light cone for the same state. The most intense peak residing on $k_x = 0, \ k_y < 0$ shows directional radiation with a polar angle of $-16.0\tcdegree$ from the $k_z$ axis.}
\end{figure*} 

By element-wise examination, we also have $\hat{\rm H}_{\rm EP3S}^2 = \hat{\rm O}$ (i.e. $m = 2$) when $b = e^{i \frac{2\pi}{3}} a$ ($c = e^{i \frac{4\pi}{3}} a$). Such a case should be excluded, because it indicates the reduction of EP order from $n = 3$ to $n = 2$. \textit{Any} instances in the remaining set, $\{\hat{\rm H}_{\rm EP3S} \ | \ b \ne \, e^{i \frac{2 \pi}{3}} a\}$, are shown to have the EP3 with $\bm{\alpha} = \bm{\alpha}_{3 +}$ and $\omega = 0$, and no other $3 \times 3$ symmetric matrices do. We have hence identified all such $\hat{\rm H}_{\rm 3S}$, resolving the issue left in Ref. \cite{Heiss2008} on the search for EP3s with chiral eigenmodes. Note that solving the same problem for $\bm{\alpha}_{3 -}$ is trivial. If we start conventionally with the eigenvalues of $\hat{\rm H}_{\rm 3S}$, we need to solve the characteristic equation, $\omega^3 - (d + e + f) \omega^2 - (a^2 + b^2 + c^2 - d e - d f - e f) \omega + (a^2 f + b^2 e + c^2 d - d e f - 2 a b c) = 0$, or to analyze the conditions for nulling its coefficients. Deriving parameterized eigenvectors enables us to examine their degeneracy. Meanwhile, those processes are generally complex, especially when $N$ becomes large.

We numerically demonstrate our chiral HEP with three evanescently coupled photonic crystal (PhC) nanocavities prepared in an InGaAsP-based slab (Fig. \ref{fig:chiral_EP3}(a)). Each cavity C$j$ $(j = 1, 2, 3)$ comprises a linear defect of three holes (L3) \cite{Akahane2003}, and its TE-like ground mode is taken as a basis. Two pairs of holes on its both ends are displaced outward from the lattice-matched position by $s_{j,1}$ and $s_{j,2}$, so that its resonance frequency $\Omega_0 + \delta_j$ is tuned around 193 THz ($\approx$ 1550 nm) with its radiation loss suppressed. It also has an active region whose refractive index has a finite imaginary part ${\rm Im} \ n_j$ for control of its gain and loss $\gamma_j$, enabled by buried heterostructure technologies \cite{Takata2021,Matsuo2010,Takata2017}. The row of five holes in the middle of C1 and C2 (C2 and C3) has a reduced radius $R_1$ ($R_2$) compared to that of the PhC, which adjusts the signed evanescent coupling $\mu_{12}$ ($\mu_{23}$) to be $a$ ($-a - b$) in Eq. (\ref{eq:H3S_EP3}). Note that the array has a next-nearest-neighbor coupling $\mu_{13} = b$ that is not negligible or directly controllable. Meanwhile, its radiation-based dissipative couplings, namely $\eta_{j l}$ for ${\rm Im} \ n_j = 0$, are insignificant, because the losses of coupled modes in this case are lower than $0.33$ GHz and their difference is within $0.07$ GHz. We seek for an EP3 based on $\hat{\rm H}_{\rm EP3S}$ for $a \approx -100 \ {\rm GHz}$ and $b \approx 10 \ {\rm GHz}$ by automatically optimizing $(s_{1,2}, s_{3,2}, R_1, R_2)$ with repeated eigenfrequency simulations of the three-dimensional system based on the finite element method; see Methods. The imaginary indices $\{ {\rm Im} \ n_j \}$ have fixed ratios determined by Eq. (\ref{eq:H3S_EP3}) for $a/b = -10$. After the optimization routine, their values are updated marginally, so that the system gets closer to the degeneracy mainly by compensating for the mismatch between the $a/b$ originally expected and that estimated from the result.

The real and imaginary parts of the eigenfrequencies $\Omega := \omega + \Omega_0$ for the optimum system are shown in Fig. \ref{fig:chiral_EP3}(b) and \ref{fig:chiral_EP3}(c) as markers. The data are plotted as functions of the coefficient $c_{\rm Im}$ of the imaginary indices defined as ${\rm Im} \, (n_1, n_2, n_3) = (0.00502, -0.00871, 0.00366) \times c_{\rm Im}$. As $c_{\rm Im}$ increases from null, three branches of ${\rm Re} \, \Omega$ approach one another and exhibit singular coalescence behavior near $\Omega_{\rm EP} = \Omega_0 \approx 192.68 \ {\rm THz}$ and $c_{\rm Im} = 1$. In addition, those of ${\rm Im} \, \Omega$ linearly split from the origin first and then come closer and diverge sharply around $c_{\rm Im} = 1$. Both properties agree well with the result of an eigenvalue analysis based on Eq. (\ref{eq:H3S_EP3}) for $(a, b) = (-111.1, 9.744) \ {\rm GHz}$ shown by black curves. Nonetheless, the simulation data still have splitting of ${\rm Re} \, \Omega$ and ${\rm Im} \, \Omega$ about 10 GHz and 24 GHz around the exact EP3 that our $\hat{\rm H}_{\rm EP3S}$ guarantees, respectively. This discrepancy may come from small $\{\eta_{j l}\}$ induced by the contrast of $\{ {\rm Im} \ n_j \}$ between the cavities \cite{Takata2022}. Neither ${\rm Re} \, \Omega$ nor ${\rm Im} \, \Omega$ are vertically symmetric, indicating that the design is independent of apparent non-Hermitian symmetry.

The system for $c_{\rm Im} = 1$ is operated sufficiently near the chiral EP3, so its eigenmode distributions look almost the same and exhibit peculiar dynamics (Fig. \ref{fig:chiral_EP3}(d)). The magnetic field profile $H_z (x, y, t) \ (z = 0)$ for one of them has a maximal modal amplitude at C1 and weak fields out-of-phase in C2 and C3 at a point in time ($t = 0$). The bright cavity field pattern jumps in the sequence ${\rm C1} \rightarrow {\rm C3} \rightarrow {\rm C2} \rightarrow {\rm C1} \rightarrow \cdots$ at an interval of $\Delta t = 2 \pi / (3{\rm Re} \, \Omega)$, with the signed amplitude ratios kept. This feature indicates the chiral EP supermode with a dynamic factor, ${\rm Re} \, \bm{\alpha}(t) \propto {\rm Re} [(1, e^{i 2 \pi/3}, e^{i 4 \pi/3})^{\mathrm{T}} e^{i \Omega_{\rm EP} t}]$, exhibiting a persistent phase current in $+y$ direction.

The resultant chiral response can be detected in the Fourier intensity $|E_y (k_x, k_y)|^2$ of the primal transverse electric-field component within the light cone (Fig. \ref{fig:chiral_EP3}(e)). There is a prominent peak component on the negative side of the $k_y$ axis, which is more than three times stronger than the one with $k_y > 0$. It indicates clear directional radiation. The peak coordinate directly corresponds to a polar angle of $-16.0\tcdegree$ from the $k_z$ axis. These properties may promise a good improvement compared surface-emitting lasers around an EP2 demonstrating beam steering over a few degrees \cite{Gao2019}. 

By considering the system as a uniform linear antenna array, we can estimate the radiation direction of the EP eigenstate. Because $E_y$ fields of the cavity modes are of the fundamental order in the $y$ direction and our chiral EP has equal absolute modal amplitudes, the radiation distribution along the $k_y$ axis or correspondent azimuthal angle $\phi$ is dominated by the array factor \cite{Cheng1989}
\begin{equation}
	|F_A (\phi)| = \frac{1}{3}[1 + e^{i (-\frac{2 \pi}{\lambda} d \, {\rm cos} \phi + \xi)} +  e^{2 i (-\frac{2 \pi}{\lambda} d \, {\rm cos} \phi + \xi)}], \label{eq:ArrayFactor}
\end{equation}
where $\lambda$ is the eigen-wavelength in air, $d = 2 \sqrt{3} a$ the cavity interval, and $\xi = 2\pi/3$ the phase difference between adjacent cavity modes given by the eigenstate. We can trivially find that $|F_A (\phi)|$ is asymmetrically distributed and its most significant peak is located at $\phi = -22.3\tcdegree$. This confirms that the main features seen in Fig. \ref{fig:chiral_EP3}(e) come from the chiral eigenvector $\bm{\alpha}_{3 +}$. Note that the array factor for $\xi = -2\pi/3$ characterizes the radiation of the dual state $\bm{\alpha}_{3 -}$ and has the maximum at $\phi = +22.3\tcdegree$. The remaining discrepancy between $|E_y (k_x = 0, k_y)|$ and $|F_A (\phi)|$ may be explained by detailed modal properties of each cavity. For example, the peak angle of $|E_y|$ will be affected by the cavities' different radiative $Q$ factors depending on $\{s_{j, 1}, s_{j, 2}\}$ and asymmetric impact of $R_1 \ne R_2$ on modal fields. In addition, although $|F_A|$ has another peak at $\phi = +49.3\tcdegree$, it is suppressed in $|E_y|$ by the ultrahigh-$Q$ design of PhC cavities.

\subsection{Passive EP6 and minimally active EP7}
It is obvious that resolving constraints on more parameters is generally harder. However, if we focus on trial systems with limited numbers of parameters, we may be able to create EPs with larger order with unexpectedly simple setups. We next consider reciprocally coupled one-dimensional cavity (or equivalent waveguide) arrays described by the following $6 \times 6$ and $7 \times 7$ effective Hamiltonians
\begin{eqnarray}
	\hat{\rm H}_{\rm 6S} &=& \left(\,
	\begin{matrix}
		0 & a_6 &   &   &   &  \\
		a_6 & 0 & b_6 &   &   &  \\
		& b_6 & d_6 & c_6 &   &  \\
		&   & c_6 &-d_6 & b_6 &  \\
		&   &   & b_6 & 0 & a_6\\
		&   &   &   & a_6 & 0\\
	\end{matrix}\,\right), \label{eq:H6S} \\
	\hat{\rm H}_{\rm 7S} &=& \left(\,
	\begin{matrix}
		0 & a_7 &   &   &   &   &  \\
		a_7 & 0 & b_7 &   &   &   &   \\
		& b_7 & 0 & c_7 &   &   &   \\
		&   & c_7 & 0 & d_7 &   &   \\
		&   &   & d_7 & 0 & e_7 &   \\
		&   &   &   & e_7 & 0 & f_7 \\
		&   &   &   &   & f_7 & 0 \\
	\end{matrix}\,\right), \label{eq:H7S}
\end{eqnarray}
where all variables are generally complex. Our aim is to reduce $\hat{\rm H}$'s nonzero diagonal elements as much as possible, because they mostly require external control in experiment. We also expect that simplest solutions would only require combinations of a few values of non-diagonal or coupling terms, thereby securing scalable implementations. Although $\hat{\rm H}_{\rm 7S}$ is based solely on nearest-neighbor couplings, $\hat{\rm H}_{\rm 6S}$ has a pair of on-site potentials $\pm d_6$, since matrices of even dimension without diagonal elements tend to face the reduction of EP order, namely $m < N$. We discuss such an example in SI.

Maximally degenerate EP6s are obtained when $\hat{\rm H}_{\rm 6S}$ satisfies Eq. (\ref{nilpotence_eq}) with $m = 6$. Because we arrange that $\hat{\rm H}_{\rm 6S}$ has only four parameters, it is straightforward to pick and examine factors to vanish, such as ${\rm det} \, \hat{\rm H}_{\rm 6S} = -a^4(c^2 + d^2)$, ${\rm tr} \, \hat{\rm H}_{\rm 6S}^2 = 2[2(a^2 + b^2) + c^2 + d^2]$, and ${\rm tr} \, \hat{\rm H}_{\rm 6S}^4 = (c^2 + d^2)[8 b^2 + 3 (c^2 + d^2)]$, considering Eq. (\ref{eq:trace_condition}). As a result, a simplest sufficient condition for $\hat{\rm H}_{\rm 6S}^6 = \hat{\rm O}$ is given by
\begin{equation}
	a_6^2 + b_6^2 = 0, \qquad c_6^2 + d_6^2 = 0. \label{eq:H6S^6_zero_condition}
\end{equation}
When Eq. (\ref{eq:H6S^6_zero_condition}) is satisfied, we also see that $a_6, b_6, c_6, d_6 \ne 0$ is necessary and sufficient for $\hat{\rm H}_{\rm 6S}^5 \ne \hat{\rm O}$. Therefore, Eq. (\ref{eq:H6S^6_zero_condition}) with $a_6, b_6, c_6, d_6 \ne 0$ guarantees the existence of an EP6 of $\hat{\rm H}_{\rm 6S}$.

We have to tackle a more complicated problem for $\hat{\rm H}_{\rm 7S}$, i.e. $m = 7$. However, by examining three factors in $\hat{\rm H}_{\rm 7S}^7$ to be eliminated, namely ${\rm tr} \, \hat{\rm H}_{\rm 7S}^2$, ${\rm tr} \, \hat{\rm H}_{\rm 7S}^4$, and ${\rm tr} \, \hat{\rm H}_{\rm 7S}^6$, we can extract a somewhat narrowed but manageable sufficient condition for $\hat{\rm H}_{\rm 7S}^7 = \hat{\rm O}$ 
\begin{eqnarray}
	a_7^2 + b_7^2 = 0, \qquad c_7^2 + d_7^2 = 0, \nonumber \\
	e_7^2 + f_7^2 = 0, \qquad b_7^2 + f_7^2 = 0. \label{eq:H7S^7_zero_condition}
\end{eqnarray}
For this case, we can again simplify $\hat{\rm H}_{\rm 7S}^6$ and ensure that $\hat{\rm H}_{\rm 7S}^6 \ne \hat{\rm O}$ if and only if $a_7, b_7, c_7, d_7, e_7, f_7 \ne 0$, thereby identifying instances of $\hat{\rm H}_{\rm 7S}$ with EP7s. Because Eqs. (\ref{eq:H6S^6_zero_condition}) and (\ref{eq:H7S^7_zero_condition}) obviously need parameters with finite imaginary parts, the results here present a distinctive series of non-Hermitian matrices, which may be explored experimentally for demonstrating EP6s and EP7s.

We propose possible photonic implementations of such $\hat{\rm H}_{\rm 6S}$ and $\hat{\rm H}_{\rm 7S}$ with ring cavities and waveguides. Figure \ref{fig:passive_EP6_active_EP7}(a) illustrates a passive system of six linearly aligned and identically structured resonators that can exhibit an EP6 based on $\hat{\rm H}_{\rm 6S}$ for $(a_6, b_6, c_6, d_6) \propto (i, -1, i, 1)$. The basis for each cavity is a resonant mode that couples with waveguide modes propagating along the arrows. A pair of closely positioned cavities has an evanescent coupling $\mu \in \mathbb{R}$ for their modes with opposite circulations. With a waveguide placed just in the middle, two cavities both suffer an out-coupling loss $i \gamma \in i\mathbb{R}$. However, the interference of their output fields affects the net loss of supermodes, modeled as a dissipative coupling $i \gamma$ with the same magnitude \cite{Suh2004}. The resonant frequencies of cavities 3 and 4 may be thermo-optically detuned by $\pm \delta \in \mathbb{R}$ with, for example, electric heaters installed beneath them \cite{Hodaei2017enh}. The effective Hamiltonian of this passive system is derived by the TCMT as
\begin{equation}
	\hat{\rm H}_{\rm 6P} = \hat{\rm H}_{\rm 6S} + i\gamma \hat{\rm I}_6, \label{eq:H6P}
\end{equation}
for $(a_6, b_6, c_6, d_6) = (i \gamma, -\mu, i \gamma, \delta)$. According to Eq. (\ref{eq:H6S^6_zero_condition}), $\hat{\rm H}_{\rm 6P}$ embraces the EP6 with an eigenvector $\bm{v}_{\rm EP6} \propto (-i, 0, 1, i, 0, 1)^{\rm T}$ when $\mu = \delta = \gamma$. Notably, the loss is arranged to be uniform over the array and hence does not hamper the degeneracy in our design.

A seven-cavity counterpart, shown in Fig. \ref{fig:passive_EP6_active_EP7}(b), can be compatible with $\hat{\rm H}_{\rm 7S}$ for $(a_7, b_7, c_7, d_7, e_7, f_7) \propto (-1, i, -1, i, i, -1)$. Again, we expect to alternatively introduce evanescent and dissipative couplings for having an EP7 at the expense of uniform out-coupling loss. However, the system actually needs a cavity coupled with two waveguides, which is the fifth one here, because of the odd Hamiltonian dimension and $b_7^2 + f_7^2 = 0$. This cavity is hence assumed to be composed of a pumped gain medium that provides stimulated emission of rate $g \in \mathbb{R}$ compensating for a part of its dissipation loss of rate $2 \gamma$. The system is minimally active in this sense, and its effective Hamiltonian is described by
\begin{equation}
	\hat{\rm H}_{\rm 7A} = \hat{\rm H}_{\rm 7S} + i (\gamma - g) \hat{\rm D}_{7(5)} + i\gamma \hat{\rm I}_7, \label{eq:H7A}
\end{equation}
for $(a_7, b_7, c_7, d_7, e_7, f_7) \propto (-\mu, i\gamma, -\mu, i\gamma, i\gamma, -\mu)$, where $\hat{\rm D}_{7(5)} := {\rm diag}(0, 0, 0, 0, 1, 0, 0)$ is the diagonal matrix with its (5, 5) element being unity and other ones null. Equation (\ref{eq:H7S^7_zero_condition}) readily tells us that the EP7 is found for $\mu = g = \gamma$. Its eigenvector is $\bm{v}_{\rm EP7} \propto (-1, 0, i, 0, 1, 0, i)^{\rm T}$.
\begin{figure*}[t]
	\includegraphics[width=\linewidth]{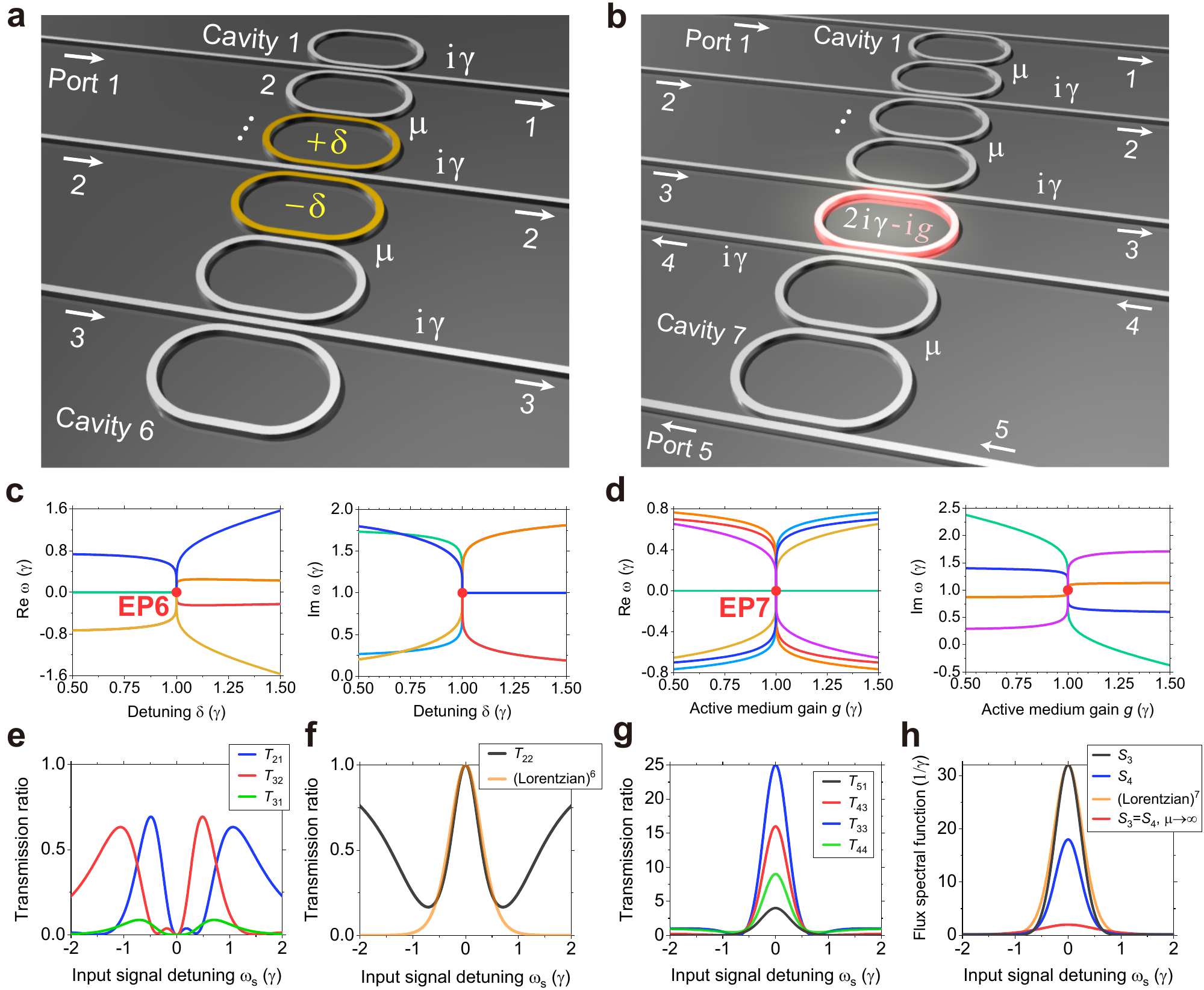}
	\caption{\label{fig:passive_EP6_active_EP7} Photonic designs and responses of passive and active HEPs. (a) System of coupled ring resonators and waveguides involving a passive EP6, and (b) another for a minimally active EP7. Evanescent and waveguide-based dissipative couplings between cavities are all $-\mu < 0$ and $i\gamma \ (\gamma > 0)$, respectively. Each waveguide also introduces an out-coupling loss $i \gamma$ of each cavity coupled. In (a), cavities 3 and 4 have resonance frequency detuning $\pm \delta$. In (b), only cavity 5 is active and its loss $2 i \gamma$ is partially compensated by a gain $-i g \ (g > 0)$. Arrows indicate the light propagation directions considered and hence determine the input and output of the waveguide ports and the directions of responding cavity modes. Complex eigenfrequency detuning of the (c) passive and (d) active systems for $\mu = \gamma$. The EP6 and EP7 are found when $\delta = \mu = \gamma$ and $g = \mu = \gamma$. (e) Internal and (f) external spectral transmission responses for the passive cavities, where $T_{lj}$ is the ratio of power transmission from port $j$ to $l$. While there is no steady transmission inside the cavities for the EP frequency or $\omega_s = 0$, the rejection signal through waveguide 2 ($T_{22}$) exhibits an induced transparency peak fit by the LF6 (orange curve). (g) Transmission ratios for the active system. $T_{51}$ is strictly described as the LF7. $T_{43}$ and $T_{33}$ show 16- and 25-fold enhancement of the transmission power at the EP7. (h) Spectral transfer functions $(S_3, S_4)$ from excitation intensity (coupled spontaneous emission) of cavity 5 to flux outputs at ports 3 and 4. Their peak enhancements are 16- and 9-fold, compared to the case of large evanescent coupling or Hermitian limit ($\mu \rightarrow \infty$) equivalent to the sole Lorentzian response of cavity 5.}
\end{figure*}

The complex eigenfrequency detuning $\omega$ from ${\rm Re} \, \Omega_{\rm EP} = \Omega_0$ for these systems have particular properties depending on their characteristic parameters, as shown in Figs. \ref{fig:passive_EP6_active_EP7}(c) and (d). When the cavity detuning $\delta$ increases and the EP6 is crossed in the passive array for $\mu = \gamma$, the number of separate branches of ${\rm Re} \, \omega$ changes from three to four as the one with ${\rm Re} \, \omega = 0$ splits. Correspondingly, ${\rm Im} \, \omega$ exhibits qualitatively opposite behavior. In the active counterpart with $\mu = \gamma$, introducing the gain $g$ results in symmetric and anti-symmetric trajectories of ${\rm Re} \, \omega$ and ${\rm Im} \, \omega$ with reference to the EP7, respectively. Remarkably, the system has an eigenstate holding ${\rm Re} \, \omega = 0$ over the process. It undergoes an abrupt transition of ${\rm Im} \, \omega$ from positive to negative values around the EP. This state is then expected to exhibit single-mode oscillation, as it exclusively acquires ${\rm Im} \, \omega < 0$ beyond $g = 7/6$.

A striking difference between the passive and active EP systems is revealed when we examine the transmittance of the light injected via their constituent waveguides. It is derived by solving the coupled-mode equations with ports and an excitation term in the frequency domain \cite{Suh2004} (see also Methods for details). We define the sides of the waveguides with inward- and outward-directed arrows as input and output ports numbered from the top, respectively. The transmittance from port $j$ to $l$ is denoted as $T_{lj}(\omega_s)$, which depends on the excitation signal frequency $\omega_s \in \mathbb{R}$ measured relative to ${\rm Re} \ \Omega_{\rm EP}$. 

For the six-cavity array, the information of internal transmission is exhausted by $T_{12}$, $T_{32}$, and $T_{31}$ depicted in Fig. \ref{fig:passive_EP6_active_EP7}(e), because of reciprocity and mirror symmetry. Importantly, all of them are strictly null at the EP frequency, i.e. $\omega_s = 0$, indicating that the passive EP6 is dark for any of the ports despite that its eigenmode has a finite out-coupling loss $i \gamma$. The TCMT also gives the system's steady internal amplitude distribution, $\bm{A}(\omega_s = 0) \propto (s_{+,1} + i s_{+,2}, \, -i s_{+,2}, \, 0, \, 0, \, -i s_{+,2}, \, i s_{+,2} + s_{+,3})$, where $s_{+,j}$ is the flux amplitude incident at port $j \ (j = 1, 2, 3)$. Indeed, $s_{+,1}$ and $s_{+,3}$ only couple with cavities 1 and 6 attributed to rejection, respectively. The fields induced by $s_{+,2}$ are orthogonal to those by $s_{+,1}$ and $s_{+,3}$ and cancel out at output ports 1 and 3. Cavities 3 and 4, which have port 2 in between, are empty for any cases. Such responses are in stark contrast to standard transmissive cavity add-drop filters and coupled-resonator optical waveguides \cite{Yariv1999}. 

The EP's response is embedded in $T_{22}$, which is rather the cavities' rejection signal back to the same waveguide as the input. It is analytically derived as
\begin{equation}
	T_{22}(\omega_s) = \frac{[ (\omega_s^2 + \gamma^2)^3 - 4 \gamma^4 \omega_s^2 ]^2}{(\omega_s^2 + \gamma^2)^6}, \label{eq:T22_EP6}
\end{equation}
and plotted in Fig. \ref{fig:passive_EP6_active_EP7}(f) using a black curve. It exhibits a characteristic peak and dip structure with unity transmittance at the EP, i.e. $T_{22}(0) = 1$. The peak domain is approximately fit by the LF6, namely $[L_{\gamma}(\omega_s)]^6$ where $L_{\gamma}(\omega_s) := \gamma^{2}/(\omega_s^2 + \gamma^2)$, shown as the orange curve. Its linewidth is $\sqrt{\sqrt[6]{2} - 1} \approx 0.350$ times narrower than that of a single-cavity Lorentzian resonance $L_{\gamma}(\omega_s)$ with the same field loss $\gamma$. This peak shape is specific to the response of an EP6 predicted in a general perturbation analysis \cite{Pick2017gen}. As such, what is observed here is a higher-order variant of induced transparency effects \cite{Waks2006}. It is caused by the destructive interference of cavity internal fields, which might rather be mentioned as the manifestation of the limited observability of passive EPs \cite{Schomerus2022,Wiersig2022}. Meanwhile, such a peculiar rejected signal seems qualitatively consistent with experimental demonstration of EPs based on reflection measurements \cite{Lin2011,Feng2013,Lawrence2014}.

The minimally active system exhibits dramatically altered properties, as shown in Fig. \ref{fig:passive_EP6_active_EP7}(g). Its internal transmission spectra all have peaks at $\omega_s = 0$, namely the EP7. This means that the system has non-vanishing internal and output field amplitudes for excitation using each port, although we omit their detailed descriptions because of complexity. The transmission through the entire array, $T_{51}$, is exactly given by the LF7
\begin{align}
	T_{51}(\omega_s) &=  4[L_{\gamma}(\omega_s)]^7 \nonumber \\
	&= \frac{4\gamma^{14}}{(\omega_s^2 + \gamma^2)^7}, \label{eq:T51_EP7_exact}
\end{align}
which has a linewidth of $\approx 0.323 \times 2 \gamma$ and is again consistent with the perturbation theory. Eq. (\ref{eq:T51_EP7_exact}) involves quadruple peak signal enhancement, because cavity 5 is pumped below the lasing threshold and thus works as an optical amplifier. Remarkably, the spectral lines for the injection and detection with ports closer to this component, such as $T_{43}$, $T_{33}$ and $T_{44}$, are even higher and a little narrower, thereby indicating stronger impact of $g$. The linewidth narrowing beyond the LF7 suggests the contribution of a factor other than the local density of states (LDOS) of EPs, namely the interference of output fields from adjacent cavities sharing ports. This Fano effect should take place as long as these cavities have non-vanishing modal amplitudes and their phase difference is unequal to $\pm \pi/2$. As with the EP6 case, one of the rejective responses, $T_{33}$, gives the highest enhancement at $\omega_s = 0$, which is 25-fold.

The cavity array with the EP7 also has an intrinsic response to the spontaneous emission from the gain medium, as seen in Fig. \ref{fig:passive_EP6_active_EP7}(h). It is evaluated by the spectral transfer function 
\begin{equation}
	S_j(\omega_s) = \frac{|s_{-, j}(\omega_s)|^2}{|c_{+, 5}(\omega_s)|^2}, \label{eq:sj_EP7}
\end{equation}
where $s_{-, j}(\omega_s)$ is the output flux amplitude at port $j$ and $c_{+, 5}(\omega_s)$ the exclusive excitation amplitude for the resonant mode of cavity 5. Here we consider white noise with unity power spectral density coming from the medium, by normalizing $S_j(\omega_s)$ with $|c_{+, 5}(\omega_s)|^2 = 1$ for a spectral range of interest. We do not introduce any excitation to other cavities, i.e. $|c_{+, l}(\omega_s)|^2 = 0 \ (l \ne 5)$. The direct outputs from the active resonator have strongly enhanced peaks at the EP frequency, namely $S_3(0) = 32/\gamma$ and $S_4(0) = 18/\gamma$. Note that $S_j(\omega_s)$ originally has a factor of two accounting for the intensity outcoupling rate $2 \gamma$, since we denote the field loss as $\gamma$. These spectral lineshapes are again slightly narrower than LF7s with the same peak values, which may also be attributed to the Fano effect. However, only every other element is nonzero in the EP7 eigenvector $\bm{v}_{\rm EP7} \propto (-1, 0, i, 0, 1, 0, i)^{\rm T}$, and it hence does not make the output fields interfere. In fact, the steady state under the excitation of cavity 5, $\bm{v}_{\rm st, 5}(\omega_s = 0) \propto (-1/3, -i/3, 2i/3, -1/3, 1, -1/2, i/2)^{\rm T}$, is distinct from $\bm{v}_{\rm EP7}$ and distributes modal amplitudes over the entire array. The components linearly independent of $\bm{v}_{\rm EP7}$ should contribute to the extra linewidth narrowing, and they stem from the balance between the constant excitation and its relaxation toward the eigenstate.

An important case for comparison is a system with large evanescent couplings represented by $\mu \rightarrow \infty$. This condition makes cavity 5 effectively isolated from the other ones, because the relative impact of dissipative couplings becomes negligible. As a result, only ports 3 and 4 have responses to the excitation, and they are described by the Lorentzian function (LF1) peaked at $\omega_s = 0$, i.e. $S_3(\omega_s), \ S_4(\omega_s) \rightarrow 2\gamma/(\omega_s^2+\gamma^2$), $S_j(\omega_s) \rightarrow 0 \ (j \ne 3,4)$. This confirms the system's equivalence to a single cavity with gain $g = \gamma$ and two output waveguides each of which gives field dissipation loss $\gamma$. Correspondingly, the peak amplifications for $S_3$ and $S_4$ in Fig. \ref{fig:passive_EP6_active_EP7}(h) are 16- and 9-fold, respectively. They are evidently EP-based, since switching between the two cases only requires modification of the energy-conserving factor, $\mu$. Unlike previous studies \cite{Pick2017gen,Takata2021}, this specific process does not accompany the coalescence of two evenly split spectral peaks. As such, it starts with twice peak LDOS, and the enhancement here should hence be half of that derived in the perturbation theory. The 16-fold change in $S_3 (0)$, while including the contribution of output interference, is close to the prediction modified for a single spectral peak indicating a factor of $\approx 31/2 = 15.5$ for an EP7. In our active system, the EP transition can be detected only by nonlinear changes in the peak intensity and linewidth of the static "zero mode", which is also seen in Fig. \ref{fig:passive_EP6_active_EP7}(d). Such features may open up extra possibilities for optical sensing, regulation, and switching. Some representative spectral transmittance and transfer functions are shown analytically in SI.

We also find drastic differences in response dynamics between the passive and active HEPs by examining the inverse Fourier transforms ($\mathcal{F}^{-1}$) of their characteristic transmission spectra, $T_{22}(\omega_s)$ [Eq. (\ref{eq:T22_EP6})] and $T_{51}(\omega_s)$ [Eq. (\ref{eq:T51_EP7_exact})]. Because $T_{l j}(\omega_s)$ is considered normalized by the constant input spectral intensity $|s_{+, j}(\omega_s)|^2 = 1$, $\mathcal{F}^{-1}[T_{l j}](t)$ is equivalent to the autocorrelation envelope of the output flux amplitude $s_{-, l}$ at port $l$ induced by impulse input to port $j$. First, $\mathcal{F}^{-1}[T_{22}](t)$ of the passive EP6 is dominated by a fast rejective response with a damping time constant $\tau$ of about $1/\gamma$, which is also obtained in the exponential decay just for a single resonator with the loss $\gamma$. In stark contrast, $\mathcal{F}^{-1}[T_{51}](t)$ for the active EP7 exhibits significantly delayed damping with $\tau = 4.92/\gamma$, because the common outcoupling decay $e^{-\gamma t}$ is counteracted up to the sixth order of $t$ by the synergy of the mode coalescence and gain. This result highlights the peculiar advantage of the minimally active HEP in binding light over the passive one; see SI for details.

Before closing this section, we mention that the design by nilpotence is also effective for periodic systems. By placing an additional condition on the couplings between unit cells given by the Bloch theorem, we can find an EP5 in the band structure of a one-dimensional lattice, for example; see SI.

\subsection{Exponential raising of EP order by induction}
The remaining problem is how we can faithfully raise the EP order over the algebraic difficulty in designing systems at scale from scratch. Here we provide the following inductively applicable theorem.\\
\\
\textbf{Theorem 1:} {\it Let $\hat{\rm H} \in \mathbb{C}^{N \times N}$ be a symmetric Hamiltonian $\hat{\rm H} = \hat{\rm H}^{\rm T}$ having an {\rm EP}$N$ with a zero eigenvalue, and suppose that a similarity transformation $\hat{\rm P}$ deriving the {\rm JNF} of $\hat{\rm H}$ can be written as a lower triangular matrix. Then, a $2N \times 2N$ symmetric Hamiltonian}
\begin{equation}
	\hat{\rm H}' = 
	\left(\,
	\begin{matrix}
		\hat{\rm H} + \hat{\rm A} & \hat{\rm B}   \\
		\hat{\rm R}\hat{\rm B}\hat{\rm R} & \ \hat{\rm R}(\hat{\rm H} - \hat{\rm A})\hat{\rm R}
	\end{matrix}\,\right), \label{Theorem1_Hprime}
\end{equation}
{\it has an {\rm EP}$2N$ with a zero eigenvalue, and we can find a similarity transformation $\hat{\rm P}'_{\rm L}$ deriving the {\rm JNF} of $\hat{\rm H}'$ in the form of a lower triangular matrix. Here, $\hat{\rm A}$, $\hat{\rm B}$ and $\hat{\rm R} \in \mathbb{C}^{N \times N}$ are}
\begin{equation}
	\hat{\rm A} = 
	\left(\,
	\begin{matrix}
		0 & \quad & \quad  \\
		\quad & \quad & \quad  \\
		\quad & \quad & A
	\end{matrix}\,\right), \
	\hat{\rm B} = 
	\left(\,
	\begin{matrix}
		\quad & \quad & 0  \\
		\quad & \quad & \quad  \\
		B & \quad & \quad 
	\end{matrix}\,\right), \
	\hat{\rm R} = 
	\left(\,
	\begin{matrix}
		0 & \quad & 1 \ \\[-7pt]
		\quad & \iddots & \quad \\[-3pt]
		1 & \quad & 0 
	\end{matrix}\,\right), \
	\label{Theorem1_ABR}
\end{equation}
{\it with $A, \ B \in \mathbb{C}$ being nonzero and satisfying $A^2 + B^2 = 0$.}\\
\\
Its proof is based on careful search for $\hat{\rm P}'_{\rm L}$ and calculation leading directly to the fact that the JNF of ${\rm H'}$ is a single JC of size $2N$ being equivalent to an EP$2N$. Details are shown in Methods.
\begin{figure*}[t]
	\includegraphics[width=\linewidth]{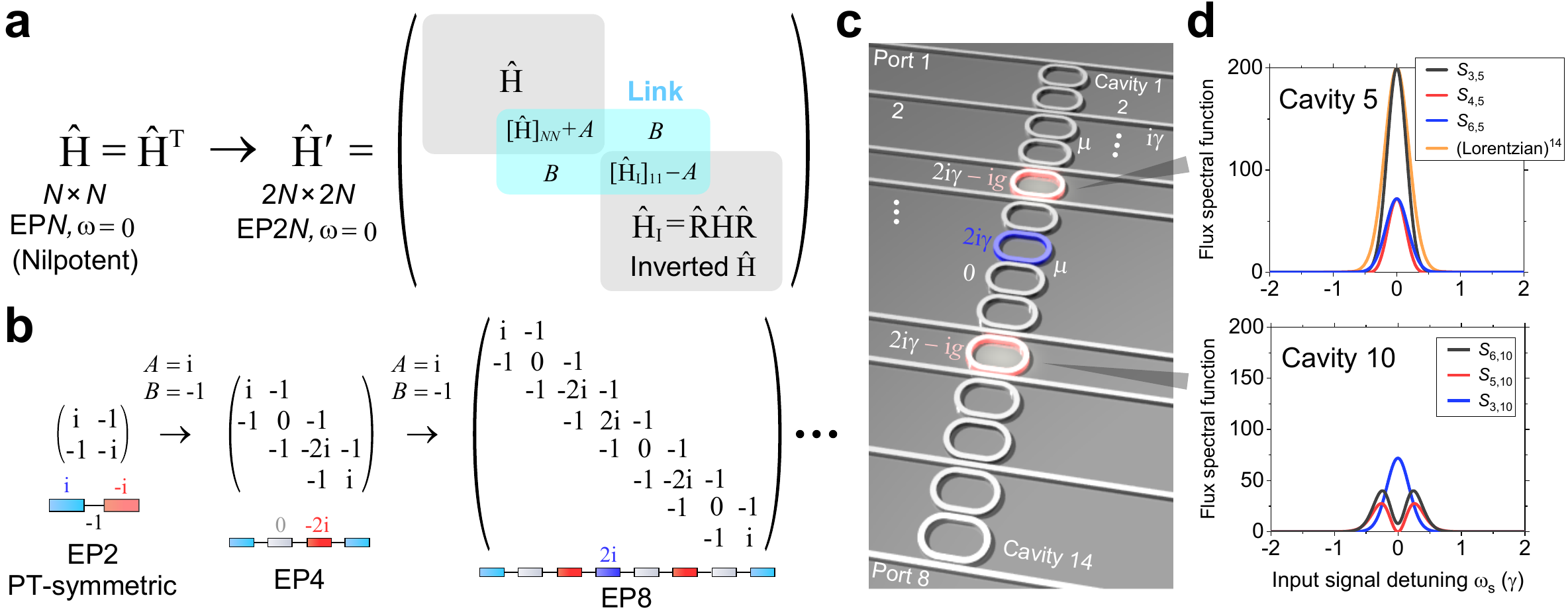}
	\caption{\label{fig:raising_EP_order} Inductive theorem for raising the EP order and its application. (a) Illustration of the procedure. The extended system $\hat{\rm H}'$ has diagonal blocks based on the source Hamiltonian $\hat{\rm H}$ and its inverted copy $\hat{\rm H}_{\rm I}= \hat{\rm R} \hat{\rm H} \hat{\rm R}$, and their boundary elements are updated as $[\hat{\rm H}']_{NN} = [\hat{\rm H}]_{NN} + A$, $[\hat{\rm H}']_{N(N+1)} = [\hat{\rm H}']_{(N+1)N} = B$, $[\hat{\rm H}']_{(N+1)(N+1)} = [\hat{\rm H}_{\rm I}]_{11} - A$, where $A, B \ne 0$ and $A^2 + B^2 = 0$. (b) A series of one-dimensional HEP systems with exponentially raised order based on a $2 \times 2$ PT-symmetric Hamiltonian. We can generate instances with limitlessly large EP order by repeating the process in (a). (c) A 14-cavity system designed by using Theor. 1 for $\hat{\rm H} = \hat{\rm H}_{\rm 7A}$ in Eq. (\ref{eq:H7A}) and $A \propto i$, $B \propto -1$. It has an EP14 when $\mu = g = \gamma$. (d) Spectral  transfer functions $\{S_{l,j}\}$ from excitation intensities of cavities $j = 5, 10$ to output fluxes at ports $l = 3,4,5,6$. The upper part of the array takes over responses of $\hat{\rm H}_{\rm 7A}$, and spontaneous emission coupled to cavity 5 can be enhanced by 20 dB at $\omega_s = 0$. However, the responses within the lower part are mostly suppressed at the EP, as the outputs based on local emission in cavity 10 tend to be frequency-split due to the evanescent coupling involving cavity 8.}
\end{figure*}

Theorem 1 already tells us the way of doubling the EP order, which is also illustrated in Fig. \ref{fig:raising_EP_order}(a). First, we prepare two subsystems: an $N \times N$ symmetric nilpotent Hamiltonian with an EP$N$ in hand and its inverted copy, $\hat{\rm H}_{\rm I}:= \hat{\rm R} \hat{\rm H} \hat{\rm R}$ ($\hat{\rm R}^{-1} = \hat{\rm R}$). Next, we make the $\hat{\rm H}$'s boundary component $[\hat{\rm H}]_{NN}$ and its twin $[\hat{\rm H}_{\rm I}]_{11} = [\hat{\rm H}]_{NN}$ of $\hat{\rm H}_{\rm I}$ have pairwise potential modulation $\pm A$ and a reciprocal coupling $B$. The sufficient condition for the extended system $\hat{\rm H}' \in \mathbb{C}^{2N \times 2N}$ to have an EP$2N$ with $\omega = 0$ is $A^2 + B^2 = 0$. Crucially, any nontrivial $2 \times 2$ symmetric nilpotent matrices are shown to cover a lower triangular $\hat{\rm P}$ and can hence be the source Hamiltonian $\hat{\rm H}$ in Theor. 1; see SI for details. This is also the case for symmetric Hamiltonians within nearest neighbor couplings, such as the nilpotent $\hat{\rm H}_{\rm 6S}$ and $\hat{\rm H}_{\rm 7S}$, and $\hat{\rm H}_{\rm EP3S}$ with $b = 0$ discussed in the last sections. By applying the process $h \in \mathbb{N}$ times to an initial system with an EP$n$, we get an EP of order $n 2^h$ in the final one, and $h$ can be arbitrarily large, or $h \rightarrow \infty$. Moreover, the parameters $A = A_j$, $B = B_j$ used in each step ($j = 1, 2, \cdots, h$) can be chosen \textit{independently}, enabling numerous variations of the extended system. 

A series of lattice systems with maximally degenerate EPs is derived from a $2 \times 2$ PT-symmetric Hamiltonian by applying our theorem, as shown in Fig. \ref{fig:raising_EP_order}(b). Here, the initial matrix has normalized elements, and we consider $A = i$, $B = -1$ throughout the inductive operations so that the result $\hat{\rm H}'$ of every round keeps uniform nearest-neighbor (evanescent) couplings. Interestingly, PT symmetry is lost after the very first extension on $\hat{\rm H}$. The systems generated here characteristically hold unity losses $[\hat{\rm H}]_{11} = [\hat{\rm H}]_{NN} = i$ on their edge elements and exhibit apparently intricate distributions of on-site gain and loss with doubled magnitude, $\pm 2 i$, in units of the coupling rate. Thus, they become hard to find out without using Theor. 1 after a few induction cycles. However, they provide manageable implementations of EP$2^{h+1}$, because the non-Hermitian potential parameters take only four signed integer levels including null. We show a couple more effective Hamiltonians of this set in SI. Note that we can obtain another sequence beginning with an anti-PT-symmetric Hamiltonian by multiplying all the matrices here by $i$.

To see how doubling EP order affects the responses of host systems, we also design a photonic system with an EP14 based on $\hat{\rm H}_{\rm 14A} = \hat{\rm H}'$ for $\hat{\rm H} = \hat{\rm H}_{\rm 7A}$ and $A \propto i$, $B \propto -1$ in Theor. 1, as depicted in Fig. \ref{fig:raising_EP_order}(c). According to the extension procedure, the boundary twin cavities 7 and 8 miss a waveguide in between unlike Fig. \ref{fig:passive_EP6_active_EP7}(b) and are evanescently coupled; the former accepts an absorption loss $2 i \gamma$, and the latter becomes transparent (lossless). Cavities 5 and 10 are also dual and have on-site potential $i(2 \gamma - g)$ originating from out-couplings and pumped active media. As expected, the EP14 is obtained for $\mu = g = \gamma$.

A notable finding here is that responses by active HEPs scale with their order but can be strongly affected by local potential profiles. Figure \ref{fig:raising_EP_order}(d) plots spectral transfer functions $\{S_{l,j}\}$ from the excitation of active cavities 5 and 10 ($j = 5, 10$) to flux outputs at ports $l = 3, 4, 5, 6$ of the system operating at the EP14. It shows that spontaneous emission coupled to cavity 5 is amplified more by the active EP14 than that of the seven-cavity array seen in Fig. \ref{fig:passive_EP6_active_EP7}(h). The peak enhancement for port 3 reaches as large as 20 dB (100-fold) compared to the single-cavity Lorentzian response $2 L_{\gamma}(\omega_s)$ obtained for $\mu \rightarrow \infty$. Slightly modified lineshapes of such peaks from the LF14 again indicate that the steady state holds the components that are linearly independent of the EP eigenstate $\bm{v}_{\rm EP14} \propto (-1, 0, i, 0, 1, 0, i, -1, 0, i, 0, -1, 0, -i)^{\rm T}$ and affect the outputs via interference. In contrast, responses to the excitation of cavity 10, such as $S_{5,10}$ and $S_{6,10}$, have spectral dips and are rather suppressed at the EP ($\omega_s = 0$). These dramatically altered properties obviously originate from the difference in the imaginary potential modulation on cavities 7 and 8, because other parts are based on two copies of $\hat{\rm H}_{\rm 7A}$ and are hence symmetric with respect to the center. In the upper half of the system, cavity 7 has significant loss as $\hat{\rm H}_{\rm 7A}$ does. Thus, its local responses remain intense at the basis resonance frequency $\omega_s = 0$ by the impact of imaginary potential. On the other hand, the evanescent coupling between cavities 8 and 9 becomes effective as cavity 8 is lossless, so the spectral functions within the rest half region tend to be frequency-split. Note that the emission output involving both cavities 5 and 10 can restore a single enhanced peak at $\omega_s = 0$, as represented by $S_{6,5}$ and $S_{3,10}$. See SI for analytic descriptions of the spectral functions and transmission spectra with enhancement.

Such system behavior can also be quantitatively confirmed by the resolvent or Green function \cite{Wiersig2022}, $\hat{\rm G}(\omega_s, \hat{\rm H}) := (\omega_s \hat{\rm I} - \hat{\rm H})^{-1}$, relating the steady intracavity responses $\bm{\alpha}(\omega_s)$ to excitation $\bm{c}(\omega_s)$ via $\bm{\alpha}(\omega_s) = -i \hat{\rm G}(\omega_s, \hat{\rm H})  \bm{c}(\omega_s)$. Notably, for the resonant excitation of the EP ($\omega_s = 0$), the upper diagonal block of $\hat{\rm G}(0, \hat{\rm H}_{\rm 14A})$ is similar to $\hat{\rm G}(0, \hat{\rm H}_{\rm 7A})$, taking over the characteristic responses. In contrast, its lower diagonal block has suppressed elements and is thus likely to indicate locally dark responses. Details are shown in SI. 

Our result tells us how the detection of EPs can be hampered. In general, external excitation can couple with the fields that are orthogonal to EP eigenstates \cite{Chen2020}. Mathematically, even such components converge at the EPs in time evolution given by the operation of $\hat{\rm H}$. However, for the case of steadily driven systems their net responses are determined by the interplay between sustained excitation and internal fields. As such, any local system profiles, such as excitation, couplings, and on-site potential, can affect the observability of EPs. Although there are some theoretical clues in this sense \cite{Schomerus2022,Wiersig2022,Hashemi2022}, we still miss a general guiding principle to predict the similarity (or difference) between eigenmodes and steady states under excitation and then identify exact conditions for intriguing non-Hermitian phenomena. This work provides fundamental tools and a practical prospect to scale up systems with maximally degenerate EPs exhibiting intrinsic responses.

\section{Discussion}\label{sec:discussion}
In summary, we have developed a design concept for HEPs with large order, implementability, and unconventional responses based on the nilpotence of matrices and mathematical induction. The nilpotence for an index $m$ is always sufficient for an EP$m$ regardless of the presence or absence of apparent non-Hermitian symmetry, and it covers all possible cases for maximally degenerate EPs with $m = N$. This approach is applicable to arbitrary trial system matrices and readily lets us find conditions for practical parameterized Hamiltonians to have HEPs. We have explored photonic HEPs with various particular responses, namely an EP3 with 3-vector chirality or directional radiation in PhC cavities, an EP6 in passive ring resonators and waveguides exhibiting EP-induced transparency, and an EP7 in a similar but minimally active system enabling enhanced spectral responses, which also involve the help of Fano effect. The other technical leap is our inductive theorem to exponentially raise the EP order starting out from a known system with a maximally degenerate EP. We have shown a series of implementable Hamiltonians with such HEPs based on the $2 \times 2$ PT-symmetric one. We have also found an EP14 with further magnified responses in the extension of our active system with the EP7. Our work will trigger extensive searches for HEPs with unconventional phenomena in various physical systems, such as photonic \cite{Lawrence2014,Peng2016,Gao2019,Brandstetter2014,Miao2016,Liao2023,Takata2021,Hodaei2017enh}, plasmonic \cite{Park2020}, phononic \cite{Tang2020,delPino2022,Patil2022,Chen2020}, and electronic ones \cite{Kononchuk2022,XiaoZ2019}. It will also help practically resolve fundamental issues such as the bound of observable non-Hermitian responses for systems with steady excitation. Efficient designs of HEPs incorporating amplifications will be essential to their extensions to nonlinear counterparts under steady oscillation \cite{Bai2024} and applications such as mode locking towards on-chip ultrafast light sources \cite{Imamura2024}, and tailored linear and nonlinear signal operations in neuromorphic computing \cite{Deng2021}.

\appendix

\section*{Methods}
\subsection{Correspondence between nilpotence index and EP order}
As shown in Fig. \ref{fig:JNF_nilpotence}(a), any square matrix $\hat{\rm H} \in \mathbb{C}^{N \times N}$ can be similarity-transformed to its JNF, $\hat{\rm J}_{\rm H} = \hat{\rm P}^{-1} \hat{\rm H} \hat{\rm P}$. Here, $\hat{\rm P} \ne \hat{\rm O}$ is granted because $\hat{\rm P}$ is regular. Thus, $\hat{\rm H}$ can always be written as the inverse transformation of $\hat{\rm J}_{\rm H}$, namely $\hat{\rm H} = \hat{\rm P} \hat{\rm J}_{\rm H} \hat{\rm P}^{-1}$. By taking the $m$th power of both sides of this equation and supposing the nilpotence of $\hat{\rm H}$ with the index of $m$, $\hat{\rm H}^{m} = \hat{\rm O}$ and $\hat{\rm H}^{m - 1} \ne \hat{\rm O}$, we have $\hat{\rm P} \hat{\rm J}_{\rm H}^m \hat{\rm P}^{-1} = \hat{\rm O}$ and hence $\hat{\rm J}_{\rm H}^m = \hat{\rm O}$. Similarly, we have $\hat{\rm P} \hat{\rm J}_{\rm H}^{m-1} \hat{\rm P}^{-1} = \hat{\rm H}^{m - 1}$ and then find $\hat{\rm J}_{\rm H}^{m-1} \ne \hat{\rm O}$; otherwise, we have $\hat{\rm H}^{m - 1} = \hat{\rm O}$, which trivially contradicts with the nilpotence.

By the definition and nilpotence, $\hat{\rm J}_{\rm H}$ is a block diagonal matrix composed of the JCs $\{\hat{\rm J}(0, d_j)\}$ (again, see Fig. \ref{fig:JNF_nilpotence}(a)). Therefore, we have $\hat{\rm J}_{\rm H}^m = [\hat{\rm J}(0, d_1)]^m \oplus [\hat{\rm J}(0, d_2)]^m \oplus \cdots \oplus [\hat{\rm J}(0, d_u)]^m = \hat{\rm O}$, i.e. $[\hat{\rm J}(0, d_j)]^m = \hat{\rm O}$ for all of $j = 1, 2, \ldots, u$. On the other hand, there exists a JC $\hat{\rm J}(0, d_l)$ in $\hat{\rm J}_{\rm H}$ such that $[\hat{\rm J}(0, d_l)]^{m - 1} \ne \hat{\rm O}$ because of $\hat{\rm J}_{\rm H}^{m-1} \ne \hat{\rm O}$. In addition, an element-wise calculation lets us obtain $[\hat{\rm J}(0, k)]^k = \hat{\rm O}$, and $\hat{\rm J}(0, k), [\hat{\rm J}(0, k)]^2, \ldots, [\hat{\rm J}(0, k)]^{k - 1} \ne \hat{\rm O}$ for any $k \in \mathbb{N}$. Consequently, we reach $m = d_l = {\rm max}\{d_j\}$, which means that $m$ is the largest order of $\hat{\rm H}$'s EPs \cite{Miyake2022}. 

\subsection{Temporal coupled-mode theory}
In this work, we consider systems of $N$ on-chip optical resonators. They are assumed to be based on high-$Q$ defect cavities in a photonic insulator or ring cavities with the impact of reflection neglected. Suppose we place each resonator in a solitary environment and obtain an eigenmode $\bm{\Phi}_j ({\bf r} - {\bf r}_j)$ of the spatial Maxwell equations, with ${\bf r}_j$ denoting the position of cavity $j$ ($j = 1, 2, \cdots, N$) in Cartesian coordinates. The TCMT \cite{Takata2022,Yariv1999} tells that the electromagnetic fields of the entire coupled system can approximately be described by a linear combination of such cavity modes, namely $\bm{\Phi}({\bf r}, t) = \sum_{i = j}^{N} A_j(t) e^{i \Omega_0 t} \bm{\Phi}_j ({\bf r} - {\bf r}_j)$, where $A_j(t) \in \mathbb{C}$ is the modal amplitude for cavity $j$, $\Omega_0 \in \mathbb{R}$ the average of eigenfrequencies of the cavity modes, and $e^{i \Omega_0 t}$ the temporal phase factor. This ansatz lets the wave equation reduce to the following series of linear differential equations for $\{A_j\}$
\begin{equation}
	-i \frac{\textrm{d}A_{j}}{\textrm{d}t} = \left( \delta_j + i \gamma_j \right) A_{j} - \sum_{k = 1, k \ne j}^N \mu_{j k} A_k, \label{eq:CME}
\end{equation}
where $\delta_j \in \mathbb{R}$, $\gamma_j \in \mathbb{R}$, and $\mu_{j k} \in \mathbb{C}$ denote cavity $j$'s resonance frequency detuning with reference to $\Omega_0$, modal loss ($> 0$) or gain ($< 0$), and coupling by the mode of cavity $k$, respectively. The modal amplitudes can be written as a vector $\bm{A} = (A_1, \cdots, A_N)^\mathrm{T}$ and further decomposed into $\bm{A} = \bm{\alpha} e^{i \omega t}$, with $\bm{\alpha} = (\alpha_1, \cdots, \alpha_N)^\mathrm{T} \in \mathbb{C}^N$ and $\omega \in \mathbb{C}$ being the amplitudes and detuning of steady responses. As a result, we have the eigenvalue problem with $\bm{\alpha}$ for the effective Hamiltonian $\hat{\rm H}$, namely Eq. (\ref{eigenvalue_eq}), where $[\hat{\rm H}]_{j j} = \delta_j + i \gamma_j$ and $[\hat{\rm H}]_{j k} = -\mu_{j k} \ (j \ne k)$. The coupling terms $\{\mu_{j k}\}$ are often limited to those among nearest-neighbor elements, since they decay exponentially with spatial intervals between cavities. Because the basis here is a series of separate modes, reciprocity also implies $\mu_{j k} = \mu_{k j}$. This $\hat{\rm H}$ describes the system's internal dynamics and predicts accurately its response to pulsed excitation.

When the system has steady excitation by light coupling via waveguides, i.e. well-defined ports, Eq. (\ref{eq:CME}) can be extended in vector form as \cite{Suh2004}
\begin{eqnarray}
	\frac{\textrm{d} \bm{A}}{\textrm{d}t} &=& i \hat{\rm H} \bm{A} + \hat{\rm K}_{\rm I} \bm{s}_{+}, \label{eq:TCME_amplitudes}\\
	\bm{s}_{-} &=& \hat{\rm C} \bm{s}_{+} + \hat{\rm K}_{\rm O} \bm{A}, \label{eq:TCME_ports}
\end{eqnarray}
where $\bm{s}_{+} = (s_{+,j}) \in \mathbb{C}^{n_{\rm w}}$ and $\bm{s}_{-} = (s_{-,j}) \in \mathbb{C}^{n_{\rm w}}$ denote input and output flux amplitudes of the ports, respectively, with $n_{\rm w}$ being the number of waveguides. Here, we assign a specific propagation direction for each port as done in Figs. \ref{fig:passive_EP6_active_EP7}(a), \ref{fig:passive_EP6_active_EP7}(b), and \ref{fig:raising_EP_order}(c), so that we can consider a series of basis cavity modes coupled over the entire array. Since we neglect the impact of reflection, we do not deal with the other set of modes propagating backwards. 

We determine a proper description of the couplings among cavities and ports for a passive setup, before introducing on-site gain as needed. The input and output processes are denoted by $\hat{\rm K}_{\rm I} \in \mathbb{C}^{N \times n_{\rm w}}$ and $\hat{\rm K}_{\rm O} \in \mathbb{C}^{n_{\rm w} \times N}$, respectively. The matrix $\hat{\rm C} \in \mathbb{C}^{n_{\rm w} \times n_{\rm w}}$ describes direct scattering processes of the waveguides (i.e. guided propagation) without the impact of resonances. We assume that the waveguides are identical and do not have any propagation loss or crosstalk; $\hat{\rm C}$ is hence diagonal unitary. Although there is still a phase degree of freedom, we take $\hat{\rm C} = \hat{\rm I}_{n_{\rm w}}$ for simplicity. In this case, time-reversal symmetry, energy conservation, and Lorentz reciprocity impose the following constraints \cite{Suh2004}
\begin{gather}
	\hat{\rm K}_{\rm I} = -\hat{\rm K}_{\rm I}^{*}, \ \hat{\rm K}_{\rm O} = -\hat{\rm K}_{\rm O}^{*}, \label{eq:TCME_TRS}\\
	\hat{\rm K}_{\rm O}^{\dagger}\hat{\rm K}_{\rm O} = 2 \hat{\rm \Gamma}, \label{eq:TCME_EC}\\
	\hat{\rm K}_{\rm O} = \hat{\rm K}_{\rm I}^{\rm T}, \label{eq:TCME_LR}
\end{gather}
where $\hat{\rm \Gamma}$ is the portion of ${\rm Im} \ \hat{\rm H}$ that is attributed to the outcouplings. The elements of $\hat{\rm K}_{\rm O}$ corresponding to the couplings between cavities and waveguides are only finite. They are supposed to be all identical and hence denoted by as a single parameter, say $\kappa$. Based on Eq. (\ref{eq:TCME_EC}), we have the desired imaginary coupling profile of $\hat{\rm H}$ and an additional constraint $|\kappa|^2 = 2 \gamma$. By combining this with Eq. (\ref{eq:TCME_TRS}), we find that $\kappa$ is purely imaginary and that we can choose $\kappa = i \sqrt{2 \gamma}$. We trivially obtain $\hat{\rm K}_{\rm I}$ by using Eq. (\ref{eq:TCME_LR}).

To study spectral system responses, we Fourier-transform Eqs. (\ref{eq:TCME_amplitudes}) and (\ref{eq:TCME_ports}), namely apply $\bm{A}, \ \bm{s}_{+}, \ \bm{s}_{-} \rightarrow \bm{A}(\omega_s), \ \bm{s}_{+}(\omega_s), \ \bm{s}_{-}(\omega_s)$ and $\textrm{d} \bm{A}/\textrm{d} t \rightarrow i \omega_s \bm{A}(\omega_s)$. Since we focus on a narrow spectral range compared with the optical resonance frequency, the matrix elements are assumed to be constant. We basically consider a single-channel input at port $j$ ($s_{+,l} = 0 \ (l \ne j)$) and solve Eq. (\ref{eq:TCME_amplitudes}) in the frequency domain for $\bm{A}(\omega_s)$. By substituting the solution into Eq. (\ref{eq:TCME_ports}), we obtain the flux transmission ratio $T_{l j} (\omega_s) = |s_{-,l}(\omega_s)|^2/|s_{+,j}(\omega_s)|^2$ for output port $l$. For the spectral transfer function $S_{l j} (\omega_s)$ of spontaneous emission, we selectively excite cavity $j$ with unity flux amplitude regardless of the waveguide coupling profile, i.e. $\hat{\rm K}_{\rm I} \bm{s}_{+}(\omega_s) \rightarrow \bm{c}_{+}(\omega_s)$, $c_{+,j}(\omega_s)  = 1$, $c_{+,l}(\omega_s)  = 0 \ (l \ne j)$. We then solve the same linear problem to have $S_{l j} (\omega_s) = |s_{-,l}(\omega_s)|^2$ that is in units of $1/\gamma$.

\subsection{Simulation of photonic crystal cavities}
A triangular-lattice PhC of air holes is patterned on a 250 nm-thick slab composed of an InGaAsP-based heterostructure. The lattice constant and hole radius of this background PhC is 395 nm and $R_0 = 100 \ {\rm nm}$, respectively. As indicated in Fig. \ref{fig:chiral_EP3}(a), radius modulations from $R_0$ are also carried out on the holes at the upper and lower ends of the array to suppress undesired cavity frequency detuning. For introducing the pumping of C$j$, we define a cuboid region in it with dimensions of $1385 \ {\rm nm} \times 350 \ {\rm nm} \times 250 \ {\rm nm}$ and an imaginary part of its material refractive index $n_{{\rm i}, j}$. Such a configuration can be achieved by buried heterostructure technologies or fine control of optical pumping on devices with quantum wells. For simplicity, the entire slab material is assumed to have a real part of the index $n_{\rm r} = 3.41$. The system eigenmodes are simulated by a commercial solver based on the finite-element method (COMSOL Multiphysics).

We first examine the eigenfrequencies of the system with uniform couplings of about $- 100 \ {\rm GHz}$ and no imaginary material index. Although the array has a linear alignment, it is found to have a nonnegligible next-nearest-neighbor coupling $b \approx 10 \ {\rm GHz}$, which is estimated by its unevenly spaced eigenvalues. Next, considering that the hole shift (radius variation) also affects slightly the modal shape and hence the couplings (detuning), we adopt automated optimization of $(s_{1,2}, s_{3,2}, R_1, R_2)$ toward a triple degeneracy of eigenvalues (i.e. an EP3 for Eq. (\ref{eq:H3S_EP3})) with $a/b \approx -10$ based on the Nelder-Mead algorithm. The initial parameter values are $(s_{1,2}, s_{3,2}, R_1, R_2)_{\rm ini} =$ (94 nm, 77.4 nm, 85.2 nm, 94.75 nm), other fixed variables are $(s_{2, 2}, s_{1, 1}, s_{2,1}, s_{3,1}) =$ (85 nm, 129 nm, 123 nm, 116 nm), and they are determined by the results of single- and two-cavity simulations. The imaginary indices here, $(n_{{\rm i}, 1}, n_{{\rm i}, 2}, n_{{\rm i}, 3}) = (0.55, -0.95, 0.4)\times n_{\rm Im}$, have constant coefficients according to the diagonal elements of Eq. (\ref{eq:H3S_EP3}) for $(a, b) = (-1, 0.1)$ and a common factor $n_{\rm Im}$ distinct for different trials of optimization. A good result is obtained for $n_{\rm Im} = 0.00915$ with the optimized parameters $(s_{1,2}, s_{3,2}, R_1, R_2)_{\rm opt} =$ (95.727 nm, 73.796 nm, 85.203 nm, 94.815 nm). We then marginally vary the imaginary indices $(n_{{\rm i}, 1}, n_{{\rm i}, 2}, n_{{\rm i}, 3})$ to compensate for the discrepancy of $a/b$ in our original estimation and that of the optimized system.

\subsection{Proof of Theorem 1}
We prove that the JNF of $\hat{\rm H}'$ defined in Eqs. (\ref{Theorem1_Hprime}) and (\ref{Theorem1_ABR}) is $\hat{\rm J}_{\rm H'} = \hat{\rm J}(0, 2N)$ by applying the similarity transformation
\begin{equation}
	\hat{\rm P}' = 
	\left(\,
	\begin{matrix}
		A \hat{\rm P} & \hat{\rm M} \\
		B \hat{\rm P}_{\rm I} & \hat{\rm O}
	\end{matrix}\,\right), \label{eq:Pprime}
\end{equation}
where $\hat{\rm J}_{\rm H} = \hat{\rm P}^{-1} \hat{\rm H} \hat{\rm P} = \hat{\rm J}(0, N)$, $\hat{\rm P}_{\rm I} = \hat{\rm R} \hat{\rm P}$, $A, B \in \mathbb{C}$ satisfy $A^2 + B^2 = 0$, and $\hat{\rm M} \in \mathbb{C}^{N \times N}$ is a regular matrix. Here, we see that the inverse matrix of $\hat{\rm P}'$ is
\begin{equation}
	\hat{\rm P}'^{-1} = 
	\left(\,
	\begin{matrix}
		\hat{\rm O} & (1/B) \hat{\rm P}_{\rm I}^{-1} \\
		\hat{\rm M}^{-1} & (A/B) \hat{\rm M}^{-1} \hat{\rm R}
	\end{matrix}\,\right), \label{eq:PprimeInverse}
\end{equation}
namely $\hat{\rm P} \hat{\rm P}'^{-1} = \hat{\rm I}$ and thus $\hat{\rm P}'$ is surely regular, regardless of the details of $\hat{\rm M}$.

We derive an $\hat{\rm M}$ satisfying $\hat{\rm P}'^{-1} \hat{\rm H}' \hat{\rm P}' = \hat{\rm J}(0, 2N)$. A block-wise calculation gives
\begin{equation}
	\hat{\rm P}'^{-1} \hat{\rm H}' \hat{\rm P}' = 
	\left(\,
	\begin{matrix}
		\hat{\rm J}(0, N) & (1/A)\hat{\rm P}^{-1}\hat{\rm A}\hat{\rm M} \\
		\hat{\rm O} & \hat{\rm M}^{-1} \hat{\rm H} \hat{\rm M}
	\end{matrix}\,\right), \label{eq:STPprime}
\end{equation}
where we have used $(1/B)\hat{\rm B}\hat{\rm R} = (1/A)\hat{\rm A}$ and $(1/A)\hat{\rm A}\hat{\rm R} = (1/B)\hat{\rm B}$. Thus, the necessary and sufficient condition for $\hat{\rm P}'^{-1} \hat{\rm H}' \hat{\rm P}' = \hat{\rm J}(0, 2N)$ is
\begin{eqnarray}
	\frac{1}{A} \hat{\rm A} \hat{\rm M} &=& \frac{1}{B} \hat{\rm P} \hat{\rm B}, \label{eq:condition_doubing_EP_order_1}\\
	\hat{\rm M}^{-1} \hat{\rm H} \hat{\rm M} &=& \hat{\rm J}(0, N). \label{eq:condition_doubing_EP_order_2}
\end{eqnarray}
Since $\hat{\rm H}$ is symmetric and $\hat{\rm P}$ is lower triangular, we can find that the following $\hat{\rm M}$ always satisfies Eqs. (\ref{eq:condition_doubing_EP_order_1}) and (\ref{eq:condition_doubing_EP_order_2}),
\begin{equation}
	\hat{\rm M} = \frac{[\hat{\rm P}]_{N N}}{[(\hat{\rm P}^{\rm T})^{-1} \hat{\rm R}]_{N 1}} (\hat{\rm P}^{\rm T})^{-1} \hat{\rm R}, \label{eq:M_for_doubling_EP_order}
\end{equation}
namely $\hat{\rm H}'$ has a maximally degenerate EP of order $2N$. It is noteworthy that $[\hat{\rm P}]_{N N} \ne 0$ and $[(\hat{\rm P}^{\rm T})^{-1} \hat{\rm R}]_{N 1} \ne 0$ are guaranteed because $\hat{\rm P}$ is regular and lower triangular.

Since $\hat{\rm H}$ is symmetric, we also see that
\begin{equation}
	\hat{\rm P}'_{\rm L} := (\hat{\rm P}'^{\rm T})^{-1} \hat{\rm R}, \label{eq:PprimeL_for_doubling_EP_order}
\end{equation}
is regular, lower triangular, and is a similarity transformation that gives the JNF of $\hat{\rm H}'$, i.e. $\hat{\rm P}_{\rm L}'^{-1} \hat{\rm H}' \hat{\rm P}'_{\rm L} = \hat{\rm J}(0, 2N)$. \qed

Because it is trivial that $\hat{\rm H}'$ is symmetric, we can apply the entire process described above recursively to $\hat{\rm H}'$ and $\hat{\rm P}'_{\rm L}$. By repeating this, we can obtain a countably infinite series of $N2^h \times N2^h$ matrices with EPs of order $N2^h$ ($h = 1, 2, \dots$). 


\begin{thebibliography}{70}%
	\makeatletter
	\providecommand \@ifxundefined [1]{%
		\@ifx{#1\undefined}
	}%
	\providecommand \@ifnum [1]{%
		\ifnum #1\expandafter \@firstoftwo
		\else \expandafter \@secondoftwo
		\fi
	}%
	\providecommand \@ifx [1]{%
		\ifx #1\expandafter \@firstoftwo
		\else \expandafter \@secondoftwo
		\fi
	}%
	\providecommand \natexlab [1]{#1}%
	\providecommand \enquote  [1]{``#1''}%
	\providecommand \bibnamefont  [1]{#1}%
	\providecommand \bibfnamefont [1]{#1}%
	\providecommand \citenamefont [1]{#1}%
	\providecommand \href@noop [0]{\@secondoftwo}%
	\providecommand \href [0]{\begingroup \@sanitize@url \@href}%
	\providecommand \@href[1]{\@@startlink{#1}\@@href}%
	\providecommand \@@href[1]{\endgroup#1\@@endlink}%
	\providecommand \@sanitize@url [0]{\catcode `\\12\catcode `\$12\catcode
		`\&12\catcode `\#12\catcode `\^12\catcode `\_12\catcode `\%12\relax}%
	\providecommand \@@startlink[1]{}%
	\providecommand \@@endlink[0]{}%
	\providecommand \url  [0]{\begingroup\@sanitize@url \@url }%
	\providecommand \@url [1]{\endgroup\@href {#1}{\urlprefix }}%
	\providecommand \urlprefix  [0]{URL }%
	\providecommand \Eprint [0]{\href }%
	\providecommand \doibase [0]{https://doi.org/}%
	\providecommand \selectlanguage [0]{\@gobble}%
	\providecommand \bibinfo  [0]{\@secondoftwo}%
	\providecommand \bibfield  [0]{\@secondoftwo}%
	\providecommand \translation [1]{[#1]}%
	\providecommand \BibitemOpen [0]{}%
	\providecommand \bibitemStop [0]{}%
	\providecommand \bibitemNoStop [0]{.\EOS\space}%
	\providecommand \EOS [0]{\spacefactor3000\relax}%
	\providecommand \BibitemShut  [1]{\csname bibitem#1\endcsname}%
	\let\auto@bib@innerbib\@empty
	\bibitem [{\citenamefont {Georgi}(1993)}]{Georgi1993}%
	\BibitemOpen
	\bibfield  {author} {\bibinfo {author} {\bibfnamefont {H.}~\bibnamefont
			{Georgi}},\ }\href@noop {} {\emph {\bibinfo {title} {The Physics of Waves}}}\
	(\bibinfo  {publisher} {Prentice-Hall, Englewood Cliffs},\ \bibinfo {year}
	{1993})\ \Eprint
	{https://arxiv.org/abs/\url{https://sites.harvard.edu/hgeorgi/}}
	{\url{https://sites.harvard.edu/hgeorgi/}} \BibitemShut {NoStop}%
	\bibitem [{\citenamefont {Moiseyev}(2011)}]{Moiseyev2011}%
	\BibitemOpen
	\bibfield  {author} {\bibinfo {author} {\bibfnamefont {N.}~\bibnamefont
			{Moiseyev}},\ }\href {https://doi.org/10.1017/CBO9780511976186} {\emph
		{\bibinfo {title} {Non-{H}ermitian Quantum Mechanics}}}\ (\bibinfo
	{publisher} {Cambridge University Press, Cambridge},\ \bibinfo {year}
	{2011})\BibitemShut {NoStop}%
	\bibitem [{\citenamefont {Bender}\ and\ \citenamefont
		{Boettcher}(1998)}]{Bender1998}%
	\BibitemOpen
	\bibfield  {author} {\bibinfo {author} {\bibfnamefont {C.~M.}\ \bibnamefont
			{Bender}}\ and\ \bibinfo {author} {\bibfnamefont {S.}~\bibnamefont
			{Boettcher}},\ }\bibfield  {title} {\bibinfo {title} {Real spectra in
			non-{H}ermitian {H}amiltonians having $\mathcal{PT}$ symmetry},\ }\href
	{https://doi.org/10.1103/PhysRevLett.80.5243} {\bibfield  {journal} {\bibinfo
			{journal} {Phys. Rev. Lett.}\ }\textbf {\bibinfo {volume} {80}},\ \bibinfo
		{pages} {5243} (\bibinfo {year} {1998})}\BibitemShut {NoStop}%
	\bibitem [{\citenamefont {Makris}\ \emph {et~al.}(2008)\citenamefont {Makris},
		\citenamefont {El-Ganainy}, \citenamefont {Christodoulides},\ and\
		\citenamefont {Musslimani}}]{Makris2008}%
	\BibitemOpen
	\bibfield  {author} {\bibinfo {author} {\bibfnamefont {K.~G.}\ \bibnamefont
			{Makris}}, \bibinfo {author} {\bibfnamefont {R.}~\bibnamefont {El-Ganainy}},
		\bibinfo {author} {\bibfnamefont {D.~N.}\ \bibnamefont {Christodoulides}},\
		and\ \bibinfo {author} {\bibfnamefont {Z.~H.}\ \bibnamefont {Musslimani}},\
	}\bibfield  {title} {\bibinfo {title} {Beam dynamics in
			$\mathcal{PT}$-symmetric optical lattices},\ }\href
	{https://doi.org/10.1103/PhysRevLett.100.103904} {\bibfield  {journal}
		{\bibinfo  {journal} {Phys. Rev. Lett.}\ }\textbf {\bibinfo {volume} {100}},\
		\bibinfo {pages} {103904} (\bibinfo {year} {2008})}\BibitemShut {NoStop}%
	\bibitem [{\citenamefont {Miri}\ and\ \citenamefont {Al\`u}(2019)}]{Miri2019}%
	\BibitemOpen
	\bibfield  {author} {\bibinfo {author} {\bibfnamefont {M.-A.}\ \bibnamefont
			{Miri}}\ and\ \bibinfo {author} {\bibfnamefont {A.}~\bibnamefont {Al\`u}},\
	}\bibfield  {title} {\bibinfo {title} {Exceptional points in optics and
			photonics},\ }\href {https://doi.org/10.1126/science.aar7709} {\bibfield
		{journal} {\bibinfo  {journal} {Science}\ }\textbf {\bibinfo {volume}
			{363}},\ \bibinfo {pages} {eaar7709} (\bibinfo {year} {2019})}\BibitemShut
	{NoStop}%
	\bibitem [{\citenamefont {{\"O}zdemir}\ \emph {et~al.}(2019)\citenamefont
		{{\"O}zdemir}, \citenamefont {Rotter}, \citenamefont {Nori},\ and\
		\citenamefont {Yang}}]{Ozdemir2019par}%
	\BibitemOpen
	\bibfield  {author} {\bibinfo {author} {\bibfnamefont {{\c{S}}.}~\bibnamefont
			{{\"O}zdemir}}, \bibinfo {author} {\bibfnamefont {S.}~\bibnamefont {Rotter}},
		\bibinfo {author} {\bibfnamefont {F.}~\bibnamefont {Nori}},\ and\ \bibinfo
		{author} {\bibfnamefont {L.}~\bibnamefont {Yang}},\ }\bibfield  {title}
	{\bibinfo {title} {Parity--time symmetry and exceptional points in
			photonics},\ }\href {https://doi.org/10.1038/s41563-019-0304-9} {\bibfield
		{journal} {\bibinfo  {journal} {Nature Materials}\ }\textbf {\bibinfo
			{volume} {18}},\ \bibinfo {pages} {783} (\bibinfo {year} {2019})}\BibitemShut
	{NoStop}%
	\bibitem [{\citenamefont {Parto}\ \emph {et~al.}(2021)\citenamefont {Parto},
		\citenamefont {Liu}, \citenamefont {Bahari}, \citenamefont {Khajavikhan},\
		and\ \citenamefont {Christodoulides}}]{Parto2021}%
	\BibitemOpen
	\bibfield  {author} {\bibinfo {author} {\bibfnamefont {M.}~\bibnamefont
			{Parto}}, \bibinfo {author} {\bibfnamefont {Y.~G.~N.}\ \bibnamefont {Liu}},
		\bibinfo {author} {\bibfnamefont {B.}~\bibnamefont {Bahari}}, \bibinfo
		{author} {\bibfnamefont {M.}~\bibnamefont {Khajavikhan}},\ and\ \bibinfo
		{author} {\bibfnamefont {D.~N.}\ \bibnamefont {Christodoulides}},\ }\bibfield
	{title} {\bibinfo {title} {Non-{H}ermitian and topological photonics: optics
			at an exceptional point},\ }\href {https://doi.org/10.1515/nanoph-2020-0434}
	{\bibfield  {journal} {\bibinfo  {journal} {Nanophotonics}\ }\textbf
		{\bibinfo {volume} {10}},\ \bibinfo {pages} {403} (\bibinfo {year}
		{2021})}\BibitemShut {NoStop}%
	\bibitem [{\citenamefont {Kato}(1995)}]{Kato1995}%
	\BibitemOpen
	\bibfield  {author} {\bibinfo {author} {\bibfnamefont {T.}~\bibnamefont
			{Kato}},\ }\href {https://doi.org/10.1007/978-3-642-66282-9} {\emph {\bibinfo
			{title} {Perturbation theory for linear operators}}}\ (\bibinfo  {publisher}
	{Springer, Berlin, Heidelberg},\ \bibinfo {year} {1995})\BibitemShut
	{NoStop}%
	\bibitem [{\citenamefont {Berry}(2004)}]{Berry2004}%
	\BibitemOpen
	\bibfield  {author} {\bibinfo {author} {\bibfnamefont {M.}~\bibnamefont
			{Berry}},\ }\bibfield  {title} {\bibinfo {title} {Physics of nonhermitian
			degeneracies},\ }\href {https://doi.org/10.1023/B:CJOP.0000044002.05657.04}
	{\bibfield  {journal} {\bibinfo  {journal} {Czech. J. Phys.}\ }\textbf
		{\bibinfo {volume} {54}},\ \bibinfo {pages} {1039} (\bibinfo {year}
		{2004})}\BibitemShut {NoStop}%
	\bibitem [{\citenamefont {Heiss}(2012)}]{Heiss2012}%
	\BibitemOpen
	\bibfield  {author} {\bibinfo {author} {\bibfnamefont {W.~D.}\ \bibnamefont
			{Heiss}},\ }\bibfield  {title} {\bibinfo {title} {The physics of exceptional
			points},\ }\href {https://doi.org/10.1088/1751-8113/45/44/444016} {\bibfield
		{journal} {\bibinfo  {journal} {J. Phys. A: Math. Theor.}\ }\textbf {\bibinfo
			{volume} {45}},\ \bibinfo {pages} {444016} (\bibinfo {year}
		{2012})}\BibitemShut {NoStop}%
	\bibitem [{\citenamefont {Dembowski}\ \emph {et~al.}(2001)\citenamefont
		{Dembowski}, \citenamefont {Gr\"af}, \citenamefont {Harney}, \citenamefont
		{Heine}, \citenamefont {Heiss}, \citenamefont {Rehfeld},\ and\ \citenamefont
		{Richter}}]{Dembowski2001}%
	\BibitemOpen
	\bibfield  {author} {\bibinfo {author} {\bibfnamefont {C.}~\bibnamefont
			{Dembowski}}, \bibinfo {author} {\bibfnamefont {H.-D.}\ \bibnamefont
			{Gr\"af}}, \bibinfo {author} {\bibfnamefont {H.~L.}\ \bibnamefont {Harney}},
		\bibinfo {author} {\bibfnamefont {A.}~\bibnamefont {Heine}}, \bibinfo
		{author} {\bibfnamefont {W.~D.}\ \bibnamefont {Heiss}}, \bibinfo {author}
		{\bibfnamefont {H.}~\bibnamefont {Rehfeld}},\ and\ \bibinfo {author}
		{\bibfnamefont {A.}~\bibnamefont {Richter}},\ }\bibfield  {title} {\bibinfo
		{title} {Experimental observation of the topological structure of exceptional
			points},\ }\href {https://doi.org/10.1103/PhysRevLett.86.787} {\bibfield
		{journal} {\bibinfo  {journal} {Phys. Rev. Lett.}\ }\textbf {\bibinfo
			{volume} {86}},\ \bibinfo {pages} {787} (\bibinfo {year} {2001})}\BibitemShut
	{NoStop}%
	\bibitem [{\citenamefont {Lin}\ \emph {et~al.}(2011)\citenamefont {Lin},
		\citenamefont {Ramezani}, \citenamefont {Eichelkraut}, \citenamefont
		{Kottos}, \citenamefont {Cao},\ and\ \citenamefont
		{Christodoulides}}]{Lin2011}%
	\BibitemOpen
	\bibfield  {author} {\bibinfo {author} {\bibfnamefont {Z.}~\bibnamefont
			{Lin}}, \bibinfo {author} {\bibfnamefont {H.}~\bibnamefont {Ramezani}},
		\bibinfo {author} {\bibfnamefont {T.}~\bibnamefont {Eichelkraut}}, \bibinfo
		{author} {\bibfnamefont {T.}~\bibnamefont {Kottos}}, \bibinfo {author}
		{\bibfnamefont {H.}~\bibnamefont {Cao}},\ and\ \bibinfo {author}
		{\bibfnamefont {D.~N.}\ \bibnamefont {Christodoulides}},\ }\bibfield  {title}
	{\bibinfo {title} {Unidirectional invisibility induced by {PT}-symmetric
			periodic structures},\ }\href
	{https://doi.org/10.1103/PhysRevLett.106.213901} {\bibfield  {journal}
		{\bibinfo  {journal} {Phys. Rev. Lett.}\ }\textbf {\bibinfo {volume} {106}},\
		\bibinfo {pages} {213901} (\bibinfo {year} {2011})}\BibitemShut {NoStop}%
	\bibitem [{\citenamefont {Feng}\ \emph {et~al.}(2013)\citenamefont {Feng},
		\citenamefont {Xu}, \citenamefont {Fegadolli}, \citenamefont {Lu},
		\citenamefont {Oliveira}, \citenamefont {Almeida}, \citenamefont {Chen},\
		and\ \citenamefont {Scherer}}]{Feng2013}%
	\BibitemOpen
	\bibfield  {author} {\bibinfo {author} {\bibfnamefont {L.}~\bibnamefont
			{Feng}}, \bibinfo {author} {\bibfnamefont {Y.-L.}\ \bibnamefont {Xu}},
		\bibinfo {author} {\bibfnamefont {W.~S.}\ \bibnamefont {Fegadolli}}, \bibinfo
		{author} {\bibfnamefont {M.-H.}\ \bibnamefont {Lu}}, \bibinfo {author}
		{\bibfnamefont {J.~E.~B.}\ \bibnamefont {Oliveira}}, \bibinfo {author}
		{\bibfnamefont {V.~R.}\ \bibnamefont {Almeida}}, \bibinfo {author}
		{\bibfnamefont {Y.-F.}\ \bibnamefont {Chen}},\ and\ \bibinfo {author}
		{\bibfnamefont {A.}~\bibnamefont {Scherer}},\ }\bibfield  {title} {\bibinfo
		{title} {Experimental demonstration of a unidirectional reflectionless
			parity-time metamaterial at optical frequencies},\ }\href
	{https://doi.org/10.1038/nmat3495} {\bibfield  {journal} {\bibinfo  {journal}
			{Nat. Mater.}\ }\textbf {\bibinfo {volume} {12}},\ \bibinfo {pages} {108}
		(\bibinfo {year} {2013})}\BibitemShut {NoStop}%
	\bibitem [{\citenamefont {Lawrence}\ \emph {et~al.}(2014)\citenamefont
		{Lawrence}, \citenamefont {Xu}, \citenamefont {Zhang}, \citenamefont {Cong},
		\citenamefont {Han}, \citenamefont {Zhang},\ and\ \citenamefont
		{Zhang}}]{Lawrence2014}%
	\BibitemOpen
	\bibfield  {author} {\bibinfo {author} {\bibfnamefont {M.}~\bibnamefont
			{Lawrence}}, \bibinfo {author} {\bibfnamefont {N.}~\bibnamefont {Xu}},
		\bibinfo {author} {\bibfnamefont {X.}~\bibnamefont {Zhang}}, \bibinfo
		{author} {\bibfnamefont {L.}~\bibnamefont {Cong}}, \bibinfo {author}
		{\bibfnamefont {J.}~\bibnamefont {Han}}, \bibinfo {author} {\bibfnamefont
			{W.}~\bibnamefont {Zhang}},\ and\ \bibinfo {author} {\bibfnamefont
			{S.}~\bibnamefont {Zhang}},\ }\bibfield  {title} {\bibinfo {title}
		{Manifestation of {$PT$} symmetry breaking in polarization space with
			terahertz metasurfaces},\ }\href
	{https://doi.org/10.1103/PhysRevLett.113.093901} {\bibfield  {journal}
		{\bibinfo  {journal} {Phys. Rev. Lett.}\ }\textbf {\bibinfo {volume} {113}},\
		\bibinfo {pages} {093901} (\bibinfo {year} {2014})}\BibitemShut {NoStop}%
	\bibitem [{\citenamefont {Peng}\ \emph {et~al.}(2016)\citenamefont {Peng},
		\citenamefont {{\"O}zdemir}, \citenamefont {Liertzer}, \citenamefont {Chen},
		\citenamefont {Kramer}, \citenamefont {Y{\i}lmaz}, \citenamefont {Wiersig},
		\citenamefont {Rotter},\ and\ \citenamefont {Yang}}]{Peng2016}%
	\BibitemOpen
	\bibfield  {author} {\bibinfo {author} {\bibfnamefont {B.}~\bibnamefont
			{Peng}}, \bibinfo {author} {\bibfnamefont {{\c S}.~K.}\ \bibnamefont
			{{\"O}zdemir}}, \bibinfo {author} {\bibfnamefont {M.}~\bibnamefont
			{Liertzer}}, \bibinfo {author} {\bibfnamefont {W.}~\bibnamefont {Chen}},
		\bibinfo {author} {\bibfnamefont {J.}~\bibnamefont {Kramer}}, \bibinfo
		{author} {\bibfnamefont {H.}~\bibnamefont {Y{\i}lmaz}}, \bibinfo {author}
		{\bibfnamefont {J.}~\bibnamefont {Wiersig}}, \bibinfo {author} {\bibfnamefont
			{S.}~\bibnamefont {Rotter}},\ and\ \bibinfo {author} {\bibfnamefont
			{L.}~\bibnamefont {Yang}},\ }\bibfield  {title} {\bibinfo {title} {Chiral
			modes and directional lasing at exceptional points},\ }\href
	{https://doi.org/10.1073/pnas.1603318113} {\bibfield  {journal} {\bibinfo
			{journal} {Proc. Natl. Acad. Sci.}\ }\textbf {\bibinfo {volume} {113}},\
		\bibinfo {pages} {6845} (\bibinfo {year} {2016})}\BibitemShut {NoStop}%
	\bibitem [{\citenamefont {Leykam}\ \emph {et~al.}(2017)\citenamefont {Leykam},
		\citenamefont {Bliokh}, \citenamefont {Huang}, \citenamefont {Chong},\ and\
		\citenamefont {Nori}}]{Leykam2017}%
	\BibitemOpen
	\bibfield  {author} {\bibinfo {author} {\bibfnamefont {D.}~\bibnamefont
			{Leykam}}, \bibinfo {author} {\bibfnamefont {K.~Y.}\ \bibnamefont {Bliokh}},
		\bibinfo {author} {\bibfnamefont {C.}~\bibnamefont {Huang}}, \bibinfo
		{author} {\bibfnamefont {Y.~D.}\ \bibnamefont {Chong}},\ and\ \bibinfo
		{author} {\bibfnamefont {F.}~\bibnamefont {Nori}},\ }\bibfield  {title}
	{\bibinfo {title} {Edge modes, degeneracies, and topological numbers in
			non-{H}ermitian systems},\ }\href
	{https://doi.org/10.1103/PhysRevLett.118.040401} {\bibfield  {journal}
		{\bibinfo  {journal} {Phys. Rev. Lett.}\ }\textbf {\bibinfo {volume} {118}},\
		\bibinfo {pages} {040401} (\bibinfo {year} {2017})}\BibitemShut {NoStop}%
	\bibitem [{\citenamefont {Gao}\ \emph {et~al.}(2019)\citenamefont {Gao},
		\citenamefont {Thompson}, \citenamefont {Dave}, \citenamefont {Fryslie},\
		and\ \citenamefont {Choquette}}]{Gao2019}%
	\BibitemOpen
	\bibfield  {author} {\bibinfo {author} {\bibfnamefont {Z.}~\bibnamefont
			{Gao}}, \bibinfo {author} {\bibfnamefont {B.~J.}\ \bibnamefont {Thompson}},
		\bibinfo {author} {\bibfnamefont {H.}~\bibnamefont {Dave}}, \bibinfo {author}
		{\bibfnamefont {S.~T.~M.}\ \bibnamefont {Fryslie}},\ and\ \bibinfo {author}
		{\bibfnamefont {K.~D.}\ \bibnamefont {Choquette}},\ }\bibfield  {title}
	{\bibinfo {title} {Non-{H}ermiticity and exceptional points in coherently
			coupled vertical cavity laser diode arrays},\ }\href
	{https://doi.org/10.1063/1.5083084} {\bibfield  {journal} {\bibinfo
			{journal} {Appl. Phys. Lett.}\ }\textbf {\bibinfo {volume} {114}},\ \bibinfo
		{pages} {061103} (\bibinfo {year} {2019})}\BibitemShut {NoStop}%
	\bibitem [{\citenamefont {Wang}\ \emph {et~al.}(2020)\citenamefont {Wang},
		\citenamefont {Jiang}, \citenamefont {Zhao}, \citenamefont {Zhang},
		\citenamefont {Hsu}, \citenamefont {Peng}, \citenamefont {Stone},
		\citenamefont {Jiang},\ and\ \citenamefont {Yang}}]{Wang2020}%
	\BibitemOpen
	\bibfield  {author} {\bibinfo {author} {\bibfnamefont {C.}~\bibnamefont
			{Wang}}, \bibinfo {author} {\bibfnamefont {X.}~\bibnamefont {Jiang}},
		\bibinfo {author} {\bibfnamefont {G.}~\bibnamefont {Zhao}}, \bibinfo {author}
		{\bibfnamefont {M.}~\bibnamefont {Zhang}}, \bibinfo {author} {\bibfnamefont
			{C.~W.}\ \bibnamefont {Hsu}}, \bibinfo {author} {\bibfnamefont
			{B.}~\bibnamefont {Peng}}, \bibinfo {author} {\bibfnamefont {A.~D.}\
			\bibnamefont {Stone}}, \bibinfo {author} {\bibfnamefont {L.}~\bibnamefont
			{Jiang}},\ and\ \bibinfo {author} {\bibfnamefont {L.}~\bibnamefont {Yang}},\
	}\bibfield  {title} {\bibinfo {title} {Electromagnetically induced
			transparency at a chiral exceptional point},\ }\href
	{https://doi.org/10.1038/s41567-019-0746-7} {\bibfield  {journal} {\bibinfo
			{journal} {Nat. Phys.}\ }\textbf {\bibinfo {volume} {16}},\ \bibinfo {pages}
		{334} (\bibinfo {year} {2020})}\BibitemShut {NoStop}%
	\bibitem [{\citenamefont {Tang}\ \emph {et~al.}(2020)\citenamefont {Tang},
		\citenamefont {Jiang}, \citenamefont {Ding}, \citenamefont {Xiao},
		\citenamefont {Zhang}, \citenamefont {Chan},\ and\ \citenamefont
		{Ma}}]{Tang2020}%
	\BibitemOpen
	\bibfield  {author} {\bibinfo {author} {\bibfnamefont {W.}~\bibnamefont
			{Tang}}, \bibinfo {author} {\bibfnamefont {X.}~\bibnamefont {Jiang}},
		\bibinfo {author} {\bibfnamefont {K.}~\bibnamefont {Ding}}, \bibinfo {author}
		{\bibfnamefont {Y.-X.}\ \bibnamefont {Xiao}}, \bibinfo {author}
		{\bibfnamefont {Z.-Q.}\ \bibnamefont {Zhang}}, \bibinfo {author}
		{\bibfnamefont {C.~T.}\ \bibnamefont {Chan}},\ and\ \bibinfo {author}
		{\bibfnamefont {G.}~\bibnamefont {Ma}},\ }\bibfield  {title} {\bibinfo
		{title} {Exceptional nexus with a hybrid topological invariant},\ }\href
	{https://doi.org/10.1126/science.abd8872} {\bibfield  {journal} {\bibinfo
			{journal} {Science}\ }\textbf {\bibinfo {volume} {370}},\ \bibinfo {pages}
		{1077} (\bibinfo {year} {2020})}\BibitemShut {NoStop}%
	\bibitem [{\citenamefont {del Pino}\ \emph {et~al.}(2022)\citenamefont {del
			Pino}, \citenamefont {Slim},\ and\ \citenamefont {Verhagen}}]{delPino2022}%
	\BibitemOpen
	\bibfield  {author} {\bibinfo {author} {\bibfnamefont {J.}~\bibnamefont {del
				Pino}}, \bibinfo {author} {\bibfnamefont {J.~J.}\ \bibnamefont {Slim}},\ and\
		\bibinfo {author} {\bibfnamefont {E.}~\bibnamefont {Verhagen}},\ }\bibfield
	{title} {\bibinfo {title} {Non-{H}ermitian chiral phononics through
			optomechanically induced squeezing},\ }\href
	{https://doi.org/10.1038/s41586-022-04609-0} {\bibfield  {journal} {\bibinfo
			{journal} {Nature}\ }\textbf {\bibinfo {volume} {606}},\ \bibinfo {pages}
		{82} (\bibinfo {year} {2022})}\BibitemShut {NoStop}%
	\bibitem [{\citenamefont {Patil}\ \emph {et~al.}(2022)\citenamefont {Patil},
		\citenamefont {H{\"o}ller}, \citenamefont {Henry}, \citenamefont {Guria},
		\citenamefont {Zhang}, \citenamefont {Jiang}, \citenamefont {Kralj},
		\citenamefont {Read},\ and\ \citenamefont {Harris}}]{Patil2022}%
	\BibitemOpen
	\bibfield  {author} {\bibinfo {author} {\bibfnamefont {Y.~S.~S.}\
			\bibnamefont {Patil}}, \bibinfo {author} {\bibfnamefont {J.}~\bibnamefont
			{H{\"o}ller}}, \bibinfo {author} {\bibfnamefont {P.~A.}\ \bibnamefont
			{Henry}}, \bibinfo {author} {\bibfnamefont {C.}~\bibnamefont {Guria}},
		\bibinfo {author} {\bibfnamefont {Y.}~\bibnamefont {Zhang}}, \bibinfo
		{author} {\bibfnamefont {L.}~\bibnamefont {Jiang}}, \bibinfo {author}
		{\bibfnamefont {N.}~\bibnamefont {Kralj}}, \bibinfo {author} {\bibfnamefont
			{N.}~\bibnamefont {Read}},\ and\ \bibinfo {author} {\bibfnamefont {J.~G.~E.}\
			\bibnamefont {Harris}},\ }\bibfield  {title} {\bibinfo {title} {Measuring the
			knot of non-{H}ermitian degeneracies and non-commuting braids},\ }\href
	{https://doi.org/10.1038/s41586-022-04796-w} {\bibfield  {journal} {\bibinfo
			{journal} {Nature}\ }\textbf {\bibinfo {volume} {607}},\ \bibinfo {pages}
		{271} (\bibinfo {year} {2022})}\BibitemShut {NoStop}%
	\bibitem [{\citenamefont {Feng}\ \emph {et~al.}(2014)\citenamefont {Feng},
		\citenamefont {Wong}, \citenamefont {Ma}, \citenamefont {Wang},\ and\
		\citenamefont {Zhang}}]{Feng2014sin}%
	\BibitemOpen
	\bibfield  {author} {\bibinfo {author} {\bibfnamefont {L.}~\bibnamefont
			{Feng}}, \bibinfo {author} {\bibfnamefont {Z.~J.}\ \bibnamefont {Wong}},
		\bibinfo {author} {\bibfnamefont {R.-M.}\ \bibnamefont {Ma}}, \bibinfo
		{author} {\bibfnamefont {Y.}~\bibnamefont {Wang}},\ and\ \bibinfo {author}
		{\bibfnamefont {X.}~\bibnamefont {Zhang}},\ }\bibfield  {title} {\bibinfo
		{title} {Single-mode laser by parity-time symmetry breaking},\ }\href
	{https://doi.org/10.1126/science.1258479} {\bibfield  {journal} {\bibinfo
			{journal} {Science}\ }\textbf {\bibinfo {volume} {346}},\ \bibinfo {pages}
		{972} (\bibinfo {year} {2014})}\BibitemShut {NoStop}%
	\bibitem [{\citenamefont {Hodaei}\ \emph {et~al.}(2014)\citenamefont {Hodaei},
		\citenamefont {Miri}, \citenamefont {Heinrich}, \citenamefont
		{Christodoulides},\ and\ \citenamefont {Khajavikhan}}]{Hodaei2014}%
	\BibitemOpen
	\bibfield  {author} {\bibinfo {author} {\bibfnamefont {H.}~\bibnamefont
			{Hodaei}}, \bibinfo {author} {\bibfnamefont {M.-A.}\ \bibnamefont {Miri}},
		\bibinfo {author} {\bibfnamefont {M.}~\bibnamefont {Heinrich}}, \bibinfo
		{author} {\bibfnamefont {D.~N.}\ \bibnamefont {Christodoulides}},\ and\
		\bibinfo {author} {\bibfnamefont {M.}~\bibnamefont {Khajavikhan}},\
	}\bibfield  {title} {\bibinfo {title} {Parity-time-symmetric microring
			lasers},\ }\href {https://doi.org/10.1126/science.1258480} {\bibfield
		{journal} {\bibinfo  {journal} {Science}\ }\textbf {\bibinfo {volume}
			{346}},\ \bibinfo {pages} {975} (\bibinfo {year} {2014})}\BibitemShut
	{NoStop}%
	\bibitem [{\citenamefont {Brandstetter}\ \emph {et~al.}(2014)\citenamefont
		{Brandstetter}, \citenamefont {Liertzer}, \citenamefont {Deutsch},
		\citenamefont {Klang}, \citenamefont {Sch{\"o}berl}, \citenamefont
		{T{\"u}reci}, \citenamefont {Strasser}, \citenamefont {Unterrainer},\ and\
		\citenamefont {Rotter}}]{Brandstetter2014}%
	\BibitemOpen
	\bibfield  {author} {\bibinfo {author} {\bibfnamefont {M.}~\bibnamefont
			{Brandstetter}}, \bibinfo {author} {\bibfnamefont {M.}~\bibnamefont
			{Liertzer}}, \bibinfo {author} {\bibfnamefont {C.}~\bibnamefont {Deutsch}},
		\bibinfo {author} {\bibfnamefont {P.}~\bibnamefont {Klang}}, \bibinfo
		{author} {\bibfnamefont {J.}~\bibnamefont {Sch{\"o}berl}}, \bibinfo {author}
		{\bibfnamefont {H.~E.}\ \bibnamefont {T{\"u}reci}}, \bibinfo {author}
		{\bibfnamefont {G.}~\bibnamefont {Strasser}}, \bibinfo {author}
		{\bibfnamefont {K.}~\bibnamefont {Unterrainer}},\ and\ \bibinfo {author}
		{\bibfnamefont {S.}~\bibnamefont {Rotter}},\ }\bibfield  {title} {\bibinfo
		{title} {Reversing the pump dependence of a laser at an exceptional point},\
	}\href {https://doi.org/10.1038/ncomms5034} {\bibfield  {journal} {\bibinfo
			{journal} {Nat. Commun.}\ }\textbf {\bibinfo {volume} {5}},\ \bibinfo {pages}
		{4034} (\bibinfo {year} {2014})}\BibitemShut {NoStop}%
	\bibitem [{\citenamefont {Miao}\ \emph {et~al.}(2016)\citenamefont {Miao},
		\citenamefont {Zhang}, \citenamefont {Sun}, \citenamefont {Walasik},
		\citenamefont {Longhi}, \citenamefont {Litchinitser},\ and\ \citenamefont
		{Feng}}]{Miao2016}%
	\BibitemOpen
	\bibfield  {author} {\bibinfo {author} {\bibfnamefont {P.}~\bibnamefont
			{Miao}}, \bibinfo {author} {\bibfnamefont {Z.}~\bibnamefont {Zhang}},
		\bibinfo {author} {\bibfnamefont {J.}~\bibnamefont {Sun}}, \bibinfo {author}
		{\bibfnamefont {W.}~\bibnamefont {Walasik}}, \bibinfo {author} {\bibfnamefont
			{S.}~\bibnamefont {Longhi}}, \bibinfo {author} {\bibfnamefont {N.~M.}\
			\bibnamefont {Litchinitser}},\ and\ \bibinfo {author} {\bibfnamefont
			{L.}~\bibnamefont {Feng}},\ }\bibfield  {title} {\bibinfo {title} {Orbital
			angular momentum microlaser},\ }\href
	{https://doi.org/10.1126/science.aaf8533} {\bibfield  {journal} {\bibinfo
			{journal} {Science}\ }\textbf {\bibinfo {volume} {353}},\ \bibinfo {pages}
		{464} (\bibinfo {year} {2016})}\BibitemShut {NoStop}%
	\bibitem [{\citenamefont {Liao}\ \emph {et~al.}(2023)\citenamefont {Liao},
		\citenamefont {Zhong}, \citenamefont {Du}, \citenamefont {Liu}, \citenamefont
		{Li}, \citenamefont {Wu}, \citenamefont {Deng}, \citenamefont {Lu},
		\citenamefont {Wang}, \citenamefont {Chan}, \citenamefont {Song},
		\citenamefont {Wang}, \citenamefont {Liu}, \citenamefont {Hu},\ and\
		\citenamefont {Gong}}]{Liao2023}%
	\BibitemOpen
	\bibfield  {author} {\bibinfo {author} {\bibfnamefont {K.}~\bibnamefont
			{Liao}}, \bibinfo {author} {\bibfnamefont {Y.}~\bibnamefont {Zhong}},
		\bibinfo {author} {\bibfnamefont {Z.}~\bibnamefont {Du}}, \bibinfo {author}
		{\bibfnamefont {G.}~\bibnamefont {Liu}}, \bibinfo {author} {\bibfnamefont
			{C.}~\bibnamefont {Li}}, \bibinfo {author} {\bibfnamefont {X.}~\bibnamefont
			{Wu}}, \bibinfo {author} {\bibfnamefont {C.}~\bibnamefont {Deng}}, \bibinfo
		{author} {\bibfnamefont {C.}~\bibnamefont {Lu}}, \bibinfo {author}
		{\bibfnamefont {X.}~\bibnamefont {Wang}}, \bibinfo {author} {\bibfnamefont
			{C.~T.}\ \bibnamefont {Chan}}, \bibinfo {author} {\bibfnamefont
			{Q.}~\bibnamefont {Song}}, \bibinfo {author} {\bibfnamefont {S.}~\bibnamefont
			{Wang}}, \bibinfo {author} {\bibfnamefont {X.}~\bibnamefont {Liu}}, \bibinfo
		{author} {\bibfnamefont {X.}~\bibnamefont {Hu}},\ and\ \bibinfo {author}
		{\bibfnamefont {Q.}~\bibnamefont {Gong}},\ }\bibfield  {title} {\bibinfo
		{title} {On-chip integrated exceptional surface microlaser},\ }\href
	{https://doi.org/10.1126/sciadv.adf3470} {\bibfield  {journal} {\bibinfo
			{journal} {Sci. Adv.}\ }\textbf {\bibinfo {volume} {9}},\ \bibinfo {pages}
		{eadf3470} (\bibinfo {year} {2023})}\BibitemShut {NoStop}%
	\bibitem [{\citenamefont {Yoo}\ \emph {et~al.}(2011)\citenamefont {Yoo},
		\citenamefont {Sim},\ and\ \citenamefont {Schomerus}}]{Yoo2011}%
	\BibitemOpen
	\bibfield  {author} {\bibinfo {author} {\bibfnamefont {G.}~\bibnamefont
			{Yoo}}, \bibinfo {author} {\bibfnamefont {H.-S.}\ \bibnamefont {Sim}},\ and\
		\bibinfo {author} {\bibfnamefont {H.}~\bibnamefont {Schomerus}},\ }\bibfield
	{title} {\bibinfo {title} {Quantum noise and mode nonorthogonality in
			non-{H}ermitian $\mathcal{PT}$-symmetric optical resonators},\ }\href
	{https://doi.org/10.1103/PhysRevA.84.063833} {\bibfield  {journal} {\bibinfo
			{journal} {Phys. Rev. A}\ }\textbf {\bibinfo {volume} {84}},\ \bibinfo
		{pages} {063833} (\bibinfo {year} {2011})}\BibitemShut {NoStop}%
	\bibitem [{\citenamefont {Makris}\ \emph {et~al.}(2015)\citenamefont {Makris},
		\citenamefont {Musslimani}, \citenamefont {Christodoulides},\ and\
		\citenamefont {Rotter}}]{Makris2015}%
	\BibitemOpen
	\bibfield  {author} {\bibinfo {author} {\bibfnamefont {K.~G.}\ \bibnamefont
			{Makris}}, \bibinfo {author} {\bibfnamefont {Z.~H.}\ \bibnamefont
			{Musslimani}}, \bibinfo {author} {\bibfnamefont {D.~N.}\ \bibnamefont
			{Christodoulides}},\ and\ \bibinfo {author} {\bibfnamefont {S.}~\bibnamefont
			{Rotter}},\ }\bibfield  {title} {\bibinfo {title} {Constant-intensity waves
			and their modulation instability in non-{H}ermitian potentials},\ }\href
	{https://doi.org/10.1038/ncomms8257} {\bibfield  {journal} {\bibinfo
			{journal} {Nat. Commun.}\ }\textbf {\bibinfo {volume} {6}},\ \bibinfo {pages}
		{7257} (\bibinfo {year} {2015})}\BibitemShut {NoStop}%
	\bibitem [{\citenamefont {Lin}\ \emph {et~al.}(2016)\citenamefont {Lin},
		\citenamefont {Pick}, \citenamefont {Lon\v{c}ar},\ and\ \citenamefont
		{Rodriguez}}]{Lin2016}%
	\BibitemOpen
	\bibfield  {author} {\bibinfo {author} {\bibfnamefont {Z.}~\bibnamefont
			{Lin}}, \bibinfo {author} {\bibfnamefont {A.}~\bibnamefont {Pick}}, \bibinfo
		{author} {\bibfnamefont {M.}~\bibnamefont {Lon\v{c}ar}},\ and\ \bibinfo
		{author} {\bibfnamefont {A.~W.}\ \bibnamefont {Rodriguez}},\ }\bibfield
	{title} {\bibinfo {title} {Enhanced spontaneous emission at third-order
			{D}irac exceptional points in inverse-designed photonic crystals},\ }\href
	{https://doi.org/10.1103/PhysRevLett.117.107402} {\bibfield  {journal}
		{\bibinfo  {journal} {Phys. Rev. Lett.}\ }\textbf {\bibinfo {volume} {117}},\
		\bibinfo {pages} {107402} (\bibinfo {year} {2016})}\BibitemShut {NoStop}%
	\bibitem [{\citenamefont {Pick}\ \emph {et~al.}(2017)\citenamefont {Pick},
		\citenamefont {Zhen}, \citenamefont {Miller}, \citenamefont {Hsu},
		\citenamefont {Hernandez}, \citenamefont {Rodriguez}, \citenamefont
		{Solja\v{c}i\'{c}},\ and\ \citenamefont {Johnson}}]{Pick2017gen}%
	\BibitemOpen
	\bibfield  {author} {\bibinfo {author} {\bibfnamefont {A.}~\bibnamefont
			{Pick}}, \bibinfo {author} {\bibfnamefont {B.}~\bibnamefont {Zhen}}, \bibinfo
		{author} {\bibfnamefont {O.~D.}\ \bibnamefont {Miller}}, \bibinfo {author}
		{\bibfnamefont {C.~W.}\ \bibnamefont {Hsu}}, \bibinfo {author} {\bibfnamefont
			{F.}~\bibnamefont {Hernandez}}, \bibinfo {author} {\bibfnamefont {A.~W.}\
			\bibnamefont {Rodriguez}}, \bibinfo {author} {\bibfnamefont {M.}~\bibnamefont
			{Solja\v{c}i\'{c}}},\ and\ \bibinfo {author} {\bibfnamefont {S.~G.}\
			\bibnamefont {Johnson}},\ }\bibfield  {title} {\bibinfo {title} {General
			theory of spontaneous emission near exceptional points},\ }\href
	{https://doi.org/10.1364/OE.25.012325} {\bibfield  {journal} {\bibinfo
			{journal} {Opt. Express}\ }\textbf {\bibinfo {volume} {25}},\ \bibinfo
		{pages} {12325} (\bibinfo {year} {2017})}\BibitemShut {NoStop}%
	\bibitem [{\citenamefont {Takata}\ \emph {et~al.}(2021)\citenamefont {Takata},
		\citenamefont {Nozaki}, \citenamefont {Kuramochi}, \citenamefont {Matsuo},
		\citenamefont {Takeda}, \citenamefont {Fujii}, \citenamefont {Kita},
		\citenamefont {Shinya},\ and\ \citenamefont {Notomi}}]{Takata2021}%
	\BibitemOpen
	\bibfield  {author} {\bibinfo {author} {\bibfnamefont {K.}~\bibnamefont
			{Takata}}, \bibinfo {author} {\bibfnamefont {K.}~\bibnamefont {Nozaki}},
		\bibinfo {author} {\bibfnamefont {E.}~\bibnamefont {Kuramochi}}, \bibinfo
		{author} {\bibfnamefont {S.}~\bibnamefont {Matsuo}}, \bibinfo {author}
		{\bibfnamefont {K.}~\bibnamefont {Takeda}}, \bibinfo {author} {\bibfnamefont
			{T.}~\bibnamefont {Fujii}}, \bibinfo {author} {\bibfnamefont
			{S.}~\bibnamefont {Kita}}, \bibinfo {author} {\bibfnamefont {A.}~\bibnamefont
			{Shinya}},\ and\ \bibinfo {author} {\bibfnamefont {M.}~\bibnamefont
			{Notomi}},\ }\bibfield  {title} {\bibinfo {title} {Observing exceptional
			point degeneracy of radiation with electrically pumped photonic crystal
			coupled-nanocavity lasers},\ }\href {https://doi.org/10.1364/OPTICA.412596}
	{\bibfield  {journal} {\bibinfo  {journal} {Optica}\ }\textbf {\bibinfo
			{volume} {8}},\ \bibinfo {pages} {184} (\bibinfo {year} {2021})}\BibitemShut
	{NoStop}%
	\bibitem [{\citenamefont {Yulaev}\ \emph {et~al.}(2022)\citenamefont {Yulaev},
		\citenamefont {Kim}, \citenamefont {Li}, \citenamefont {Westly},
		\citenamefont {Roxworthy}, \citenamefont {Srinivasan},\ and\ \citenamefont
		{Aksyuk}}]{Yulaev2022}%
	\BibitemOpen
	\bibfield  {author} {\bibinfo {author} {\bibfnamefont {A.}~\bibnamefont
			{Yulaev}}, \bibinfo {author} {\bibfnamefont {S.}~\bibnamefont {Kim}},
		\bibinfo {author} {\bibfnamefont {Q.}~\bibnamefont {Li}}, \bibinfo {author}
		{\bibfnamefont {D.~A.}\ \bibnamefont {Westly}}, \bibinfo {author}
		{\bibfnamefont {B.~J.}\ \bibnamefont {Roxworthy}}, \bibinfo {author}
		{\bibfnamefont {K.}~\bibnamefont {Srinivasan}},\ and\ \bibinfo {author}
		{\bibfnamefont {V.~A.}\ \bibnamefont {Aksyuk}},\ }\bibfield  {title}
	{\bibinfo {title} {Exceptional points in lossy media lead to deep polynomial
			wave penetration with spatially uniform power loss},\ }\href
	{https://doi.org/10.1038/s41565-022-01114-3} {\bibfield  {journal} {\bibinfo
			{journal} {Nat. Nanotechnol.}\ }\textbf {\bibinfo {volume} {17}},\ \bibinfo
		{pages} {583} (\bibinfo {year} {2022})}\BibitemShut {NoStop}%
	\bibitem [{\citenamefont {Wiersig}(2014)}]{Wiersig2014}%
	\BibitemOpen
	\bibfield  {author} {\bibinfo {author} {\bibfnamefont {J.}~\bibnamefont
			{Wiersig}},\ }\bibfield  {title} {\bibinfo {title} {Enhancing the sensitivity
			of frequency and energy splitting detection by using exceptional points:
			Application to microcavity sensors for single-particle detection},\ }\href
	{https://doi.org/10.1103/PhysRevLett.112.203901} {\bibfield  {journal}
		{\bibinfo  {journal} {Phys. Rev. Lett.}\ }\textbf {\bibinfo {volume} {112}},\
		\bibinfo {pages} {203901} (\bibinfo {year} {2014})}\BibitemShut {NoStop}%
	\bibitem [{\citenamefont {Hodaei}\ \emph {et~al.}(2017)\citenamefont {Hodaei},
		\citenamefont {Hassan}, \citenamefont {Wittek}, \citenamefont
		{Garcia-Gracia}, \citenamefont {El-Ganainy}, \citenamefont
		{Christodoulides},\ and\ \citenamefont {Khajavikhan}}]{Hodaei2017enh}%
	\BibitemOpen
	\bibfield  {author} {\bibinfo {author} {\bibfnamefont {H.}~\bibnamefont
			{Hodaei}}, \bibinfo {author} {\bibfnamefont {A.~U.}\ \bibnamefont {Hassan}},
		\bibinfo {author} {\bibfnamefont {S.}~\bibnamefont {Wittek}}, \bibinfo
		{author} {\bibfnamefont {H.}~\bibnamefont {Garcia-Gracia}}, \bibinfo {author}
		{\bibfnamefont {R.}~\bibnamefont {El-Ganainy}}, \bibinfo {author}
		{\bibfnamefont {D.~N.}\ \bibnamefont {Christodoulides}},\ and\ \bibinfo
		{author} {\bibfnamefont {M.}~\bibnamefont {Khajavikhan}},\ }\bibfield
	{title} {\bibinfo {title} {Enhanced sensitivity at higher-order exceptional
			points},\ }\href {https://doi.org/10.1038/nature23280} {\bibfield  {journal}
		{\bibinfo  {journal} {Nature}\ }\textbf {\bibinfo {volume} {548}},\ \bibinfo
		{pages} {187} (\bibinfo {year} {2017})}\BibitemShut {NoStop}%
	\bibitem [{\citenamefont {Chen}\ \emph {et~al.}(2017)\citenamefont {Chen},
		\citenamefont {{\"O}zdemir}, \citenamefont {Zhao}, \citenamefont {Wiersig},\
		and\ \citenamefont {Yang}}]{Chen2017exc}%
	\BibitemOpen
	\bibfield  {author} {\bibinfo {author} {\bibfnamefont {W.}~\bibnamefont
			{Chen}}, \bibinfo {author} {\bibfnamefont {{\c{S}}.~K.}\ \bibnamefont
			{{\"O}zdemir}}, \bibinfo {author} {\bibfnamefont {G.}~\bibnamefont {Zhao}},
		\bibinfo {author} {\bibfnamefont {J.}~\bibnamefont {Wiersig}},\ and\ \bibinfo
		{author} {\bibfnamefont {L.}~\bibnamefont {Yang}},\ }\bibfield  {title}
	{\bibinfo {title} {Exceptional points enhance sensing in an optical
			microcavity},\ }\href {https://doi.org/10.1038/nature23281} {\bibfield
		{journal} {\bibinfo  {journal} {Nature}\ }\textbf {\bibinfo {volume} {548}},\
		\bibinfo {pages} {192} (\bibinfo {year} {2017})}\BibitemShut {NoStop}%
	\bibitem [{\citenamefont {Park}\ \emph {et~al.}(2020)\citenamefont {Park},
		\citenamefont {Ndao}, \citenamefont {Cai}, \citenamefont {Hsu}, \citenamefont
		{Kodigala}, \citenamefont {Lepetit}, \citenamefont {Lo},\ and\ \citenamefont
		{Kant\'{e}}}]{Park2020}%
	\BibitemOpen
	\bibfield  {author} {\bibinfo {author} {\bibfnamefont {J.-H.}\ \bibnamefont
			{Park}}, \bibinfo {author} {\bibfnamefont {A.}~\bibnamefont {Ndao}}, \bibinfo
		{author} {\bibfnamefont {W.}~\bibnamefont {Cai}}, \bibinfo {author}
		{\bibfnamefont {L.}~\bibnamefont {Hsu}}, \bibinfo {author} {\bibfnamefont
			{A.}~\bibnamefont {Kodigala}}, \bibinfo {author} {\bibfnamefont
			{T.}~\bibnamefont {Lepetit}}, \bibinfo {author} {\bibfnamefont {Y.-H.}\
			\bibnamefont {Lo}},\ and\ \bibinfo {author} {\bibfnamefont {B.}~\bibnamefont
			{Kant\'{e}}},\ }\bibfield  {title} {\bibinfo {title}
		{Symmetry-breaking-induced plasmonic exceptional points and nanoscale
			sensing},\ }\href {https://doi.org/10.1038/s41567-020-0796-x} {\bibfield
		{journal} {\bibinfo  {journal} {Nat. Phys.}\ }\textbf {\bibinfo {volume}
			{16}},\ \bibinfo {pages} {462} (\bibinfo {year} {2020})}\BibitemShut
	{NoStop}%
	\bibitem [{\citenamefont {Kononchuk}\ \emph {et~al.}(2022)\citenamefont
		{Kononchuk}, \citenamefont {Cai}, \citenamefont {Ellis}, \citenamefont
		{Thevamaran},\ and\ \citenamefont {Kottos}}]{Kononchuk2022}%
	\BibitemOpen
	\bibfield  {author} {\bibinfo {author} {\bibfnamefont {R.}~\bibnamefont
			{Kononchuk}}, \bibinfo {author} {\bibfnamefont {J.}~\bibnamefont {Cai}},
		\bibinfo {author} {\bibfnamefont {F.}~\bibnamefont {Ellis}}, \bibinfo
		{author} {\bibfnamefont {R.}~\bibnamefont {Thevamaran}},\ and\ \bibinfo
		{author} {\bibfnamefont {T.}~\bibnamefont {Kottos}},\ }\bibfield  {title}
	{\bibinfo {title} {Exceptional-point-based accelerometers with enhanced
			signal-to-noise ratio},\ }\href {https://doi.org/10.1038/s41586-022-04904-w}
	{\bibfield  {journal} {\bibinfo  {journal} {Nature}\ }\textbf {\bibinfo
			{volume} {607}},\ \bibinfo {pages} {697} (\bibinfo {year}
		{2022})}\BibitemShut {NoStop}%
	\bibitem [{\citenamefont {Doppler}\ \emph {et~al.}(2016)\citenamefont
		{Doppler}, \citenamefont {Mailybaev}, \citenamefont {B{\"o}hm}, \citenamefont
		{Kuhl}, \citenamefont {Girschik}, \citenamefont {Libisch}, \citenamefont
		{Milburn}, \citenamefont {Rabl}, \citenamefont {Moiseyev},\ and\
		\citenamefont {Rotter}}]{Doppler2016}%
	\BibitemOpen
	\bibfield  {author} {\bibinfo {author} {\bibfnamefont {J.}~\bibnamefont
			{Doppler}}, \bibinfo {author} {\bibfnamefont {A.~A.}\ \bibnamefont
			{Mailybaev}}, \bibinfo {author} {\bibfnamefont {J.}~\bibnamefont {B{\"o}hm}},
		\bibinfo {author} {\bibfnamefont {U.}~\bibnamefont {Kuhl}}, \bibinfo {author}
		{\bibfnamefont {A.}~\bibnamefont {Girschik}}, \bibinfo {author}
		{\bibfnamefont {F.}~\bibnamefont {Libisch}}, \bibinfo {author} {\bibfnamefont
			{T.~J.}\ \bibnamefont {Milburn}}, \bibinfo {author} {\bibfnamefont
			{P.}~\bibnamefont {Rabl}}, \bibinfo {author} {\bibfnamefont {N.}~\bibnamefont
			{Moiseyev}},\ and\ \bibinfo {author} {\bibfnamefont {S.}~\bibnamefont
			{Rotter}},\ }\bibfield  {title} {\bibinfo {title} {Dynamically encircling an
			exceptional point for asymmetric mode switching},\ }\href
	{https://doi.org/10.1038/nature18605} {\bibfield  {journal} {\bibinfo
			{journal} {Nature}\ }\textbf {\bibinfo {volume} {537}},\ \bibinfo {pages}
		{76} (\bibinfo {year} {2016})}\BibitemShut {NoStop}%
	\bibitem [{\citenamefont {Schumer}\ \emph {et~al.}(2022)\citenamefont
		{Schumer}, \citenamefont {Liu}, \citenamefont {Leshin}, \citenamefont {Ding},
		\citenamefont {Alahmadi}, \citenamefont {Hassan}, \citenamefont {Nasari},
		\citenamefont {Rotter}, \citenamefont {Christodoulides}, \citenamefont
		{LiKamWa},\ and\ \citenamefont {Khajavikhan}}]{Schumer2022}%
	\BibitemOpen
	\bibfield  {author} {\bibinfo {author} {\bibfnamefont {A.}~\bibnamefont
			{Schumer}}, \bibinfo {author} {\bibfnamefont {Y.~G.~N.}\ \bibnamefont {Liu}},
		\bibinfo {author} {\bibfnamefont {J.}~\bibnamefont {Leshin}}, \bibinfo
		{author} {\bibfnamefont {L.}~\bibnamefont {Ding}}, \bibinfo {author}
		{\bibfnamefont {Y.}~\bibnamefont {Alahmadi}}, \bibinfo {author}
		{\bibfnamefont {A.~U.}\ \bibnamefont {Hassan}}, \bibinfo {author}
		{\bibfnamefont {H.}~\bibnamefont {Nasari}}, \bibinfo {author} {\bibfnamefont
			{S.}~\bibnamefont {Rotter}}, \bibinfo {author} {\bibfnamefont {D.~N.}\
			\bibnamefont {Christodoulides}}, \bibinfo {author} {\bibfnamefont
			{P.}~\bibnamefont {LiKamWa}},\ and\ \bibinfo {author} {\bibfnamefont
			{M.}~\bibnamefont {Khajavikhan}},\ }\bibfield  {title} {\bibinfo {title}
		{Topological modes in a laser cavity through exceptional state transfer},\
	}\href {https://doi.org/10.1126/science.abl6571} {\bibfield  {journal}
		{\bibinfo  {journal} {Science}\ }\textbf {\bibinfo {volume} {375}},\ \bibinfo
		{pages} {884} (\bibinfo {year} {2022})}\BibitemShut {NoStop}%
	\bibitem [{\citenamefont {Graefe}\ \emph {et~al.}(2008)\citenamefont {Graefe},
		\citenamefont {G{\"u}nther}, \citenamefont {Korsch},\ and\ \citenamefont
		{Niederle}}]{Graefe2008}%
	\BibitemOpen
	\bibfield  {author} {\bibinfo {author} {\bibfnamefont {E.~M.}\ \bibnamefont
			{Graefe}}, \bibinfo {author} {\bibfnamefont {U.}~\bibnamefont {G{\"u}nther}},
		\bibinfo {author} {\bibfnamefont {H.~J.}\ \bibnamefont {Korsch}},\ and\
		\bibinfo {author} {\bibfnamefont {A.~E.}\ \bibnamefont {Niederle}},\
	}\bibfield  {title} {\bibinfo {title} {A non-{H}ermitian symmetric
			{B}ose-{H}ubbard model: eigenvalue rings from unfolding higher-order
			exceptional points},\ }\href {https://doi.org/10.1088/1751-8113/41/25/255206}
	{\bibfield  {journal} {\bibinfo  {journal} {J. Phys. A: Math. Theor.}\
		}\textbf {\bibinfo {volume} {41}},\ \bibinfo {pages} {255206} (\bibinfo
		{year} {2008})}\BibitemShut {NoStop}%
	\bibitem [{\citenamefont {Heiss}(2008)}]{Heiss2008}%
	\BibitemOpen
	\bibfield  {author} {\bibinfo {author} {\bibfnamefont {W.~D.}\ \bibnamefont
			{Heiss}},\ }\bibfield  {title} {\bibinfo {title} {Chirality of wavefunctions
			for three coalescing levels},\ }\href
	{https://doi.org/10.1088/1751-8113/41/24/244010} {\bibfield  {journal}
		{\bibinfo  {journal} {J. Phys. A: Math. Theor.}\ }\textbf {\bibinfo {volume}
			{41}},\ \bibinfo {pages} {244010} (\bibinfo {year} {2008})}\BibitemShut
	{NoStop}%
	\bibitem [{\citenamefont {Nada}\ \emph {et~al.}(2017)\citenamefont {Nada},
		\citenamefont {Othman},\ and\ \citenamefont {Capolino}}]{Nada2017}%
	\BibitemOpen
	\bibfield  {author} {\bibinfo {author} {\bibfnamefont {M.~Y.}\ \bibnamefont
			{Nada}}, \bibinfo {author} {\bibfnamefont {M.~A.~K.}\ \bibnamefont
			{Othman}},\ and\ \bibinfo {author} {\bibfnamefont {F.}~\bibnamefont
			{Capolino}},\ }\bibfield  {title} {\bibinfo {title} {Theory of coupled
			resonator optical waveguides exhibiting high-order exceptional points of
			degeneracy},\ }\href {https://doi.org/10.1103/PhysRevB.96.184304} {\bibfield
		{journal} {\bibinfo  {journal} {Phys. Rev. B}\ }\textbf {\bibinfo {volume}
			{96}},\ \bibinfo {pages} {184304} (\bibinfo {year} {2017})}\BibitemShut
	{NoStop}%
	\bibitem [{\citenamefont {Xiao}\ \emph
		{et~al.}(2019{\natexlab{a}})\citenamefont {Xiao}, \citenamefont {Zhang},
		\citenamefont {Hang},\ and\ \citenamefont {Chan}}]{XiaoYX2019}%
	\BibitemOpen
	\bibfield  {author} {\bibinfo {author} {\bibfnamefont {Y.-X.}\ \bibnamefont
			{Xiao}}, \bibinfo {author} {\bibfnamefont {Z.-Q.}\ \bibnamefont {Zhang}},
		\bibinfo {author} {\bibfnamefont {Z.~H.}\ \bibnamefont {Hang}},\ and\
		\bibinfo {author} {\bibfnamefont {C.~T.}\ \bibnamefont {Chan}},\ }\bibfield
	{title} {\bibinfo {title} {Anisotropic exceptional points of arbitrary
			order},\ }\href {https://doi.org/10.1103/PhysRevB.99.241403} {\bibfield
		{journal} {\bibinfo  {journal} {Phys. Rev. B}\ }\textbf {\bibinfo {volume}
			{99}},\ \bibinfo {pages} {241403} (\bibinfo {year}
		{2019}{\natexlab{a}})}\BibitemShut {NoStop}%
	\bibitem [{\citenamefont {Xiao}\ \emph
		{et~al.}(2019{\natexlab{b}})\citenamefont {Xiao}, \citenamefont {Li},
		\citenamefont {Kottos},\ and\ \citenamefont {Al\`u}}]{XiaoZ2019}%
	\BibitemOpen
	\bibfield  {author} {\bibinfo {author} {\bibfnamefont {Z.}~\bibnamefont
			{Xiao}}, \bibinfo {author} {\bibfnamefont {H.}~\bibnamefont {Li}}, \bibinfo
		{author} {\bibfnamefont {T.}~\bibnamefont {Kottos}},\ and\ \bibinfo {author}
		{\bibfnamefont {A.}~\bibnamefont {Al\`u}},\ }\bibfield  {title} {\bibinfo
		{title} {Enhanced sensing and nondegraded thermal noise performance based on
			$\mathcal{P}\mathcal{T}$-symmetric electronic circuits with a sixth-order
			exceptional point},\ }\href {https://doi.org/10.1103/PhysRevLett.123.213901}
	{\bibfield  {journal} {\bibinfo  {journal} {Phys. Rev. Lett.}\ }\textbf
		{\bibinfo {volume} {123}},\ \bibinfo {pages} {213901} (\bibinfo {year}
		{2019}{\natexlab{b}})}\BibitemShut {NoStop}%
	\bibitem [{\citenamefont {Zhong}\ \emph {et~al.}(2020)\citenamefont {Zhong},
		\citenamefont {Kou}, \citenamefont {{\"O}zdemir},\ and\ \citenamefont
		{El-Ganainy}}]{Zhong2020}%
	\BibitemOpen
	\bibfield  {author} {\bibinfo {author} {\bibfnamefont {Q.}~\bibnamefont
			{Zhong}}, \bibinfo {author} {\bibfnamefont {J.}~\bibnamefont {Kou}}, \bibinfo
		{author} {\bibfnamefont {{\c S}.~K.}\ \bibnamefont {{\"O}zdemir}},\ and\
		\bibinfo {author} {\bibfnamefont {R.}~\bibnamefont {El-Ganainy}},\ }\bibfield
	{title} {\bibinfo {title} {Hierarchical construction of higher-order
			exceptional points},\ }\href {https://doi.org/10.1103/PhysRevLett.125.203602}
	{\bibfield  {journal} {\bibinfo  {journal} {Phys. Rev. Lett.}\ }\textbf
		{\bibinfo {volume} {125}},\ \bibinfo {pages} {203602} (\bibinfo {year}
		{2020})}\BibitemShut {NoStop}%
	\bibitem [{\citenamefont {Yang}\ \emph {et~al.}(2021)\citenamefont {Yang},
		\citenamefont {Mao}, \citenamefont {Qin}, \citenamefont {Wang}, \citenamefont
		{Zhang}, \citenamefont {Ruan},\ and\ \citenamefont {Long}}]{Yang2021}%
	\BibitemOpen
	\bibfield  {author} {\bibinfo {author} {\bibfnamefont {H.}~\bibnamefont
			{Yang}}, \bibinfo {author} {\bibfnamefont {X.}~\bibnamefont {Mao}}, \bibinfo
		{author} {\bibfnamefont {G.-Q.}\ \bibnamefont {Qin}}, \bibinfo {author}
		{\bibfnamefont {M.}~\bibnamefont {Wang}}, \bibinfo {author} {\bibfnamefont
			{H.}~\bibnamefont {Zhang}}, \bibinfo {author} {\bibfnamefont
			{D.}~\bibnamefont {Ruan}},\ and\ \bibinfo {author} {\bibfnamefont {G.-L.}\
			\bibnamefont {Long}},\ }\bibfield  {title} {\bibinfo {title} {Scalable
			higher-order exceptional surface with passive resonators},\ }\href
	{https://doi.org/10.1364/OL.435843} {\bibfield  {journal} {\bibinfo
			{journal} {Opt. Lett.}\ }\textbf {\bibinfo {volume} {46}},\ \bibinfo {pages}
		{4025} (\bibinfo {year} {2021})}\BibitemShut {NoStop}%
	\bibitem [{\citenamefont {Mandal}\ and\ \citenamefont
		{Bergholtz}(2021)}]{Mandal2021}%
	\BibitemOpen
	\bibfield  {author} {\bibinfo {author} {\bibfnamefont {I.}~\bibnamefont
			{Mandal}}\ and\ \bibinfo {author} {\bibfnamefont {E.~J.}\ \bibnamefont
			{Bergholtz}},\ }\bibfield  {title} {\bibinfo {title} {Symmetry and
			higher-order exceptional points},\ }\href
	{https://doi.org/10.1103/PhysRevLett.127.186601} {\bibfield  {journal}
		{\bibinfo  {journal} {Phys. Rev. Lett.}\ }\textbf {\bibinfo {volume} {127}},\
		\bibinfo {pages} {186601} (\bibinfo {year} {2021})}\BibitemShut {NoStop}%
	\bibitem [{\citenamefont {Delplace}\ \emph {et~al.}(2021)\citenamefont
		{Delplace}, \citenamefont {Yoshida},\ and\ \citenamefont
		{Hatsugai}}]{Delplace2021}%
	\BibitemOpen
	\bibfield  {author} {\bibinfo {author} {\bibfnamefont {P.}~\bibnamefont
			{Delplace}}, \bibinfo {author} {\bibfnamefont {T.}~\bibnamefont {Yoshida}},\
		and\ \bibinfo {author} {\bibfnamefont {Y.}~\bibnamefont {Hatsugai}},\
	}\bibfield  {title} {\bibinfo {title} {Symmetry-protected multifold
			exceptional points and their topological characterization},\ }\href
	{https://doi.org/10.1103/PhysRevLett.127.186602} {\bibfield  {journal}
		{\bibinfo  {journal} {Phys. Rev. Lett.}\ }\textbf {\bibinfo {volume} {127}},\
		\bibinfo {pages} {186602} (\bibinfo {year} {2021})}\BibitemShut {NoStop}%
	\bibitem [{\citenamefont {Sayyad}\ and\ \citenamefont
		{Kunst}(2022)}]{Sayyad2022}%
	\BibitemOpen
	\bibfield  {author} {\bibinfo {author} {\bibfnamefont {S.}~\bibnamefont
			{Sayyad}}\ and\ \bibinfo {author} {\bibfnamefont {F.~K.}\ \bibnamefont
			{Kunst}},\ }\bibfield  {title} {\bibinfo {title} {Realizing exceptional
			points of any order in the presence of symmetry},\ }\href
	{https://doi.org/10.1103/PhysRevResearch.4.023130} {\bibfield  {journal}
		{\bibinfo  {journal} {Phys. Rev. Res.}\ }\textbf {\bibinfo {volume} {4}},\
		\bibinfo {pages} {023130} (\bibinfo {year} {2022})}\BibitemShut {NoStop}%
	\bibitem [{\citenamefont {Kaltsas}\ \emph {et~al.}(2022)\citenamefont
		{Kaltsas}, \citenamefont {Komis},\ and\ \citenamefont
		{Makris}}]{Kaltsas2022}%
	\BibitemOpen
	\bibfield  {author} {\bibinfo {author} {\bibfnamefont {D.}~\bibnamefont
			{Kaltsas}}, \bibinfo {author} {\bibfnamefont {I.}~\bibnamefont {Komis}},\
		and\ \bibinfo {author} {\bibfnamefont {K.~G.}\ \bibnamefont {Makris}},\
	}\bibfield  {title} {\bibinfo {title} {Higher order exceptional points in
			infinite lattices},\ }\href {https://doi.org/10.1364/OL.459398} {\bibfield
		{journal} {\bibinfo  {journal} {Opt. Lett.}\ }\textbf {\bibinfo {volume}
			{47}},\ \bibinfo {pages} {4447} (\bibinfo {year} {2022})}\BibitemShut
	{NoStop}%
	\bibitem [{\citenamefont {Kullig}\ \emph {et~al.}(2023)\citenamefont {Kullig},
		\citenamefont {Grom}, \citenamefont {Klembt},\ and\ \citenamefont
		{Wiersig}}]{Kullig2023}%
	\BibitemOpen
	\bibfield  {author} {\bibinfo {author} {\bibfnamefont {J.}~\bibnamefont
			{Kullig}}, \bibinfo {author} {\bibfnamefont {D.}~\bibnamefont {Grom}},
		\bibinfo {author} {\bibfnamefont {S.}~\bibnamefont {Klembt}},\ and\ \bibinfo
		{author} {\bibfnamefont {J.}~\bibnamefont {Wiersig}},\ }\bibfield  {title}
	{\bibinfo {title} {Higher-order exceptional points in waveguide-coupled
			microcavities: perturbation induced frequency splitting and mode patterns},\
	}\href {https://doi.org/10.1364/PRJ.496414} {\bibfield  {journal} {\bibinfo
			{journal} {Photon. Res.}\ }\textbf {\bibinfo {volume} {11}},\ \bibinfo
		{pages} {A54} (\bibinfo {year} {2023})}\BibitemShut {NoStop}%
	\bibitem [{\citenamefont {Miyake}(2022)}]{Miyake2022}%
	\BibitemOpen
	\bibfield  {author} {\bibinfo {author} {\bibfnamefont {T.}~\bibnamefont
			{Miyake}},\ }\href {https://doi.org/10.1007/978-981-16-6994-1} {\emph
		{\bibinfo {title} {Linear Algebra: From the beginnings to the Jordan normal
				forms}}}\ (\bibinfo  {publisher} {Springer Singapore},\ \bibinfo {year}
	{2022})\BibitemShut {NoStop}%
	\bibitem [{\citenamefont {Smith}\ \emph {et~al.}(2004)\citenamefont {Smith},
		\citenamefont {Chang}, \citenamefont {Fuller}, \citenamefont {Rosenberger},\
		and\ \citenamefont {Boyd}}]{Smith2004}%
	\BibitemOpen
	\bibfield  {author} {\bibinfo {author} {\bibfnamefont {D.~D.}\ \bibnamefont
			{Smith}}, \bibinfo {author} {\bibfnamefont {H.}~\bibnamefont {Chang}},
		\bibinfo {author} {\bibfnamefont {K.~A.}\ \bibnamefont {Fuller}}, \bibinfo
		{author} {\bibfnamefont {A.~T.}\ \bibnamefont {Rosenberger}},\ and\ \bibinfo
		{author} {\bibfnamefont {R.~W.}\ \bibnamefont {Boyd}},\ }\bibfield  {title}
	{\bibinfo {title} {Coupled-resonator-induced transparency},\ }\href
	{https://doi.org/10.1103/PhysRevA.69.063804} {\bibfield  {journal} {\bibinfo
			{journal} {Phys. Rev. A}\ }\textbf {\bibinfo {volume} {69}},\ \bibinfo
		{pages} {063804} (\bibinfo {year} {2004})}\BibitemShut {NoStop}%
	\bibitem [{\citenamefont {Waks}\ and\ \citenamefont
		{Vuckovic}(2006)}]{Waks2006}%
	\BibitemOpen
	\bibfield  {author} {\bibinfo {author} {\bibfnamefont {E.}~\bibnamefont
			{Waks}}\ and\ \bibinfo {author} {\bibfnamefont {J.}~\bibnamefont
			{Vuckovic}},\ }\bibfield  {title} {\bibinfo {title} {Dipole induced
			transparency in drop-filter cavity-waveguide systems},\ }\href
	{https://doi.org/10.1103/PhysRevLett.96.153601} {\bibfield  {journal}
		{\bibinfo  {journal} {Phys. Rev. Lett.}\ }\textbf {\bibinfo {volume} {96}},\
		\bibinfo {pages} {153601} (\bibinfo {year} {2006})}\BibitemShut {NoStop}%
	\bibitem [{\citenamefont {Wiersig}(2023)}]{Wiersig2023}%
	\BibitemOpen
	\bibfield  {author} {\bibinfo {author} {\bibfnamefont {J.}~\bibnamefont
			{Wiersig}},\ }\bibfield  {title} {\bibinfo {title} {Petermann factors and
			phase rigidities near exceptional points},\ }\href
	{https://doi.org/10.1103/PhysRevResearch.5.033042} {\bibfield  {journal}
		{\bibinfo  {journal} {Phys. Rev. Res.}\ }\textbf {\bibinfo {volume} {5}},\
		\bibinfo {pages} {033042} (\bibinfo {year} {2023})}\BibitemShut {NoStop}%
	\bibitem [{\citenamefont {Schomerus}(2024)}]{Schomerus2024}%
	\BibitemOpen
	\bibfield  {author} {\bibinfo {author} {\bibfnamefont {H.}~\bibnamefont
			{Schomerus}},\ }\bibfield  {title} {\bibinfo {title} {Eigenvalue sensitivity
			from eigenstate geometry near and beyond arbitrary-order exceptional
			points},\ }\href {https://doi.org/10.1103/PhysRevResearch.6.013044}
	{\bibfield  {journal} {\bibinfo  {journal} {Phys. Rev. Res.}\ }\textbf
		{\bibinfo {volume} {6}},\ \bibinfo {pages} {013044} (\bibinfo {year}
		{2024})}\BibitemShut {NoStop}%
	\bibitem [{\citenamefont {Suh}\ \emph {et~al.}(2004)\citenamefont {Suh},
		\citenamefont {Wang},\ and\ \citenamefont {Fan}}]{Suh2004}%
	\BibitemOpen
	\bibfield  {author} {\bibinfo {author} {\bibfnamefont {W.}~\bibnamefont
			{Suh}}, \bibinfo {author} {\bibfnamefont {Z.}~\bibnamefont {Wang}},\ and\
		\bibinfo {author} {\bibfnamefont {S.}~\bibnamefont {Fan}},\ }\bibfield
	{title} {\bibinfo {title} {Temporal coupled-mode theory and the presence of
			non-orthogonal modes in lossless multimode cavities},\ }\href
	{https://doi.org/10.1109/JQE.2004.834773} {\bibfield  {journal} {\bibinfo
			{journal} {IEEE J. Quantum Electron.}\ }\textbf {\bibinfo {volume} {40}},\
		\bibinfo {pages} {1511} (\bibinfo {year} {2004})}\BibitemShut {NoStop}%
	\bibitem [{\citenamefont {Takata}\ \emph {et~al.}(2022)\citenamefont {Takata},
		\citenamefont {Roberts}, \citenamefont {Shinya},\ and\ \citenamefont
		{Notomi}}]{Takata2022}%
	\BibitemOpen
	\bibfield  {author} {\bibinfo {author} {\bibfnamefont {K.}~\bibnamefont
			{Takata}}, \bibinfo {author} {\bibfnamefont {N.}~\bibnamefont {Roberts}},
		\bibinfo {author} {\bibfnamefont {A.}~\bibnamefont {Shinya}},\ and\ \bibinfo
		{author} {\bibfnamefont {M.}~\bibnamefont {Notomi}},\ }\bibfield  {title}
	{\bibinfo {title} {Imaginary couplings in non-{H}ermitian coupled-mode
			theory: Effects on exceptional points of optical resonators},\ }\href
	{https://doi.org/10.1103/PhysRevA.105.013523} {\bibfield  {journal} {\bibinfo
			{journal} {Phys. Rev. A}\ }\textbf {\bibinfo {volume} {105}},\ \bibinfo
		{pages} {013523} (\bibinfo {year} {2022})}\BibitemShut {NoStop}%
	\bibitem [{\citenamefont {Akahane}\ \emph {et~al.}(2003)\citenamefont
		{Akahane}, \citenamefont {Asano}, \citenamefont {Song},\ and\ \citenamefont
		{Noda}}]{Akahane2003}%
	\BibitemOpen
	\bibfield  {author} {\bibinfo {author} {\bibfnamefont {Y.}~\bibnamefont
			{Akahane}}, \bibinfo {author} {\bibfnamefont {T.}~\bibnamefont {Asano}},
		\bibinfo {author} {\bibfnamefont {B.-S.}\ \bibnamefont {Song}},\ and\
		\bibinfo {author} {\bibfnamefont {S.}~\bibnamefont {Noda}},\ }\bibfield
	{title} {\bibinfo {title} {High-{Q} photonic nanocavity in a two-dimensional
			photonic crystal},\ }\href {https://doi.org/10.1038/nature02063} {\bibfield
		{journal} {\bibinfo  {journal} {Nature}\ }\textbf {\bibinfo {volume} {425}},\
		\bibinfo {pages} {944} (\bibinfo {year} {2003})}\BibitemShut {NoStop}%
	\bibitem [{\citenamefont {Matsuo}\ \emph {et~al.}(2010)\citenamefont {Matsuo},
		\citenamefont {Shinya}, \citenamefont {Kakitsuka}, \citenamefont {Nozaki},
		\citenamefont {Segawa}, \citenamefont {Sato}, \citenamefont {Kawaguchi},\
		and\ \citenamefont {Notomi}}]{Matsuo2010}%
	\BibitemOpen
	\bibfield  {author} {\bibinfo {author} {\bibfnamefont {S.}~\bibnamefont
			{Matsuo}}, \bibinfo {author} {\bibfnamefont {A.}~\bibnamefont {Shinya}},
		\bibinfo {author} {\bibfnamefont {T.}~\bibnamefont {Kakitsuka}}, \bibinfo
		{author} {\bibfnamefont {K.}~\bibnamefont {Nozaki}}, \bibinfo {author}
		{\bibfnamefont {T.}~\bibnamefont {Segawa}}, \bibinfo {author} {\bibfnamefont
			{T.}~\bibnamefont {Sato}}, \bibinfo {author} {\bibfnamefont {Y.}~\bibnamefont
			{Kawaguchi}},\ and\ \bibinfo {author} {\bibfnamefont {M.}~\bibnamefont
			{Notomi}},\ }\bibfield  {title} {\bibinfo {title} {High-speed ultracompact
			buried heterostructure photonic-crystal laser with 13 f{J} of energy consumed
			per bit transmitted},\ }\href {https://doi.org/10.1038/nphoton.2010.177}
	{\bibfield  {journal} {\bibinfo  {journal} {Nat. Photon.}\ }\textbf {\bibinfo
			{volume} {4}},\ \bibinfo {pages} {648} (\bibinfo {year} {2010})}\BibitemShut
	{NoStop}%
	\bibitem [{\citenamefont {Takata}\ and\ \citenamefont
		{Notomi}(2017)}]{Takata2017}%
	\BibitemOpen
	\bibfield  {author} {\bibinfo {author} {\bibfnamefont {K.}~\bibnamefont
			{Takata}}\ and\ \bibinfo {author} {\bibfnamefont {M.}~\bibnamefont
			{Notomi}},\ }\bibfield  {title} {\bibinfo {title}
		{$\mathcal{P}\mathcal{T}$-symmetric coupled-resonator waveguide based on
			buried heterostructure nanocavities},\ }\href
	{https://doi.org/10.1103/PhysRevApplied.7.054023} {\bibfield  {journal}
		{\bibinfo  {journal} {Phys. Rev. Applied}\ }\textbf {\bibinfo {volume} {7}},\
		\bibinfo {pages} {054023} (\bibinfo {year} {2017})}\BibitemShut {NoStop}%
	\bibitem [{\citenamefont {Cheng}(1989)}]{Cheng1989}%
	\BibitemOpen
	\bibfield  {author} {\bibinfo {author} {\bibfnamefont {D.~K.}\ \bibnamefont
			{Cheng}},\ }\href@noop {} {\emph {\bibinfo {title} {Field and Wave
				Electromagnetics}}},\ \bibinfo {edition} {2nd}\ ed.\ (\bibinfo  {publisher}
	{Addison-Wesley},\ \bibinfo {year} {1989})\BibitemShut {NoStop}%
	\bibitem [{\citenamefont {Yariv}\ \emph {et~al.}(1999)\citenamefont {Yariv},
		\citenamefont {Xu}, \citenamefont {Lee},\ and\ \citenamefont
		{Scherer}}]{Yariv1999}%
	\BibitemOpen
	\bibfield  {author} {\bibinfo {author} {\bibfnamefont {A.}~\bibnamefont
			{Yariv}}, \bibinfo {author} {\bibfnamefont {Y.}~\bibnamefont {Xu}}, \bibinfo
		{author} {\bibfnamefont {R.~K.}\ \bibnamefont {Lee}},\ and\ \bibinfo {author}
		{\bibfnamefont {A.}~\bibnamefont {Scherer}},\ }\bibfield  {title} {\bibinfo
		{title} {Coupled-resonator optical waveguide{:} a proposal and analysis},\
	}\href {https://doi.org/10.1364/OL.24.000711} {\bibfield  {journal} {\bibinfo
			{journal} {Opt. Lett.}\ }\textbf {\bibinfo {volume} {24}},\ \bibinfo {pages}
		{711} (\bibinfo {year} {1999})}\BibitemShut {NoStop}%
	\bibitem [{\citenamefont {Schomerus}(2022)}]{Schomerus2022}%
	\BibitemOpen
	\bibfield  {author} {\bibinfo {author} {\bibfnamefont {H.}~\bibnamefont
			{Schomerus}},\ }\bibfield  {title} {\bibinfo {title} {Fundamental constraints
			on the observability of non-{H}ermitian effects in passive systems},\ }\href
	{https://doi.org/10.1103/PhysRevA.106.063509} {\bibfield  {journal} {\bibinfo
			{journal} {Phys. Rev. A}\ }\textbf {\bibinfo {volume} {106}},\ \bibinfo
		{pages} {063509} (\bibinfo {year} {2022})}\BibitemShut {NoStop}%
	\bibitem [{\citenamefont {Wiersig}(2022)}]{Wiersig2022}%
	\BibitemOpen
	\bibfield  {author} {\bibinfo {author} {\bibfnamefont {J.}~\bibnamefont
			{Wiersig}},\ }\bibfield  {title} {\bibinfo {title} {Response strengths of
			open systems at exceptional points},\ }\href
	{https://doi.org/10.1103/PhysRevResearch.4.023121} {\bibfield  {journal}
		{\bibinfo  {journal} {Phys. Rev. Res.}\ }\textbf {\bibinfo {volume} {4}},\
		\bibinfo {pages} {023121} (\bibinfo {year} {2022})}\BibitemShut {NoStop}%
	\bibitem [{\citenamefont {Chen}\ \emph {et~al.}(2020)\citenamefont {Chen},
		\citenamefont {Liu}, \citenamefont {Luan}, \citenamefont {Liu}, \citenamefont
		{Wang}, \citenamefont {Zhu}, \citenamefont {Li}, \citenamefont {Gu},
		\citenamefont {Liang}, \citenamefont {Gao}, \citenamefont {Lu}, \citenamefont
		{Ge}, \citenamefont {Zhang}, \citenamefont {Zhu},\ and\ \citenamefont
		{Ma}}]{Chen2020}%
	\BibitemOpen
	\bibfield  {author} {\bibinfo {author} {\bibfnamefont {H.-Z.}\ \bibnamefont
			{Chen}}, \bibinfo {author} {\bibfnamefont {T.}~\bibnamefont {Liu}}, \bibinfo
		{author} {\bibfnamefont {H.-Y.}\ \bibnamefont {Luan}}, \bibinfo {author}
		{\bibfnamefont {R.-J.}\ \bibnamefont {Liu}}, \bibinfo {author} {\bibfnamefont
			{X.-Y.}\ \bibnamefont {Wang}}, \bibinfo {author} {\bibfnamefont {X.-F.}\
			\bibnamefont {Zhu}}, \bibinfo {author} {\bibfnamefont {Y.-B.}\ \bibnamefont
			{Li}}, \bibinfo {author} {\bibfnamefont {Z.-M.}\ \bibnamefont {Gu}}, \bibinfo
		{author} {\bibfnamefont {S.-J.}\ \bibnamefont {Liang}}, \bibinfo {author}
		{\bibfnamefont {H.}~\bibnamefont {Gao}}, \bibinfo {author} {\bibfnamefont
			{L.}~\bibnamefont {Lu}}, \bibinfo {author} {\bibfnamefont {L.}~\bibnamefont
			{Ge}}, \bibinfo {author} {\bibfnamefont {S.}~\bibnamefont {Zhang}}, \bibinfo
		{author} {\bibfnamefont {J.}~\bibnamefont {Zhu}},\ and\ \bibinfo {author}
		{\bibfnamefont {R.-M.}\ \bibnamefont {Ma}},\ }\bibfield  {title} {\bibinfo
		{title} {Revealing the missing dimension at an exceptional point},\ }\href
	{https://doi.org/10.1038/s41567-020-0807-y} {\bibfield  {journal} {\bibinfo
			{journal} {Nat. Phys.}\ }\textbf {\bibinfo {volume} {16}},\ \bibinfo {pages}
		{571} (\bibinfo {year} {2020})}\BibitemShut {NoStop}%
	\bibitem [{\citenamefont {Hashemi}\ \emph {et~al.}(2022)\citenamefont
		{Hashemi}, \citenamefont {Busch}, \citenamefont {Christodoulides},
		\citenamefont {Ozdemir},\ and\ \citenamefont {El-Ganainy}}]{Hashemi2022}%
	\BibitemOpen
	\bibfield  {author} {\bibinfo {author} {\bibfnamefont {A.}~\bibnamefont
			{Hashemi}}, \bibinfo {author} {\bibfnamefont {K.}~\bibnamefont {Busch}},
		\bibinfo {author} {\bibfnamefont {D.~N.}\ \bibnamefont {Christodoulides}},
		\bibinfo {author} {\bibfnamefont {S.~K.}\ \bibnamefont {Ozdemir}},\ and\
		\bibinfo {author} {\bibfnamefont {R.}~\bibnamefont {El-Ganainy}},\ }\bibfield
	{title} {\bibinfo {title} {Linear response theory of open systems with
			exceptional points},\ }\href {https://doi.org/10.1038/s41467-022-30715-8}
	{\bibfield  {journal} {\bibinfo  {journal} {Nat. Commun.}\ }\textbf {\bibinfo
			{volume} {13}},\ \bibinfo {pages} {3281} (\bibinfo {year}
		{2022})}\BibitemShut {NoStop}%
	\bibitem [{\citenamefont {Bai}\ \emph {et~al.}(2024)\citenamefont {Bai},
		\citenamefont {Liu}, \citenamefont {Fang}, \citenamefont {Li}, \citenamefont
		{Lin}, \citenamefont {Wan},\ and\ \citenamefont {Xiao}}]{Bai2024}%
	\BibitemOpen
	\bibfield  {author} {\bibinfo {author} {\bibfnamefont {K.}~\bibnamefont
			{Bai}}, \bibinfo {author} {\bibfnamefont {T.-R.}\ \bibnamefont {Liu}},
		\bibinfo {author} {\bibfnamefont {L.}~\bibnamefont {Fang}}, \bibinfo {author}
		{\bibfnamefont {J.-Z.}\ \bibnamefont {Li}}, \bibinfo {author} {\bibfnamefont
			{C.}~\bibnamefont {Lin}}, \bibinfo {author} {\bibfnamefont {D.}~\bibnamefont
			{Wan}},\ and\ \bibinfo {author} {\bibfnamefont {M.}~\bibnamefont {Xiao}},\
	}\bibfield  {title} {\bibinfo {title} {Observation of nonlinear exceptional
			points with a complete basis in dynamics},\ }\href
	{https://doi.org/10.1103/PhysRevLett.132.073802} {\bibfield  {journal}
		{\bibinfo  {journal} {Phys. Rev. Lett.}\ }\textbf {\bibinfo {volume} {132}},\
		\bibinfo {pages} {073802} (\bibinfo {year} {2024})}\BibitemShut {NoStop}%
	\bibitem [{\citenamefont {Imamura}\ \emph {et~al.}(2024)\citenamefont
		{Imamura}, \citenamefont {Fujii}, \citenamefont {Nakashima},\ and\
		\citenamefont {Tanabe}}]{Imamura2024}%
	\BibitemOpen
	\bibfield  {author} {\bibinfo {author} {\bibfnamefont {R.}~\bibnamefont
			{Imamura}}, \bibinfo {author} {\bibfnamefont {S.}~\bibnamefont {Fujii}},
		\bibinfo {author} {\bibfnamefont {A.}~\bibnamefont {Nakashima}},\ and\
		\bibinfo {author} {\bibfnamefont {T.}~\bibnamefont {Tanabe}},\ }\bibfield
	{title} {\bibinfo {title} {Exceptional point proximity-driven mode-locking in
			coupled microresonators},\ }\href {https://doi.org/10.1364/OE.524556}
	{\bibfield  {journal} {\bibinfo  {journal} {Opt. Express}\ }\textbf {\bibinfo
			{volume} {32}},\ \bibinfo {pages} {22280} (\bibinfo {year}
		{2024})}\BibitemShut {NoStop}%
	\bibitem [{\citenamefont {Deng}\ and\ \citenamefont
		{Khajavikhan}(2021)}]{Deng2021}%
	\BibitemOpen
	\bibfield  {author} {\bibinfo {author} {\bibfnamefont {H.}~\bibnamefont
			{Deng}}\ and\ \bibinfo {author} {\bibfnamefont {M.}~\bibnamefont
			{Khajavikhan}},\ }\bibfield  {title} {\bibinfo {title} {Parity--time
			symmetric optical neural networks},\ }\href
	{https://doi.org/10.1364/OPTICA.435525} {\bibfield  {journal} {\bibinfo
			{journal} {Optica}\ }\textbf {\bibinfo {volume} {8}},\ \bibinfo {pages}
		{1328} (\bibinfo {year} {2021})}\BibitemShut {NoStop}%
\end{thebibliography}

\begin{thebibliography}{3}%
	\makeatletter
	\providecommand \@ifxundefined [1]{%
		\@ifx{#1\undefined}
	}%
	\providecommand \@ifnum [1]{%
		\ifnum #1\expandafter \@firstoftwo
		\else \expandafter \@secondoftwo
		\fi
	}%
	\providecommand \@ifx [1]{%
		\ifx #1\expandafter \@firstoftwo
		\else \expandafter \@secondoftwo
		\fi
	}%
	\providecommand \natexlab [1]{#1}%
	\providecommand \enquote  [1]{``#1''}%
	\providecommand \bibnamefont  [1]{#1}%
	\providecommand \bibfnamefont [1]{#1}%
	\providecommand \citenamefont [1]{#1}%
	\providecommand \href@noop [0]{\@secondoftwo}%
	\providecommand \href [0]{\begingroup \@sanitize@url \@href}%
	\providecommand \@href[1]{\@@startlink{#1}\@@href}%
	\providecommand \@@href[1]{\endgroup#1\@@endlink}%
	\providecommand \@sanitize@url [0]{\catcode `\\12\catcode `\$12\catcode
		`\&12\catcode `\#12\catcode `\^12\catcode `\_12\catcode `\%12\relax}%
	\providecommand \@@startlink[1]{}%
	\providecommand \@@endlink[0]{}%
	\providecommand \url  [0]{\begingroup\@sanitize@url \@url }%
	\providecommand \@url [1]{\endgroup\@href {#1}{\urlprefix }}%
	\providecommand \urlprefix  [0]{URL }%
	\providecommand \Eprint [0]{\href }%
	\providecommand \doibase [0]{https://doi.org/}%
	\providecommand \selectlanguage [0]{\@gobble}%
	\providecommand \bibinfo  [0]{\@secondoftwo}%
	\providecommand \bibfield  [0]{\@secondoftwo}%
	\providecommand \translation [1]{[#1]}%
	\providecommand \BibitemOpen [0]{}%
	\providecommand \bibitemStop [0]{}%
	\providecommand \bibitemNoStop [0]{.\EOS\space}%
	\providecommand \EOS [0]{\spacefactor3000\relax}%
	\providecommand \BibitemShut  [1]{\csname bibitem#1\endcsname}%
	\let\auto@bib@innerbib\@empty
	\bibitem [{\citenamefont {Miyake}(2022)}]{Miyake2022_S}%
	\BibitemOpen
	\bibfield  {author} {\bibinfo {author} {\bibfnamefont {T.}~\bibnamefont
			{Miyake}},\ }\href {https://doi.org/10.1007/978-981-16-6994-1} {\emph
		{\bibinfo {title} {Linear Algebra: From the beginnings to the Jordan normal
				forms}}}\ (\bibinfo  {publisher} {Springer Singapore},\ \bibinfo {year}
	{2022})\BibitemShut {NoStop}%
	\bibitem [{\citenamefont {Bossart}\ and\ \citenamefont
		{Fleury}(2021)}]{Bossart2021_S}%
	\BibitemOpen
	\bibfield  {author} {\bibinfo {author} {\bibfnamefont {A.}~\bibnamefont
			{Bossart}}\ and\ \bibinfo {author} {\bibfnamefont {R.}~\bibnamefont
			{Fleury}},\ }\bibfield  {title} {\bibinfo {title} {Non-{H}ermitian time
			evolution: From static to parametric instability},\ }\href
	{https://doi.org/10.1103/PhysRevA.104.042225} {\bibfield  {journal} {\bibinfo
			{journal} {Phys. Rev. A}\ }\textbf {\bibinfo {volume} {104}},\ \bibinfo
		{pages} {042225} (\bibinfo {year} {2021})}\BibitemShut {NoStop}%
	\bibitem [{\citenamefont {Pick}\ \emph {et~al.}(2017)\citenamefont {Pick},
		\citenamefont {Zhen}, \citenamefont {Miller}, \citenamefont {Hsu},
		\citenamefont {Hernandez}, \citenamefont {Rodriguez}, \citenamefont
		{Solja\v{c}i\'{c}},\ and\ \citenamefont {Johnson}}]{Pick2017gen_S}%
	\BibitemOpen
	\bibfield  {author} {\bibinfo {author} {\bibfnamefont {A.}~\bibnamefont
			{Pick}}, \bibinfo {author} {\bibfnamefont {B.}~\bibnamefont {Zhen}}, \bibinfo
		{author} {\bibfnamefont {O.~D.}\ \bibnamefont {Miller}}, \bibinfo {author}
		{\bibfnamefont {C.~W.}\ \bibnamefont {Hsu}}, \bibinfo {author} {\bibfnamefont
			{F.}~\bibnamefont {Hernandez}}, \bibinfo {author} {\bibfnamefont {A.~W.}\
			\bibnamefont {Rodriguez}}, \bibinfo {author} {\bibfnamefont {M.}~\bibnamefont
			{Solja\v{c}i\'{c}}},\ and\ \bibinfo {author} {\bibfnamefont {S.~G.}\
			\bibnamefont {Johnson}},\ }\bibfield  {title} {\bibinfo {title} {General
			theory of spontaneous emission near exceptional points},\ }\href
	{https://doi.org/10.1364/OE.25.012325} {\bibfield  {journal} {\bibinfo
			{journal} {Opt. Express}\ }\textbf {\bibinfo {volume} {25}},\ \bibinfo
		{pages} {12325} (\bibinfo {year} {2017})}\BibitemShut {NoStop}%
\end{thebibliography}

%

\begin{acknowledgments}
	We thank Shota Kita and Hisashi Sumikura for fruitful discussion. \textbf{Funding:} This work was supported by JSPS KAKENHI under Grant Number JP20H05641. \textbf{Author contributions:} K.T. conceived the idea, developed the theoretical framework, performed the analysis and simulation, and wrote the manuscript. A.M. helped with the design concept and theoretical demonstration. M.N. added to the methodology. A.S. supported the simulation design. M.N. and A.S. supervised the project. All authors discussed the results and commented on the manuscript. \textbf{Competing interests:} The authors declare no conflicts of interest. \textbf{Data and materials availability:} All data underlying the results presented in this paper can be obtained from the authors upon reasonable request.
\end{acknowledgments}


%

\clearpage

}

\setcounter{equation}{0}
\setcounter{figure}{0}
\setcounter{affil}{0}
\renewcommand{\thefigure}{S\arabic{figure}}
\renewcommand{\theequation}{S\arabic{equation}}
\setcounter{MaxMatrixCols}{32}

\title{Higher-order exceptional points unveiled by nilpotence and mathematical induction: Supplemental information}



\author{Kenta Takata}
\altaffiliation[Currently at ]{NTT Research, Inc., Sunnyvale, California 94085, USA}
\affiliation{Nanophotonics Center, NTT, Inc., Atsugi, Kanagawa 243-0198, Japan}
\affiliation{NTT Basic Research Laboratories, NTT, Inc., Atsugi, Kanagawa 243-0198, Japan}

\author{Adam Mock}
\altaffiliation[Currently at ]{Central Michigan University, Mount Pleasant, Michigan 48859, USA}
\affiliation{NTT Basic Research Laboratories, NTT, Inc., Atsugi, Kanagawa 243-0198, Japan}

\author{Masaya Notomi}
\affiliation{Nanophotonics Center, NTT, Inc., Atsugi, Kanagawa 243-0198, Japan}
\affiliation{NTT Basic Research Laboratories, NTT, Inc., Atsugi, Kanagawa 243-0198, Japan}
\affiliation{Department of Physics, Institute of Science Tokyo, Ookayama, Meguro 152-8550, Japan}

\author{Akihiko Shinya}
\affiliation{Nanophotonics Center, NTT, Inc., Atsugi, Kanagawa 243-0198, Japan}
\affiliation{NTT Basic Research Laboratories, NTT, Inc., Atsugi, Kanagawa 243-0198, Japan}



\date{\today}


\maketitle


\section{Mathematical backgrounds}
We start by reviewing basic concepts of linear algebra related to the Jordan normal form (JNF) \cite{Miyake2022_S}. Throughout this work, vectors and matrices are denoted as bold characters and regular characters with hats, respectively, and $N \in \mathbb{N}$ is used for the dimension of square matrices.
\subsection{Jordan cells (boxes)}
Jordan cells (JCs) or boxes are building blocks of matrices in the JNF. They are defined as the following semi-diagonal square matrices
\begin{eqnarray}
	\hat{\rm J}(\omega, d) & = & 
	\left.\left(\,
	\begin{matrix}
		\omega & 1 & \quad & 0 \\[-6pt]
		\quad & \omega & \ddots & \quad \\[-6pt]
		\quad & \quad & \ddots & 1 \\[0pt]
		0 & \quad & \quad & \omega  
	\end{matrix}\,\right) \right\} d \quad (d \ge 2), \nonumber \\ 
	\hat{\rm J}(\omega, 1) & = & \omega, \label{Jordan cell}
\end{eqnarray}
where $d \in \mathbb{N}$ and $\omega \in \mathbb{C}$ are the dimension and eigenvalue of $\hat{\rm J}(\omega, d)$, respectively. For clarity, the element one row above each $\omega$ is 1, and other nondiagonal elements are all null for every $\hat{\rm J}(\omega, d)$ with $d \ge 2$. $\hat{\rm J}(\omega, 1)$ is just a complex number but works as a $1 \times 1$ matrix. It is trivial to show that the characteristic polynomial of $\hat{\rm J}(\omega, d)$ is $p_{{\rm J}(\omega, d)}(x) = (x - \omega)^d$, and its eigenvalue equation $p_{{\rm J}(\omega, d)}(x) = 0$ hence has a $d$-fold root $x = \omega$. 

\subsection{Jordan normal (canonical) form}
An $N \times N$ matrix $\hat{\rm J}$ described in the block-diagonal form composed of JCs, namely
\begin{equation}
	\hat{\rm J} = 
	\left(\,
	\begin{matrix}
		\hat{\rm J}(\omega_1, d_1) & \quad & \quad & 0 \\[0pt]
		\quad & \hat{\rm J}(\omega_2, d_2) & \quad & \quad \\[-6pt]
		\quad & \quad & \ddots & \quad \\[0pt]
		0 & \quad & \quad & \hat{\rm J}(\omega_u, d_u)
	\end{matrix}\,\right), \label{Jordan matrix}
\end{equation}
is called a Jordan matrix. Here, we have $d_1 + d_2 + \cdots + d_u = N$ for a $u \in \mathbb{N}$, and the eigenvalues of $\hat{\rm J}$ are $\{\omega_1, \omega_2, \cdots, \omega_u\}$, which are $\{d_1, d_2, \cdots, d_u\}$-fold, respectively.

For any square matrix $\hat{\rm H} \in \mathbb{C}^{N \times N}$, there is a similarity transformation, i.e. $\hat{\rm J}_{\rm H} = \hat{\rm P}^{-1} \hat{\rm H} \hat{\rm P}$ with $\hat{\rm P}$ being regular, where $\hat{\rm J}_{\rm H}$ is a Jordan matrix. As such, $\hat{\rm J}_{\rm H}$ is termed the JNF of $\hat{\rm H}$. In addition, a famous theorem tells that {\it $\hat{\rm J}_{\rm H}$ is uniquely determined up to the order (or sequence) of JCs}. Note that $\hat{\rm J}$ is diagonal when $d_1 = d_2 = \cdots = d_u = 1 \quad (u = N)$. Thus, the JNF includes the case of the standard diagonalization of matrices. Since it is impossible to diagonalize JCs with $d \ge 2$, $\hat{\rm H}$ with $\hat{\rm J}_{\rm H}$ including such JCs is said to be {\it defective}.

\subsection{Generalized eigenvectors}
The similarity transformation $\hat{\rm J}_{\rm H} = \hat{\rm P}^{-1} \hat{\rm H} \hat{\rm P}$ means the change of basis vectors. Any initially defined $\hat{\rm H} \in \mathbb{C}^{N \times N}$ implies that its basis is the $N$-dimensional fundamental vectors $\bm{e}_1 = (1, 0, \cdots, 0)^\mathrm{T}$, $\bm{e}_2 = (0, 1, \cdots, 0)^\mathrm{T}$, $\cdots$, $\bm{e}_N = (0, 0, \cdots, 1)^\mathrm{T}$, where T means the transposition. Each column of $\hat{\rm P}$ corresponds to a basis vector for $\hat{\rm J}_{\rm H}$ described as a linear combination of $\{\bm{e}_1, \bm{e}_2, \cdots, \bm{e}_N\}$. Without loss of generality, we suppose $\hat{\rm J}_{\rm H}$ is of the form shown in Eq. (\ref{Jordan matrix}) and $\hat{\rm P}$ is denoted by its column vectors as 
\begin{eqnarray}
	\hat{\rm P} &=& 
	\Big(\,
	\begin{matrix}
		\bm{v}_{1,1} \, \cdots \, \bm{v}_{1,d_1} & \bm{v}_{2,1} \, \cdots \, \bm{v}_{2,d_2} & \cdots & \bm{v}_{u,1} \, \cdots \, \bm{v}_{u,d_u}
	\end{matrix}\,\Big) \nonumber \\
	&=& \Big(\,
	\begin{matrix}
		\{\bm{v}_{i,j}\} 
	\end{matrix}\,\Big) \quad (i = 1, 2, \cdots, u, \ j = 1, 2, \cdots, d_i),
\end{eqnarray}
where $i$ corresponds to the index of JCs in $\hat{\rm J}_{\rm H}$, and $j$ is that of the column vectors within the $i$th group of them. By focusing on the $i = l$th JC of $\hat{\rm J}_{\rm H}$, we see that the modified similarity relation, $\hat{\rm P} \hat{\rm J}_{\rm H} = \hat{\rm H} \hat{\rm P}$, yields a series of equations
\begin{eqnarray}
	\hat{\rm H} \bm{v}_{l,j} &=& \omega_l \bm{v}_{l,j} + \bm{v}_{l,j-1} \quad (j = 2, 3, \cdots, d_l), \label{gen_eigenvector} \\
	\hat{\rm H} \bm{v}_{l,1} &=& \omega_l \bm{v}_{l,1}. \label{eigenvector}
\end{eqnarray}
Eqs. (\ref{gen_eigenvector}) and (\ref{eigenvector}) show that $\bm{v}_{l,1}$, namely the leftmost column vector in the group, is the only eigenvector of $\hat{\rm H}$ with the eigenvalue $\omega_l$. The other ones $\{\bm{v}_{l,2}, \cdots, \bm{v}_{l,d_l}\}$ do not satisfy the eigenvalue equation, due to the non-diagonal unity elements in $\hat{\rm J}(\omega_l, d_l)$ (see Eq. (\ref{Jordan cell})). However, by considering $\hat{\rm N}_l := \hat{\rm H} - \omega_l \hat{\rm I}_N$, where $\hat{\rm I}_N$ is the $N \times N$ identity matrix, we can simplify Eqs. (\ref{gen_eigenvector}) and (\ref{eigenvector}) as $\hat{\rm N}_l \bm{v}_{l,j} = \bm{v}_{l,j-1}$ and $\hat{\rm N}_l \bm{v}_{l,1} = \bm{0}$, respectively, where $\bm{0}$ is zero vector. We can further integrate them into
\begin{equation}
	\hat{\rm N}_l^{j} \bm{v}_{l,j} = \bm{0} \quad (j = 1, 2, \cdots, d_l). \label{nil_transformaton}
\end{equation}
In other words, multiple operations of $\hat{\rm N}_l$ link $\{\bm{v}_{l,j}\}$ in the form of a sequence that terminates with zero vector
\begin{equation}
	\bm{v}_{l,d_l} \xrightarrow{\hat{\rm N}_l} \bm{v}_{l,d_l -1} \xrightarrow{\hat{\rm N}_l} \cdots \xrightarrow{\hat{\rm N}_l} \bm{v}_{l,1} \xrightarrow{\hat{\rm N}_l} \bm{0}. \label{vec_sequence}
\end{equation}

In this sense, $\{\bm{v}_{l,j}\}$ are called generalized eigenvectors of $\hat{\rm H}$ to the eigenvalue $\omega_l$. The vector subspace spanned by $\{\bm{v}_{l,j}\}$ is termed the generalized eigenspace of $\hat{\rm H}$ with the eigenvalue $\omega_l$ and denoted as $\mathcal{W}(\hat{\rm H}, \omega_l)$. Here, $\hat{\rm N}_l$ is said to be a nilpotent transformation within $\mathcal{W}(\hat{\rm H}, \omega_l)$. The structure of Eq. (\ref{vec_sequence}) guarantees that $\{\bm{v}_{l,j}\}$ are linearly independent of each other. In addition, generalized eigenspaces with different eigenvalues are also linearly independent. As a result, the generalized eigenvectors $\{\bm{v}_{i,j}\}$ for all $\{\omega_i\}$ span the entire complex vector space, $\mathbb{C}^N$. This is why $\hat{\rm P}$ is full rank and hence has its inverse matrix, $\hat{\rm P}^{-1}$. To derive the JNF of a given matrix, we first obtain all its eigenvectors $\{\bm{v}_{l,1}\}$ by solving Eq. (\ref{eigenvector}). Next, we trace back the flow in Eq. (\ref{vec_sequence}) by using Eq. (\ref{gen_eigenvector}) and $\bm{v}_{l,1}$ for all $l = 1, 2, \cdots, u$, to have the entire $\{\bm{v}_{i,j}\}$ or $\hat{\rm P}$. Note that each JC has each vector chain, i.e. Eq. (\ref{vec_sequence}), even if some may have a common eigenvalue, $\omega_l = \omega_{l'}$ for $l \ne l'$. 

\subsection{Nilpotence}
Let us recall that every matrix has its JNF comprising JCs. Here, each JC is associated with a set of generalized eigenvectors in $\mathcal{W}(\hat{\rm H}, \omega_i)$, and the operator $\hat{\rm N}_i = \hat{\rm H} - \omega_i \hat{\rm I}_N$ including the eigenvalue $\omega_i$ links and nulls these vectors according to Eq. (\ref{vec_sequence}).

Suppose all the eigenvalues of $\hat{\rm H}$ are zero, i.e. $\omega_i = 0$ for $i = 1, \cdots , u$. This condition results in $\hat{\rm N}_i = \hat{\rm H}$ for all $i$, and thus any $N$-vectors eventually vanish by the multiple operations of $\hat{\rm H}$. In other words, the $\hat{\rm H}$ here induces Eq. (\ref{vec_sequence}) for all generalized eigenvectors and hence their linear combinations, which cover the entire vector space $\mathbb{C}^N$. This actually means that a power of $\hat{\rm H}$ itself becomes zero matrix. As such, {\it the nilpotence of $\hat{\rm H}$ is equivalent to that its eigenvalues are all zero.} According to Eq. (\ref{Jordan matrix}), every nilpotent matrix hence has a JNF composed of JCs with zero eigenvalues. As shown in Materials and Methods, this leads to the direct relation between the nilpotence and order of EPs of $\hat{\rm H}$. It can be put as the following corollary of the existence theorem on the JNF \cite{Miyake2022_S}.\\
\\
\textbf{Corollary 1:} {\it Let $\hat{\rm H}$ be an $N \times N$ complex nilpotent matrix. The nilpotence index $m$ of $\hat{\rm H}$, for which $\hat{\rm H}$ satisfies $\hat{\rm H}^{m-1} \ne \hat{\rm O}, \ \hat{\rm H}^m = \hat{\rm O}$, coincides with the largest dimension of the Jordan cells in the Jordan normal form of $\hat{\rm H}$.}\\

\section{Reduction of EP order for Hamiltonians with even dimension} \label{sec:EPOrderReduction}
A coupling-based Hamiltonian with an even dimension tends to have an EP with reduced order. Here we consider a 6$\times$6 example
\begin{equation}
	\hat{\rm H}_{\rm 6C} = \left(\,
	\begin{matrix}
		0 & a &   &   &   &  \\
		a & 0 & b &   &   &  \\
		& b & 0 & c &   &  \\
		&   & c & 0 & d &  \\
		&   &   & d & 0 & e\\
		&   &   &   & e & 0\\
	\end{matrix}\,\right), \label{eq:H6C} \\
\end{equation}
where $a, b, c, d, e \in \mathbb{C}$. We expect that it may have an EP6 and hence examine the conditions for $\hat{\rm H}_{\rm 6C}^5 \ne \hat{\rm O}, \ \hat{\rm H}_{\rm 6C}^6 = \hat{\rm O}$ with the help of ${\rm tr} \, \hat{\rm H}_{\rm 6C}^l = 0$ for $l = 1, 2, \dots, 5$. By a straightforward matrix multiplication, we find that ${\rm tr} \, \hat{\rm H}_{\rm 6C} = {\rm tr} \, \hat{\rm H}_{\rm 6C}^3 = {\rm tr} \, \hat{\rm H}_{\rm 6C}^5 = 0$ trivially holds and have
\begin{align}
	{\rm tr} \, \hat{\rm H}_{\rm 6C}^2 = 0 & \iff a^2 + b^2 + c^2 + d^2 + e^2 = 0, \label{eq:H6CNPC1} \\
	\notag \\
	{\rm tr} \, \hat{\rm H}_{\rm 6C}^4 = 0 & \iff a^4 + b^4 + c^4 + d^4 + e^4 \notag \\
	& + 2 (a^2 b^2 + b^2 c^2 + c^2 d^2 + d^2 e^2) = 0. \label{eq:H6CNPC2}
\end{align}
Equations (\ref{eq:H6CNPC1}) and (\ref{eq:H6CNPC2}) let us simplify $\hat{\rm H}_{\rm 6C}^5$ and $\hat{\rm H}_{\rm 6C}^6$ as
\begin{widetext}
	\begin{equation}
		\hat{\rm H}_{\rm 6C}^5 = \left(\,
		\begin{matrix}
			0 			& a c^2 e^2 & 0								 & -a b c e^2					 & 0							& a b c d e \\
			a c^2 e^2	& 0 		& 0								 & 0 							 & 0							& 0 \\
			0 			& 0 		& 0 							 & -c e^2 (b^2 + c^2 + d^2 + e^2)& 0							& c d e (b^2 + c^2 + d^2 + e^2) \\
			-a b c e^2	& 0			& -c e^2 (b^2 + c^2 + d^2 + e^2) & 0 							 & 0							& 0\\
			0			& 0			& 0								 & 0 							 & 0							& e [c^2 d^2 + (d^2 + e^2)^2]\\
			a b c d e 	& 0			& c d e (b^2 + c^2 + d^2 + e^2)	 & 0 							 & e [c^2 d^2 + (d^2 + e^2)^2]	& 0\\
		\end{matrix}\,\right), \label{eq:H6C5th} \\
	\end{equation}
\end{widetext}
\begin{equation}
	\hat{\rm H}_{\rm 6C}^6 = e^2 [c^2 d^2 + (d^2 + e^2)^2] \, \hat{\rm I}_6. \label{eq:H6C6th} 
\end{equation}
We see that $\hat{\rm H}_{\rm 6C}^6$ in Eq. (\ref{eq:H6C6th}) vanishes if and only if (i)$e = 0$ or (ii)$c^2 d^2 + (d^2 + e^2)^2 = 0$. For both cases, however, we also have $\hat{\rm H}_{\rm 6C}^5 = \hat{\rm O}$ and thus the EP order is reduced, as shown below.

When (i)$e = 0$, $\hat{\rm H}_{\rm 6C}^5$ in Eq. (\ref{eq:H6C5th}) trivially vanishes, since all its generally nonzero elements have $e$ as a common factor.

For (ii)$c^2 d^2 + (d^2 + e^2)^2 = 0$, Eqs. (\ref{eq:H6CNPC1}) and (\ref{eq:H6CNPC2}) are reduced by this condition to 
\begin{equation}
	c^2 (b^2 + c^2 + d^2 + e^2) = 0.
\end{equation}
Thus, $c = 0$ or $b^2 + c^2 + d^2 + e^2 = 0$ is supposed to be required for the nilpotence. However, $c = 0$ and $c^2 d^2 + (d^2 + e^2)^2 = 0$ result in $\hat{\rm H}_{\rm 6C}^5 = \hat{\rm O}$ as the case of $e = 0$. In addition, $b^2 + c^2 + d^2 + e^2 = 0$ and Eq. (\ref{eq:H6CNPC1}) give $a = 0$ and hence the same conclusion. 

We have covered all the cases for $\hat{\rm H}_{\rm 6C}^6 = \hat{\rm O}$, and they have been shown to accompany $\hat{\rm H}_{\rm 6C}^5 = \hat{\rm O}$. Therefore, possible EP order for $\hat{\rm H}_{\rm 6C}$ is five or less. In fact, this problem cannot be resolved even though we consider a ring-shaped array, namely we introduce $[\hat{\rm H}_{\rm 6C}]_{1 6} = [\hat{\rm H}_{\rm 6C}]_{6 1} = f \ne 0$ to Eq. (\ref{eq:H6C}). We also have a similar result for a 4$\times$4 Hamiltonian, but we omit details to avoid redundancy.

\section{Transmission spectra and spectral transfer functions} \label{sec:SpectralResponses}
We analytically show the spectral transmission ratios $\{T_{l j}^{(N)}(\omega_s)\}$ and transfer functions $\{S_{l,j}^{(N)}(\omega_s)\}$ of our cavity arrays operating at HEPs. For clarity, we add a superscript $N$ denoting the system size to each of them. Note that we factor out the flux outcoupling loss $2 \gamma$ of the transfer functions. For the six-cavity system with the passive EP6 ($\delta = \mu = \gamma$), we have
\begin{widetext}
	\begin{align}
		T_{11}^{(6)}(\omega_s) &= 1 - \frac{4 \gamma^4 \omega_s^2}{(\omega_s^2 + \gamma^2)^3} - \frac{8 \gamma^5 \omega_s^3}{(\omega_s^2 + \gamma^2)^4} + \frac{4 \gamma^8 \omega_s^4 + 24 \gamma^9 \omega_s^3 - 8 \gamma^7 \omega_s^5}{(\omega_s^2 + \gamma^2)^6}, \\
		T_{33}^{(6)}(\omega_s) &= 1 - \frac{4 \gamma^4 \omega_s^2}{(\omega_s^2 + \gamma^2)^3} + \frac{8 \gamma^5 \omega_s^3}{(\omega_s^2 + \gamma^2)^4} + \frac{4 \gamma^8 \omega_s^4 - 24 \gamma^9 \omega_s^3 + 8 \gamma^7 \omega_s^5}{(\omega_s^2 + \gamma^2)^6}, 
	\end{align}
	\begin{align}
		T_{12}^{(6)}(\omega_s) &= T_{21}^{(6)}(\omega_s) \nonumber \\
		&= \frac{4 \gamma^4 \omega_s^2 (\gamma^6 - 4 \gamma^5 \omega_s + \gamma^4 \omega_s^2 + 6 \gamma^3 \omega_s^3 + 3 \gamma^2 \omega_s^4 + 2 \gamma \omega_s^5 + \omega_s^6)}{(\omega_s^2 + \gamma^2)^6}, \\
		T_{32}^{(6)}(\omega_s) &= T_{23}^{(6)}(\omega_s) \nonumber \\
		&= \frac{4 \gamma^4 \omega_s^2 (\gamma^6 + 4 \gamma^5 \omega_s + \gamma^4 \omega_s^2 - 6 \gamma^3 \omega_s^3 + 3 \gamma^2 \omega_s^4 - 2 \gamma \omega_s^5 + \omega_s^6)}{(\omega_s^2 + \gamma^2)^6}, \\
		T_{31}^{(6)}(\omega_s) &= T_{13}^{(6)}(\omega_s) \nonumber \\
		&= \frac{4 \gamma^8 \omega_s^4}{(\omega_s^2 + \gamma^2)^6}.
	\end{align}
	The seven-cavity system has as many as 25 and 5 elements for $T_{l j}$ and $S_l := S_{l,5}$, respectively. We hence focus on the ones that represent significant peak enhancement for the operation at the active EP7 ($g = \mu = \gamma$). They are given by
	\begin{align}
		T_{33}^{(7)}(\omega_s) &= \frac{25 \gamma^{14} - 69 \gamma^{12} \omega_s^2 + 9 \gamma^{10} \omega_s^4 + 31 \gamma^8 \omega_s^6 + 43 \gamma^6 \omega_s^8 + 21 \gamma^4 \omega_s^{10} + 7 \gamma^2 \omega_s^{12} + \omega_s^{14}}{(\omega_s^2 + \gamma^2)^7}, \\
		T_{44}^{(7)}(\omega_s) &= \frac{9 \gamma^{14} - 21 \gamma^{12} \omega_s^2 + 37 \gamma^{10} \omega_s^4 - 17 \gamma^8 \omega_s^6 + 31 \gamma^6 \omega_s^8 + 21 \gamma^4 \omega_s^{10} + 7 \gamma^2 \omega_s^{12} + \omega_s^{14}}{(\omega_s^2 + \gamma^2)^7}, \\
		T_{55}^{(7)}(\omega_s) &= \frac{9 \gamma^{14} - 9 \gamma^{12} \omega_s^2 + 33 \gamma^{10} \omega_s^4 - 9 \gamma^8 \omega_s^6 + 19 \gamma^6 \omega_s^8 + 17 \gamma^4 \omega_s^{10} + 7 \gamma^2 \omega_s^{12} + \omega_s^{14}}{(\omega_s^2 + \gamma^2)^7}, \\
		T_{34}^{(7)}(\omega_s) &= T_{43}^{(7)}(\omega_s) \nonumber \\
		&= \frac{4 \gamma^2 (4 \gamma^{12} - 11 \gamma^{10} \omega_s^2 + 13 \gamma^{8} \omega_s^4 - 8 \gamma^6 \omega_s^6 + 2 \gamma^4 \omega_s^8 + \omega_s^{12})}{(\omega_s^2 + \gamma^2)^7}, \\
		T_{35}^{(7)}(\omega_s) &= T_{53}^{(7)}(\omega_s) \nonumber \\
		&= \frac{4 \gamma^6 (4 \gamma^{8} - 7 \gamma^{6} \omega_s^2 + 2 \gamma^{4} \omega_s^4 + \gamma^2 \omega_s^6 + \omega_s^8 )}{(\omega_s^2 + \gamma^2)^7},
	\end{align}
	\begin{align}
		S_{3}^{(7)}(\omega_s) &= 2 \gamma \frac{16 \gamma^{12} - 28 \gamma^{10} \omega_s^2 + 12 \gamma^{8} \omega_s^4 - 3 \gamma^6 \omega_s^6 + 6 \gamma^4 \omega_s^8 + \gamma^2 \omega_s^{10} + \omega_s^{12}}{(\omega_s^2 + \gamma^2)^7},\\
		S_{4}^{(7)}(\omega_s) &= 2 \gamma \frac{( \gamma^{4} - \gamma^{2} \omega_s^2 + \omega_s^4)^2 (9 \gamma^4 + 3 \gamma^2 \omega_s^2 + \omega_s^{4})}{(\omega_s^2 + \gamma^2)^7}, \\ 
		S_{5}^{(7)}(\omega_s) &= 2 \gamma \frac{\gamma^{4}( \gamma^{4} - \gamma^{2} \omega_s^2 + \omega_s^4) (9 \gamma^4 + 3 \gamma^2 \omega_s^2 + \omega_s^{4})}{(\omega_s^2 + \gamma^2)^7}.
	\end{align}
	
	The extended active system with the EP14 ($g = \mu = \gamma$) exhibits responses with more complex descriptions. This limits us again to a few examples of interest for each of $T_{l j}$, $S_{l,5}$, and $S_{l,10}$ including the ones plotted in the main text.
	\begin{align} 
		T_{18}^{(14)}(\omega_s) =& T_{81}^{(14)}(\omega_s) \nonumber \\
		=& \frac{4 \gamma^{28}}{(\omega_s^2 + \gamma^2)^{14}}, \\		
		T_{33}^{(14)}(\omega_s) = & \big(169 \gamma^{28} - 1434 \gamma^{26} \omega_s^2 + 4635 \gamma^{24} \omega_s^4 - 5648 \gamma^{22} \omega_s^6 + 585 \gamma^{20} \omega_s^8 + 2 \gamma^{18} \omega_s^{10} + 1767 \gamma^{16} \omega_s^{12} + 2184 \gamma^{14} \omega_s^{14} \nonumber \\
		& + 2695 \gamma^{12} \omega_s^{16} + 2026 \gamma^{10} \omega_s^{18} + 1053 \gamma^{8} \omega_s^{20} + 376 \gamma^{6} \omega_s^{22} + 91 \gamma^{4} \omega_s^{24} + 14 \gamma^{2} \omega_s^{26} + \omega_s^{28}\big)(\omega_s^2 + \gamma^2)^{-14}, \\
		T_{34}^{(14)}(\omega_s) = & \, T_{43}^{(14)}(\omega_s) \nonumber \\
		= & \, 4 \gamma^2 \big(4 \gamma^{8} - 7 \gamma^{6} \omega_s^2 + 2 \gamma^{4} \omega_s^4 + \gamma^{2} \omega_s^6 + \omega_s^8 \big) \big(4 \gamma^{18} - 40 \gamma^{16} \omega_s^2 + 112 \gamma^{14} \omega_s^4 - 87 \gamma^{12} \omega_s^6 \nonumber \\
		& \qquad + 153 \gamma^{10} \omega_s^8 + 44 \gamma^{8} \omega_s^{10} + 27 \gamma^{6} \omega_s^{12} + 13 \gamma^{4} \omega_s^{14} + 6 \gamma^{2} \omega_s^{16} + \omega_s^{18} \big)(\omega_s^2 + \gamma^2)^{-14}, \\
		T_{36}^{(14)}(\omega_s) = & \, T_{63}^{(14)}(\omega_s) \nonumber \\
		= &\frac{4 \gamma^{12}\big(4 \gamma^{8} - 7 \gamma^{6} \omega_s^2 + 2 \gamma^{4} \omega_s^4 + \gamma^{2} \omega_s^6 + \omega_s^8 \big)^2}{(\omega_s^2 + \gamma^2)^{14}}, \\
		S_{3,5}^{(14)}(\omega_s) = & \, 2\gamma \big(100 \gamma^{26} - 799 \gamma^{24} \omega_s^2 + 2358 \gamma^{22} \omega_s^4 - 2611 \gamma^{20} \omega_s^6 + 1588 \gamma^{18} \omega_s^8 - 818 \gamma^{16} \omega_s^{10} + 98 \gamma^{14} \omega_s^{12} \nonumber \\
		& \qquad - 30 \gamma^{12} \omega_s^{14} + 310 \gamma^{10} \omega_s^{16} + 197 \gamma^{8} \omega_s^{18} + 105 \gamma^{6} \omega_s^{20} + 34 \gamma^{4} \omega_s^{22} + 8 \gamma^{2} \omega_s^{24} + \omega_s^{26} \big)(\omega_s^2 + \gamma^2)^{-14}, \\
		S_{4,5}^{(14)}(\omega_s) = & \, 2\gamma \big(36 \gamma^{26} - 384 \gamma^{24} \omega_s^2 + 1276 \gamma^{22} \omega_s^4 - 1727 \gamma^{20} \omega_s^6 + 2607 \gamma^{18} \omega_s^8 - 947 \gamma^{16} \omega_s^{10} + 988 \gamma^{14} \omega_s^{12} \nonumber \\
		& \qquad + 482 \gamma^{12} \omega_s^{14} + 406 \gamma^{10} \omega_s^{16} + 162 \gamma^{8} \omega_s^{18} + 89 \gamma^{6} \omega_s^{20} + 32 \gamma^{4} \omega_s^{22} + 8 \gamma^{2} \omega_s^{24} + \omega_s^{26} \big)(\omega_s^2 + \gamma^2)^{-14},\\
		S_{6,5}^{(14)}(\omega_s) = & \, 2\gamma \frac {\gamma^{10} (36 \gamma^{16} - 87 \gamma^{14} \omega_s^2 + 88 \gamma^{12} \omega_s^4 - 44 \gamma^{10} \omega_s^6 + 7 \gamma^{8} \omega_s^8 - 2 \gamma^{6} \omega_s^{10} + 11 \gamma^{4} \omega_s^{12} + 3 \gamma^{2} \omega_s^{14} + \omega_s^{16} )}{(\omega_s^2 + \gamma^2)^{14}}, \\
		S_{6,10}^{(14)}(\omega_s) = & \, 2\gamma \big(4 \gamma^{26} + 793 \gamma^{24} \omega_s^2 - 1878 \gamma^{22} \omega_s^4 + 2629 \gamma^{20} \omega_s^6 - 1940 \gamma^{18} \omega_s^8 + 326 \gamma^{16} \omega_s^{10} - 130 \gamma^{14} \omega_s^{12} \nonumber \\
		& \qquad + 446 \gamma^{12} \omega_s^{14} + 210 \gamma^{10} \omega_s^{16} + 177 \gamma^{8} \omega_s^{18} + 85 \gamma^{6} \omega_s^{20} + 34 \gamma^{4} \omega_s^{22} + 8 \gamma^{2} \omega_s^{24} + \omega_s^{26} \big)(\omega_s^2 + \gamma^2)^{-14}, \\ 
		S_{5,10}^{(14)}(\omega_s) = & \, 2\gamma \omega_s^2 \big(576 \gamma^{24} - 1104 \gamma^{22} \omega_s^2 + 2305 \gamma^{20} \omega_s^4 - 2277 \gamma^{18} \omega_s^6 + 1593 \gamma^{16} \omega_s^8 - 360 \gamma^{14} \omega_s^{10} \nonumber \\
		& \qquad + 10 \gamma^{12} \omega_s^{12} - 46 \gamma^{10} \omega_s^{14} + 122 \gamma^{8} \omega_s^{16} + 85 \gamma^{6} \omega_s^{18} + 36 \gamma^{4} \omega_s^{20} + 8 \gamma^{2} \omega_s^{22} + \omega_s^{24} \big)(\omega_s^2 + \gamma^2)^{-14}, \\
		S_{3,10}^{(14)}(\omega_s) = & \, S_{6,5}^{(14)}(\omega_s).
	\end{align}
	We see that the transmission spectra can also have significantly enhanced peaks at the EP14, since we trivially have $T_{33}^{(14)}(0) = 169$ and $T_{34}^{(14)}(0) = T_{43}^{(14)}(0) = T_{36}^{(14)}(0) =  T_{63}^{(14)}(0) = 64$.
	
	\section{Dynamical responses of passive and active HEPs}
	Characteristic response dynamics of the passive and active HEPs are manifested by the inverse Fourier transforms $\mathcal{F}^{-1}[T_{22}(\omega_s)](t)$ and $\mathcal{F}^{-1}[T_{51}(\omega_s)](t)$, respectively ($T_{22}(\omega_s) := T_{22}^{(6)}(\omega_s)$ and $T_{51}(\omega_s) := T_{51}^{(7)}(\omega_s)$ are shown in the main text). They are analytically given by
	\begin{align}
		\mathcal{F}^{-1}[T_{22}^{(6)}](t) &\propto \gamma e^{-\gamma t}\left(-1 -\gamma t  + \frac{16 (\gamma t)^2}{13} - \frac{(\gamma t)^3}{13} - \frac{(\gamma t)^4}{39} + \frac{(\gamma t)^5}{195} \right), \label{eq:autocorrEP6} \\ 
		\mathcal{F}^{-1}[T_{51}^{(7)}](t) &\propto \gamma e^{-\gamma t}\left(1 +\gamma t  + \frac{5 (\gamma t)^2}{11} + \frac{4(\gamma t)^3}{33} + \frac{2(\gamma t)^4}{99} + \frac{(\gamma t)^5}{495} + \frac{(\gamma t)^6}{10395} \right), \label{eq:autocorrEP7}
	\end{align}
\end{widetext}

with positive constant coefficients omitted. Because of their correspondence to the amplitude autocorrelation functions, they have direct relevance to the time evolution operator, which is solvable for nilpotent systems \cite{Bossart2021_S}. The polynomial factors in $\mathcal{F}^{-1}[T_{22}(\omega_s)](t)$ and $\mathcal{F}^{-1}[T_{51}(\omega_s)](t)$ are finite series because of the nilpotence of $\hat{\rm H}$.

\begin{figure}[t]
	\includegraphics[width=0.9\linewidth,left]{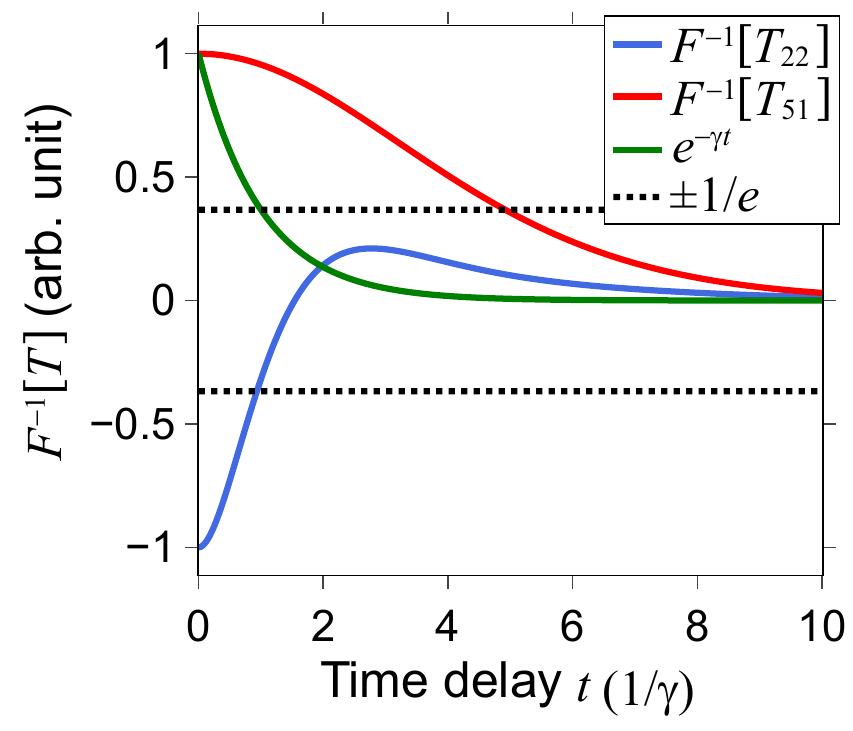}
	\caption{\label{fig:EP_response_dynamics} Transmission response dynamics of the systems with HEPs. Blue and red curves: normalized inverse Fourier transforms $\mathcal{F}^{-1}$ of the characteristic transmission spectra $T_{22}$ and $T_{51}$ for the passive EP6 and active EP7, respectively. Green curve: exponential decay $e^{-\gamma t}$ of the single cavity mode with a loss $\gamma$. Dashed lines mark $\pm 1/e$ for evaluating time constants of the dynamics. The negative initial value of $\mathcal{F}^{-1}[T_{22}](t)$ comes from the system's instant rejection, and the response peculiar to the EP6 is limited to the weak positive tail. In contrast, $\mathcal{F}^{-1}[T_{51}](t)$ exhibits both a slow onset of damping and large time constant of $\approx 4.92/\gamma$, which signify a delayed internal response fully attributed to the active EP7.}
\end{figure}

Equations (\ref{eq:autocorrEP6}) and (\ref{eq:autocorrEP7}) are plotted as the blue and red curves and contrasted in Fig. \ref{fig:EP_response_dynamics}, respectively. They are normalized by their absolute values at $t = 0$. The particularly narrowed linewidths of $T_{22}(\omega_s)$ and $T_{51}(\omega_s)$ might let us expect slow responses. However, $\mathcal{F}^{-1}[T_{22}]$ has both positive and negative terms competing with each other. Its first three terms dominate the fast response, which includes the system's instant rejection with a $\pi$ phase shift at $t = 0$ and gives a time constant of about $1/\gamma$. Hence, the major part of damping is as significant as the exponential decay for a single cavity with the loss $\gamma$ shown by the green curve. The positive terms $e^{-\gamma t}[ 16 (\gamma t)^2/13 + (\gamma t)^5/195 ]$ eventually prevail and form a long tail that exceeds the sole exponential decay for $t > 2/\gamma$. However, the signal held at this stage is weak, thereby indicating a restricted response by the passive EP6. In contrast, $\mathcal{F}^{-1}[T_{51}]$ is consistently positive and monotonically decreasing. Its damping sets in slowly and has a time constant as large as $4.92/\gamma$, because all its polynomial terms are positive and hence stand against the exponential decay factor. These features demonstrate peculiar binding of light by the minimally active EP7. We emphasize that the long-term decay rates are $\gamma$ for all the cases in Fig. \ref{fig:EP_response_dynamics}.

\section{HEPs in periodic systems}
For finding HEPs in periodic systems, we can introduce a pair of elements of $\hat{\rm H}$ such that $[\hat{\rm H}]_{j l} = [\hat{\rm H}]_{l j}^{*} \ (j \ne l)$. We study the following 5$\times$5 trial Hamiltonian
\begin{equation}
	\hat{\rm H}_{\rm 5P} = \left(\,
	\begin{matrix}
		0 & p &   &   & t  \\
		p & 0 & q &   &   \\
		& q & 0 & r &   \\
		&   & r & 0 & s \\
		t^{*}&   &   & s & 0 \\
	\end{matrix}\,\right), \label{eq:H5P} \\
\end{equation}
where $p, q, r, s, t \in \mathbb{C}$ and examine the conditions for $\hat{\rm H}_{\rm 5P}^4 \ne \hat{\rm O}$, $\hat{\rm H}_{\rm 5P}^5 = \hat{\rm O}$, and ${\rm tr} \, \hat{\rm H}_{\rm 5P}^l = 0 \ (l = 1, 2, 3, 4)$ to have an EP5. The same process as that done in Sec. \ref{sec:EPOrderReduction} lets us obtain ${\rm tr} \, \hat{\rm H}_{\rm 5P} = {\rm tr} \, \hat{\rm H}_{\rm 5P}^3 = 0$ and
\begin{align}
	{\rm tr} \, \hat{\rm H}_{\rm 5P}^2 = 0 & \iff p^2 + q^2 + r^2 + s^2 + |t|^2 = 0, \label{eq:H5PNPC1} \\
	\notag \\
	{\rm tr} \, \hat{\rm H}_{\rm 5P}^4 = 0 & \iff p^4 + q^4 + r^4 + s^4 + |t|^4 \notag \\
	& + 2 (p^2 q^2 + q^2 r^2 + r^2 s^2 + s^2 |t|^2 + |t|^2 p^2) = 0. \label{eq:H5PNPC2}
\end{align}
With Eqs. (\ref{eq:H5PNPC1}) and (\ref{eq:H5PNPC2}) at hand, we can derive
\begin{equation}
	\hat{\rm H}_{\rm 5P}^5 = 2 p q r s ({\rm Re} \, t) \, \hat{\rm I}_5, \label{eq:H5P5th} 
\end{equation}
and then confirm that $\hat{\rm H}_{\rm 5P}^4 \ne \hat{\rm O}$ and $\hat{\rm H}_{\rm 5P}^5 = \hat{\rm O}$ hold when we further have ${\rm Re} \, t = 0$ and $p, q, r, s, {\rm Im} \, t \ne 0$. Thus, it is enough to put $t = i t' \ (t' \in \mathbb{R})$ and search for an instance of $(p, q, r, s, t')$ that satisfies Eqs. (\ref{eq:H5PNPC1}) and (\ref{eq:H5PNPC2}). For example, Eq. (\ref{eq:H5PNPC2}) can be modified as
\begin{equation}
	(p^2 + q^2 + {t'}^2)^2 + (r^2 + s^2)^2 + 2(q^2 r^2 + s^2 {t'}^2) = 0, \label{eq:H5PNPC2mod} 
\end{equation}
and a simplest possibility is a combination of the parameters that nulls each term of Eq. (\ref{eq:H5PNPC2mod}). We readily see that $(p^2, q^2, r^2, s^2, {t'}^2) = (-2, 1, 1, -1, 1)$ or $(p, q, r, s, t') = (\pm \sqrt{2} i, \pm 1, \pm 1, \pm i, \pm 1)$ is one of such cases. This means that, for example, a Bloch Hamiltonian for a one-dimensional lattice
\begin{equation}
	\hat{\rm H}_{\rm 5B} = \kappa \left(\,
	\begin{matrix}
		0 & \sqrt{2}i &   &   & -e^{-i k}  \\
		\sqrt{2}i & 0 & -1 &   &   \\
		& -1 & 0 & -1 &   \\
		&   & -1 & 0 & i \\
		-e^{i k}&   &   & i & 0 \\
	\end{matrix}\,\right), \label{eq:H5B} \\
\end{equation}
where $\kappa \in \mathbb{C}$ and $k \in [-\pi, \pi)$ is the Bloch phase factor, has an EP5 when $k = \pm \pi/2$. Note that its evanescent and dissipative couplings can be switched by choosing $\kappa \in \mathbb{R}$ or $\kappa \in i \mathbb{R}$. Such a design without irrational parameters might be explored in future studies.

\section{Lower-triangular similarity transformation for 2$\times$2 systems with EPs}
Most 2$\times$2 nilpotent matrices can be similarity-transformed to their JNF ($\hat{\rm J}_{\rm H} = \hat{\rm P}^{-1} \hat{\rm H} \hat{\rm P} = \hat{\rm J}(0, 2)$) with lower-triangular matrices $\hat{\rm P}$. This fact supports the usefulness of our theorem and is shown via the following proposition.\\
\\
\textbf{Proposition 1:} {\it 2$\times$2 matrices with EPs can be specified by the components of their right eigenvectors.}\\

To prove Proposition 1 first, we consider a 2$\times$2 matrix $\hat{\rm H}_{\rm d2}$ that has an EP2 with zero eigenvalue. It is noteworthy that $\hat{\rm H}_{\rm d2} + \omega_{\rm EP} \hat{\rm I}_2$ applies equally in the case of that with an eigenvalue $\omega_{\rm EP} \in \mathbb{C}$ without changing the eigenvector. The only 2$\times$2 Jordan matrix corresponding to the EP for $\omega_{\rm EP} = 0$ is $\hat{\rm J}_{\rm EP2} = \hat{\rm J}(0, 2)$. When we denote the \textit{unnormalized} EP right eigenvector and other generalized eigenvector as $\bm{v}_{\rm EP2} = (v_1, v_2)^{\mathrm{T}} \in \mathbb{C}^2$ and $\bm{g} = (g_1, g_2)^{\mathrm{T}} \in \mathbb{C}^2$, respectively, we can construct the similarity transformation matrix 
$\hat{\rm P}_{\rm d2} = \left(\begin{smallmatrix}
	v_1 & g_1\\
	v_2 & g_2
\end{smallmatrix}\right)$
with the condition ${\rm det} \, \hat{\rm P}_{\rm d2} \ne 0$. We also assume $\bm{v}_{\rm EP2} \ne \bm{0}$, $\bm{g} \ne \bm{0}$ to avoid the trivial exception with $\hat{\rm H}_{\rm d2} = \hat{\rm O}$. By using the theorem on the existence and uniqueness of the JNF, we find that the $\hat{\rm H}_{\rm d2}$ is derived by the inverse transformation of $\hat{\rm J}_{\rm EP2}$ as
\begin{equation}
	\hat{\rm H}_{\rm d2} = \hat{\rm P}_{\rm d2} \hat{\rm J}_{\rm EP2} \hat{\rm P}_{\rm d2}^{-1} = c_{\rm d2} \left(
	\begin{matrix}
		-v_1 v_2 & v_1^2 \\
		-v_2^2 & v_1 v_2 
	\end{matrix}\right), \label{eq:IST_Hamiltonian_EP2}
\end{equation}
where $c_{\rm d2} = ({\rm det} \, \hat{\rm P}_{\rm d2})^{-1} = (v_1 g_2 - v_2 g_1)^{-1} \in \mathbb{C}$. The matrix part of Eq. (\ref{eq:IST_Hamiltonian_EP2}), namely
\begin{equation}
	\hat{\rm H}_{\rm EP2} := \left(\begin{matrix}
		-v_1 v_2 & v_1^2\\
		-v_2^2 & v_1 v_2
	\end{matrix}\right) ,\label{eq:Hamiltonian_EP2}
\end{equation}
is described by the eigenvector components $v_1$ and $v_2$ and does not depend on $g_1$ or $g_2$. In other words, any 2$\times$2 Hamiltonians having EPs with zero eigenvalue, or nilpotent matrices, can be determined by the eigenvector components up to a complex factor.

We trivially see that $\hat{\rm H}_{\rm EP2}$ also has an EP2, as confirmed by its nilpotence. Conversely, all the 2$\times$2 nilpotent matrices are exhausted by Eq. (\ref{eq:Hamiltonian_EP2}); we can provide it with an arbitrary nonzero factor $c_0 \in \mathbb{C}$ by transforming the variables as $v_1 \rightarrow \sqrt{c_0} v'_1$ and $v_2 \rightarrow \sqrt{c_0} v'_2$ with $v'_1, v'_2 \in \mathbb{C}$. We emphasize that the only constraint on $(g_1, g_2)$ is ${\rm det} \, \hat{\rm P}_{\rm d2} \ne 0$, which assures that $\bm{v}$ and $\bm{g}$ are linearly independent and hence span $\mathbb{C}^2$. $\qed$

A lower-triangular $\hat{\rm P}_{\rm d2}$ for $\hat{\rm H}_{\rm d2}$ means $g_1 = 0, \ g_2 \ne 0$. Here, $v_1 = 0$ is the condition for ${\rm det} \, \hat{\rm P}_{\rm d2} = 0$, and thus such instances of $\hat{\rm H}_{\rm d2}$ are not covered. However, they are only limited to the ones with $[\hat{\rm H}_{\rm d2}]_{1 1} = [\hat{\rm H}_{\rm d2}]_{1 2} = [\hat{\rm H}_{\rm d2}]_{2 2} = 0$, $[\hat{\rm H}_{\rm d2}]_{2 1} \ne 0$. For every other case, we can always find a lower-triangular $\hat{\rm P}_{\rm d2}$ that derives the JNF of $\hat{\rm H}_{\rm d2}$. Especially, nontrivial 2$\times$2 symmetric nipotent matrices fall into this category, since they require $v_1^2 = -v_2^2$ and hence suppose $v_1, v_2 \ne 0$.

\begin{widetext} 
	\section{Systems with maximally degenerate HEPs derived from 2$\times$2 PT-symmetric Hamiltonian} \label{sec:InductivelyDerivedSystems}
	By using Theorem 1, we further derive $16 \times 16$ and $32 \times 32$ effective Hamiltonians $\hat{\rm H}_{16{\rm I}}, \hat{\rm H}_{32{\rm I}}$ with maximally degenerate HEPs from the $2 \times 2$ PT-symmetric Hamiltonian. They read
	\begin{equation}
		\hat{\rm H}_{16{\rm I}} = 
		\left(
		\begin{matrix}
			i  & -1 &    &    &    &    &    &    &    &    &    &    &    &    &    &    \\
			-1 & 0  & -1 &    &    &    &    &    &    &    &    &    &    &    &    &    \\
			& -1 & -2i& -1 &    &    &    &    &    &    &    &    &    &    &    &    \\
			&    & -1 & 2i & -1 &    &    &    &    &    &    &    &    &    &    &    \\
			&    &    & -1 & 0  & -1 &    &    &    &    &    &    &    &    &    &    \\
			&    &    &    & -1 & -2i& -1 &    &    &    &    &    &    &    &    &    \\
			&    &    &    &    & -1 & 0  & -1 &    &    &    &    &    &    &    &    \\
			&    &    &    &    &    & -1 & 2i & -1 &    &    &    &    &    &    &    \\
			&    &    &    &    &    &    & -1 & 0  & -1 &    &    &    &    &    &    \\
			&    &    &    &    &    &    &    & -1 & 0  & -1 &    &    &    &    &    \\
			&    &    &    &    &    &    &    &    & -1 & -2i& -1 &    &    &    &    \\
			&    &    &    &    &    &    &    &    &    & -1 & 0  & -1 &    &    &    \\
			&    &    &    &    &    &    &    &    &    &    & -1 & 2i & -1 &    &    \\
			&    &    &    &    &    &    &    &    &    &    &    & -1 & -2i& -1 &    \\
			&    &    &    &    &    &    &    &    &    &    &    &    & -1 & 0  & -1 \\
			&    &    &    &    &    &    &    &    &    &    &    &    &    & -1 & i  \\
		\end{matrix} \,\right),  \label{eq:H16I}
	\end{equation}
	\begin{equation}
		\hat{\rm H}_{32{\rm I}} = 
		\left(
		\begin{smallmatrix}
			i   & -1 &    &    &    &    &    &    &    &    &    &    &    &    &    &    &	&    &    &    &    &    &    &    &    &    &    &    &    &    &    &    \\
			-1  & 0  & -1 &    &    &    &    &    &    &    &    &    &    &    &    &    &	&    &    &    &    &    &    &    &    &    &    &    &    &    &    &    \\
			& -1 & -2i& -1 &    &    &    &    &    &    &    &    &    &    &    &    &	&    &    &    &    &    &    &    &    &    &    &    &    &    &    &    \\
			&    & -1 & 2i & -1 &    &    &    &    &    &    &    &    &    &    &    &	&    &    &    &    &    &    &    &    &    &    &    &    &    &    &    \\
			&    &    & -1 & 0  & -1 &    &    &    &    &    &    &    &    &    &    &	&    &    &    &    &    &    &    &    &    &    &    &    &    &    &    \\
			&    &    &    & -1 & -2i& -1 &    &    &    &    &    &    &    &    &    &	&    &    &    &    &    &    &    &    &    &    &    &    &    &    &    \\
			&    &    &    &    & -1 & 0  & -1 &    &    &    &    &    &    &    &    &	&    &    &    &    &    &    &    &    &    &    &    &    &    &    &    \\
			&    &    &    &    &    & -1 & 2i & -1 &    &    &    &    &    &    &    &	&    &    &    &    &    &    &    &    &    &    &    &    &    &    &    \\
			&    &    &    &    &    &    & -1 & 0  & -1 &    &    &    &    &    &    &	&    &    &    &    &    &    &    &    &    &    &    &    &    &    &    \\
			&    &    &    &    &    &    &    & -1 & 0  & -1 &    &    &    &    &    &	&    &    &    &    &    &    &    &    &    &    &    &    &    &    &    \\
			&    &    &    &    &    &    &    &    & -1 & -2i& -1 &    &    &    &    &	&    &    &    &    &    &    &    &    &    &    &    &    &    &    &    \\
			&    &    &    &    &    &    &    &    &    & -1 & 0  & -1 &    &    &    &	&    &    &    &    &    &    &    &    &    &    &    &    &    &    &    \\
			&    &    &    &    &    &    &    &    &    &    & -1 & 2i & -1 &    &    &	&    &    &    &    &    &    &    &    &    &    &    &    &    &    &    \\
			&    &    &    &    &    &    &    &    &    &    &    & -1 & -2i& -1 &    &	&    &    &    &    &    &    &    &    &    &    &    &    &    &    &    \\
			&    &    &    &    &    &    &    &    &    &    &    &    & -1 & 0  & -1 &	&    &    &    &    &    &    &    &    &    &    &    &    &    &    &    \\
			&    &    &    &    &    &    &    &    &    &    &    &    &    & -1 & 2i &-1	&    &    &    &    &    &    &    &    &    &    &    &    &    &    &    \\
			&    &    &    &    &    &    &    &    &    &    &    &    &    &    & -1 & 0  & -1 &    &    &    &    &    &    &    &    &    &    &    &    &    &    \\
			&    &    &    &    &    &    &    &    &    &    &    &    &    &    &    &-1  & 0  & -1 &    &    &    &    &    &    &    &    &    &    &    &    &    \\
			&    &    &    &    &    &    &    &    &    &    &    &    &    &    &    &	& -1 & -2i& -1 &    &    &    &    &    &    &    &    &    &    &    &    \\
			&    &    &    &    &    &    &    &    &    &    &    &    &    &    &    &    &    & -1 & 2i & -1 &    &    &    &    &    &    &    &    &    &    &    \\
			&    &    &    &    &    &    &    &    &    &    &    &    &    &    &    &	&    &    & -1 & 0  & -1 &    &    &    &    &    &    &    &    &    &    \\
			&    &    &    &    &    &    &    &    &    &    &    &    &    &    &    &	&    &    &    & -1 & -2i& -1 &    &    &    &    &    &    &    &    &    \\
			&    &    &    &    &    &    &    &    &    &    &    &    &    &    &    &	&    &    &    &    & -1 & 0  & -1 &    &    &    &    &    &    &    &    \\
			&    &    &    &    &    &    &    &    &    &    &    &    &    &    &    &	&    &    &    &    &    & -1 & 0  & -1 &    &    &    &    &    &    &    \\
			&    &    &    &    &    &    &    &    &    &    &    &    &    &    &    &	&    &    &    &    &    &    & -1 & 2i & -1 &    &    &    &    &    &    \\
			&    &    &    &    &    &    &    &    &    &    &    &    &    &    &    &	&    &    &    &    &    &    &    & -1 & 0  & -1 &    &    &    &    &    \\
			&    &    &    &    &    &    &    &    &    &    &    &    &    &    &    &	&    &    &    &    &    &    &    &    & -1 & -2i& -1 &    &    &    &    \\
			&    &    &    &    &    &    &    &    &    &    &    &    &    &    &    &	&    &    &    &    &    &    &    &    &    & -1 & 0  & -1 &    &    &    \\
			&    &    &    &    &    &    &    &    &    &    &    &    &    &    &    &	&    &    &    &    &    &    &    &    &    &    & -1 & 2i & -1 &    &    \\
			&    &    &    &    &    &    &    &    &    &    &    &    &    &    &    &	&    &    &    &    &    &    &    &    &    &    &    & -1 & -2i& -1 &    \\
			&    &    &    &    &    &    &    &    &    &    &    &    &    &    &    &	&    &    &    &    &    &    &    &    &    &    &    &    & -1 & 0  & -1 \\
			&    &    &    &    &    &    &    &    &    &    &    &    &    &    &    &	&    &    &    &    &    &    &    &    &    &    &    &    &    & -1 & i  \\
		\end{smallmatrix}\,\right). \label{eq:H32I}
	\end{equation} 
	It is noteworthy that adding a small perturbation $\epsilon$ to their $(1,1)$ elements result in eigenvalues scaling with $\epsilon^{1/16}$ and $\epsilon^{1/32}$. We can trivially confirm that both $\hat{\rm H}_{16{\rm I}}$ and $\hat{\rm H}_{32{\rm I}}$ only have uniform nearest-neighbor couplings $-1$, on-site gain and loss with the twice magnitude $\pm 2i$, and the loss $i$ balanced with the coupling for the edge sites as nonzero elements. It is also straightforward to obtain larger systems of this series, namely $\{\hat{\rm H}_{64{\rm I}}, \hat{\rm H}_{128{\rm I}}, \dots \}$. However, we omit them since none of them is found to fit on a single page without reducing the font size too much.
\end{widetext}

\section{Responses of HEPs estimated by resolvent}
\begin{figure}[t]
	\includegraphics[width=0.95\linewidth,center]{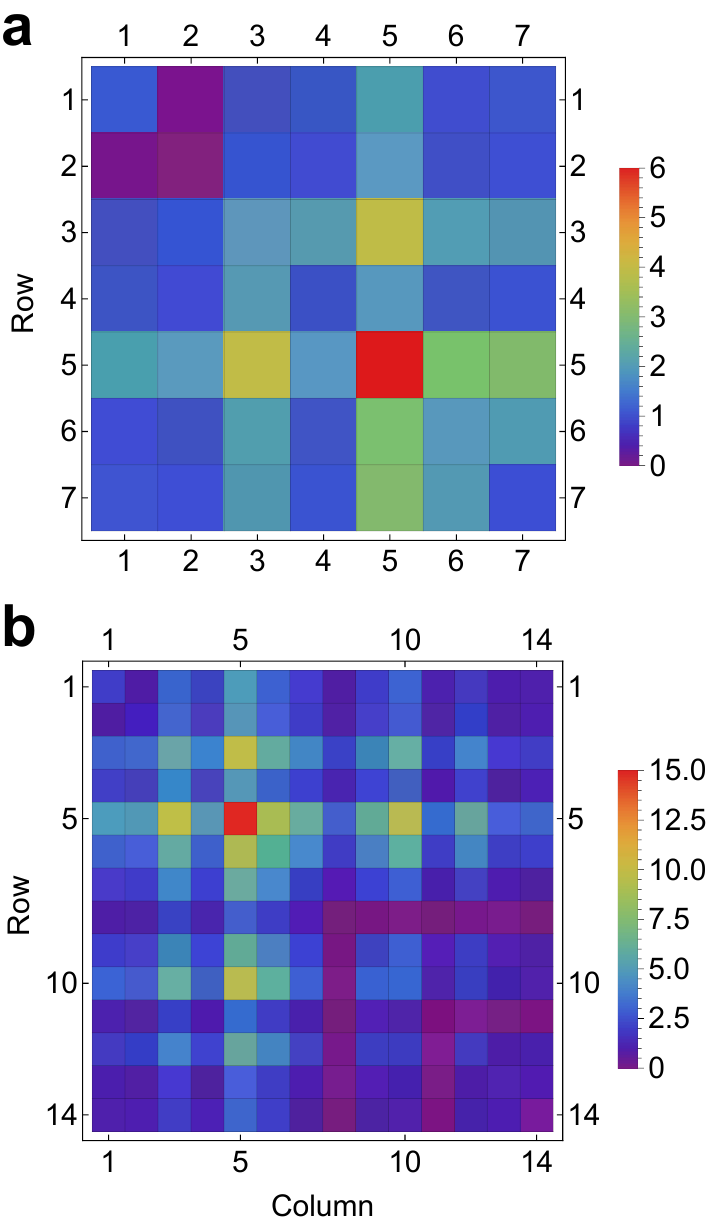}
	\caption{\label{fig:resolvent} Absolute values of resolvent elements, (a) $\{|[\hat{\rm G}(0, \hat{\rm H}_{\rm 7A})]_{j k}|\}$ and (b) $\{|[\hat{\rm G}(0, \hat{\rm H}_{\rm 14A})]_{j k}|\}$, evaluating the intra-cavity responses of the 7- and 14-cavity systems shown in Figs. 3(b) and 4(c) operating at an EP7 and EP14, respectively, to resonant excitation ($\omega_s = 0$).}
\end{figure}

The resolvent $\hat{\rm G}(\omega_s, \hat{\rm H}) = (\omega_s \hat{\rm I} - \hat{\rm H})^{-1}$ directly gives the system's response to excitation in the spectral domain. It never diverges in our analysis, since we consider steady excitation $\omega_s \in \mathbb{R}$ and the eigenvalues of our HEPs all have an imaginary part $\gamma \ne 0$ due to the uniform loss. For reference resonant excitation ($\omega_s = 0$) of our coupled photonic lattices, ${\rm Re} \ \hat{\rm G}(0, \hat{\rm H})$ just has nondiagonal nonzero elements denoting nonlocal system responses via couplings. In contrast, ${\rm Im} \ \hat{\rm G}(0, \hat{\rm H})$ can also contain finite diagonal elements, which solely measure the net in-situ responses to the excitation and reflect directly the LDOS \cite{Pick2017gen_S}. Figure \ref{fig:resolvent}(a) and (b) show the absolute values of the resolvent elements $\{|[\hat{\rm G}(0, \hat{\rm H}_{\rm 7A})]_{j k}|\}$ and $\{|[\hat{\rm G}(0, \hat{\rm H}_{\rm 14A})]_{j k}|\}$ for our source and extended systems with the active EP7 and EP14, respectively, aggregating the impact of both their real and imaginary parts. For each of $\hat{\rm H}_{\rm 7A}$ and $\hat{\rm H}_{\rm 14A}$, the local response of cavity 5 is dominant, and it obviously stems from the interplay of the HEP and amplification $g$. As seen in Fig. \ref{fig:resolvent}(a), $\hat{\rm G}(0, \hat{\rm H}_{\rm 7A})$ has visible nondiagonal elements indicating significant contribution of the active cavity, such as (3,5), (6,5), (7,5) and their index-flipped ones. Although cavities 3 and 6 also have some indirect impact of gain, many other cases rather result in restricted responses, i.e. limited element values.

Such input and output selectivity of the peculiar EP responses becomes more drastic in the 14-cavity array, namely Fig. \ref{fig:resolvent}(b). While the upper diagonal block of $|\hat{\rm G}(0, \hat{\rm H}_{\rm 14A})|$ looks quite similar to $|\hat{\rm G}(0, \hat{\rm H}_{\rm 7A})|$ as mentioned in the main text, the corresponding responses are further enhanced because of the larger EP order. In contrast, most elements of its lower diagonal block are suppressed, supporting the observation that the evanescent coupling of cavity 8 induces dark responses within the latter half of the array for $\omega_s = 0$. Nonetheless, the resolvent retains some moderate elements in the upper-right and lower-left blocks, such as $[\hat{\rm G}(0, \hat{\rm H}_{\rm 7A})]_{5 \, 10}$ and $[\hat{\rm G}(0, \hat{\rm H}_{\rm 7A})]_{10 \, 5}$, suggesting ways for the light to pass through both the upper and lower parts of the system.

\end{document}